\documentclass{article}

\PassOptionsToPackage{numbers}{natbib}

\usepackage[final]{neurips_2025}
\usepackage[inkscapelatex=false]{svg}

\usepackage[utf8]{inputenc} 
\usepackage[T1]{fontenc}    
\usepackage{hyperref}      
\usepackage{url}           
\usepackage{booktabs}      
\usepackage{amsfonts}      
\usepackage{nicefrac}       
\usepackage{microtype}    
\usepackage{xcolor}        
\usepackage{graphicx}
\usepackage{subcaption}
\usepackage{array}
\usepackage{bm}
\usepackage{siunitx}
\graphicspath{{figs/}}
\DeclareGraphicsExtensions{.pdf,.png,.jpg,.jpeg}

\title{In Silico Mapping of Visual Categorical Selectivity Across the Whole Brain}

\author{
  Ethan Hwang\\
  Zuckerman Mind Brain Behavior Institute\\
  Columbia University\\
  \texttt{eh2976@columbia.edu} \\
  \And
  Hossein Adeli \\
  Zuckerman Mind Brain Behavior Institute\\
  Columbia University\\
  \texttt{ha2366@columbia.edu}
  \And
  Wenxuan Guo \\
  Zuckerman Mind Brain Behavior Institute\\
  Columbia University\\
  \texttt{wg2361@columbia.edu}
  \And
  Andrew Luo \\
  University of Hong Kong\\
  \texttt{aluo@hku.hk}
  \And
  Nikolaus Kriegeskorte \\
  Zuckerman Mind Brain Behavior Institute\\
  Columbia University\\
  \texttt{nk2765@columbia.edu}
}

\begin{document}

\maketitle

\begin{abstract}
A fine-grained account of functional selectivity in the cortex is essential for understanding how visual information is processed and represented in the brain. Classical studies using designed experiments have identified multiple category-selective regions; however, these approaches rely on preconceived hypotheses about categories. Subsequent data-driven discovery methods have sought to address this limitation but are often limited by simple, typically linear encoding models. We propose an in silico approach for data-driven discovery of novel category-selectivity hypotheses based on an encoder–decoder transformer model. The architecture incorporates a brain-region to image-feature cross-attention mechanism, enabling nonlinear mappings between high-dimensional deep network features and semantic patterns encoded in the brain activity. We further introduce a method to characterize the selectivity of individual parcels by leveraging diffusion-based image generative models and large-scale datasets to synthesize and select images that maximally activate each parcel. Our approach reveals regions with complex, compositional selectivity involving diverse semantic concepts, which we validate in silico both within and across subjects. Using a brain encoder as a “digital twin” offers a powerful, data-driven framework for generating and testing hypotheses about visual selectivity in the human brain—hypotheses that can guide future fMRI experiments. Our code is available at: \href{https://kriegeskorte-lab.github.io/in-silico-mapping-web/}{https://kriegeskorte-lab.github.io/in-silico-mapping/}.
\end{abstract}

\section{Introduction}

Over the past few decades, researchers have extensively studied the visual hierarchy in the brain, from early cortical areas that encode low-level features to higher-level regions that represent categorical information. Neuroimaging experiments, especially studies using functional magnetic resonance imaging (fMRI), have revealed specialized cortical regions for faces, places, words, bodies, and food \citep{mccarthy1997face,kanwisher1997fusiform,allison1994human,sergent_functional_1992,cohen_visual_2000,baker_visual_2007,epstein1998cortical,downing2001cortical,schwarzlose_separate_2005,simmons_pictures_2005,Jain2023,khosla2022,van_der_laan_first_2011,peelen_selectivity_2005}. However, visual perception is more nuanced than this short list of categories. It remains unknown what additional visual concepts have dedicated regions that enable humans to make sense of the complex world. Common mapping methods depend on experimenter-curated concepts, and empirically-driven alternatives require more data and expensive fMRI experiments. For example, one could present a subject with a large set of images and then label the selectivity of each cortical parcel by the images that elicit the strongest mean parcel response. As the stimulus set grows, however, the cost of data acquisition (operating the scanner, paying subjects, bonuses)---at least with current fMRI technology and experimental paradigms---may prove prohibitive. We propose methods that rely on the state-of-the-art encoding models to generalize to concepts beyond the stimuli for which fMRI responses have been measured.  

\begin{figure}[h]
    \centering
    \includegraphics[width=1\linewidth]{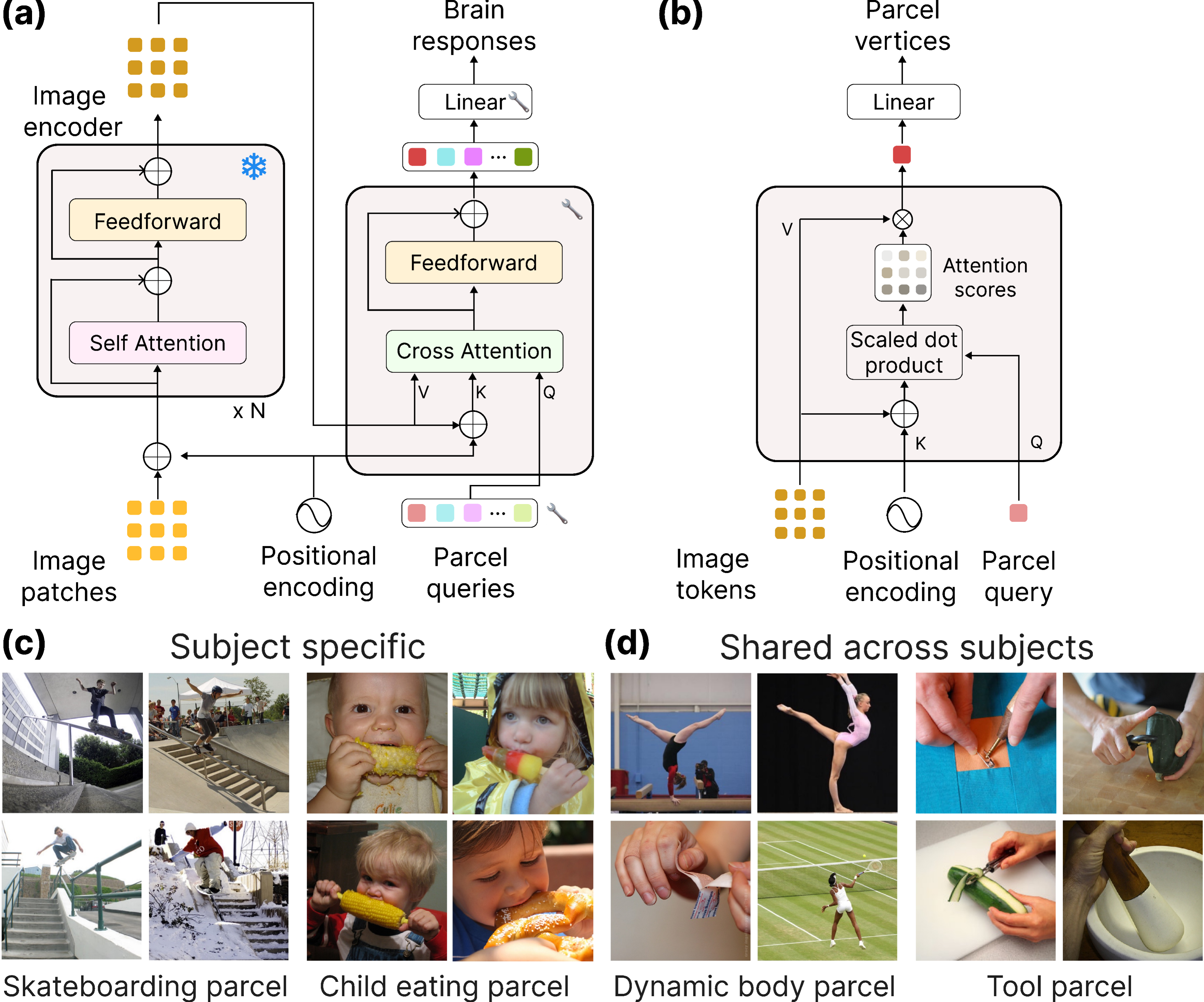}
    \caption{\textbf{Brain encoder architecture and images predicted to maximally activate selected parcels outside the visual cortex.} \textbf{(a)} Brain encoder architecture. \textbf{(b)} Cross-attention for parcel fMRI prediction. \textbf{(c)} A brain encoder ranks ImageNet \citep{deng_imagenet_2009} images by how much each image would activate a parcel. Images from two sample parcels, four images from a single subject are shown. \textbf{(d)} Images from two sample parcels, one image from each of the four subjects is shown for each parcel. All images are curated from the top 25, see Appendix~\ref{app:teaser-locations} for the parcel locations and full collage.}
    \label{fig:teaser}
\end{figure}

Our encoding model leverages recent advances in AI and large-scale neural datasets to serve as a “digital twin,” upon which we perform extensive in silico experimentation \citep{gifford2025silicodiscoveryrepresentationalrelationships,Guclu10005} to generate hypotheses of complex categorical selectivity beyond the visual cortex. The encoding model enables us to predict the neural activity of image sets far larger and more diverse than the limited number of images shown to the participant during fMRI, effectively expanding the search space for optimal visual stimuli. In addition, the encoding model is fully observable and differentiable, enabling interpretability queries using attention maps and gradient-based analyses, such as diffusion-based models that use gradients to find stimuli that elicit high activity \citep{luo_brain_2023}.

Through experimentation with large image datasets on our model, we generated hypotheses for the selectivity of many parcels outside the visual area, as well as optimal sets of images that maximally activate the parcels. These hypotheses can be tested in targeted future fMRI experiments by showing only the sets of optimal stimuli, accelerating data collection and experiment iteration, and lowering the cost of data acquisition. We demonstrate that our pipeline using brain encoders can test concepts that were not explicitly shown to the subject in the scanner, effectively enriching the diversity and size of the fMRI training set.

Key contributions of our work:

\begin{enumerate}
    \item \textbf{Massive scale:} applying in silico mapping on millions of images (ImageNet, BrainDIVE) with a transformer-based brain encoder, enabling discovery of parcel selectivity for concepts never shown in training. To the best of our knowledge, no other study has been done on this scale.
    \item \textbf{Mapping of the whole brain:} expanding beyond visual cortex and revealing human-specific semantic selectivity.
    \item \textbf{In silico verification:} our pipeline verifies selectivity hypotheses in silico with rigorous tests that evaluate how well a label can predict ground-truth activation on a held-out set within and across subjects. 
    \item \textbf{New fMRI experimental paradigm:} as datasets grow and encoding models improve, our pipeline offers a way to leverage these advances to accelerate and improve the accuracy of whole-brain mapping.
\end{enumerate}

\section{Related Work}

\paragraph{Semantic mapping.} Our work builds upon a growing body of computational modeling and machine learning research that investigates how semantic information is represented in the higher visual cortex \citep{Kriegeskorte2008matching}. Some approaches leverage large image datasets to build decoders \citep{doerig2022semantic,ferrante2023brain,takagi2022high,kay2008identifying,scotti2024mindeye2,yeung2024neural,chen2022seeing,liu2023brainclip} or models for generating optimal stimuli \citep{Cerdas2024.10.29.620889, luo_brain_2023,luo2024brainscuba,gu_neurogen_2022,ratan2021computational,matsuyama_lavca_2025}, while others use cross-domain (e.g. vision-language) mapping \citep{huth_continuous_2012,huth_natural_2016,kell_task-optimized_2018,barros-loscertales_reading_2012,nierhaus_content_2023,okumura_semantic_2024,pereira_toward_2018,simanova_modality-independent_2014, horikawa_neural_2020}. These studies face a key challenge of dataset size, since the collection of neural data is often expensive and time-consuming. Our work seeks to address this constraint by using encoders trained on large-scale datasets to perform in silico mapping. This allows us to expand the set of concepts that can be probed, beyond those stimuli shown to the subjects. 

\paragraph{Brain encoding models.} Highlighting its importance, several community-driven efforts have sought to benchmark models predicting brain responses \citep{schrimpf2018brain,turishcheva2024dynamic,gifford2023algonauts}. With the release of increasingly large neural datasets each year, researchers have introduced novel architectures and methodologies to improve the accuracy of brain encoding model, including leveraging multiple datasets and pretrained networks \citep{adeli_predicting_2023,Li2022.11.06.515387,yang2024brain,beliy2025wisdomcrowdbrainsuniversal,Yamins2016,qian2024fmripte,Freteault_alignment_2025,kriegeskorte2015deep,tang_jerry_brain_encoding,azabou2023unifiedscalableframeworkneural}. While our paper uses a cutting-edge encoding model \citep{adeli_transformer_2025}, our pipeline is ultimately encoder-agnostic, and can use any encoder that is image-computable. As researchers build better brain encoders, we expect that the space of hypotheses our pipeline could generate and their accuracy will only grow.

\paragraph{Brain-optimized stimuli.} Previous studies have introduced encoding model-based stimulus selection and empirically validated the superstimuli in non-human primates and mice \citep{bashivan2019neural, walker_neural_2020, pierzchlewicz2023energy, franke2025dual}. In this work, we extend this general approach of stimulus optimization for studying neural populations to the fMRI domain, revealing high-level human-specific selectivity beyond the visual cortex.

\section{Methods}
\label{sec:methods}

Our goal is to map the visual selectivity of parcels across the whole brain. First, we train a brain encoder to predict fMRI responses from natural scene images. Then we select visually responsive and robust parcels for experimentation to determine categorical selectivity.

\subsection{Parcellation Strategy}

\begin{figure}[h]
    \centering
    \includegraphics[width=1\linewidth]{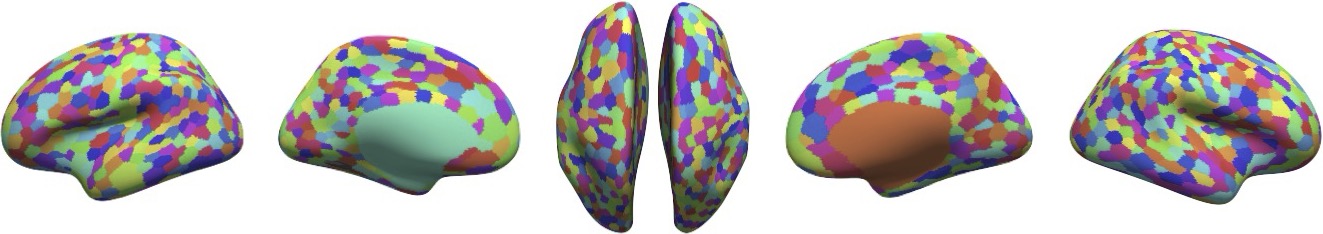}
    \caption{Schaefer-1000 parcellation}
    \label{fig:schaefer}
\end{figure}

We partitioned the 327,684 cortical vertices across the whole brain into 1,000 functional parcels using the Schaefer resting-state functional connectivity parcellation (see Figure~\ref{fig:schaefer}) \cite{schaefer_local-global_2018}.

\subsection{Brain Encoder Architecture}

Extending the work of \citet{adeli_predicting_2023}, our brain encoder predicts vertex-wise fMRI activity across the whole brain from an input image (see Figure~\ref{fig:teaser}a). An image encoder backbone (see Section~\ref{subsec:model-comparison}) first extracts patch embeddings from the image. A transformer decoder uses parcel-specific, learnable queries to attend to relevant patch embeddings via cross-attention. The decoder consists of a single transformer layer with cross-attention followed by a feedforward projection (see Figure~\ref{fig:teaser}b for cross-attention). Each output decoder token is linearly mapped to predict the fMRI responses for vertices in the corresponding parcel, and predictions from all output tokens are aggregated to obtain a whole-brain prediction. Parcel queries, the transformer decoder, and linear mappings are optimized using Adam \citep{kingma2014adam} to minimize the mean squared error between the predicted and actual fMRI responses. All other layers, including the backbone, are frozen. Separate models are trained for each subject.

To improve prediction accuracy, we ensemble multiple instances of the brain encoder. For each subject, we trained two random seeds with features from four different DINOv2 backbone layers (the 0th, 2nd, 4th, and 6th layers from the last). To predict a vertex, we take the weighted average across model predictions, scaled by softmax weights from validation set accuracy for each model on that vertex.

\subsection{Parcel selection process for further experimentation}
\label{subsec:parcel-selection}

Not all regions in the brain are visually responsive, so we selected parcels for further experimentation that satisfy three criteria: \textbf{(1) Location:} Fewer than 10\% of parcel vertices overlap with the labeled area, since we are most interested in parcels beyond the visual cortex. \textbf{(2) Visual responsiveness:} The average noise ceiling must be in the top 25\% of parcels that satisfy (1). \textbf{(3) Model prediction accuracy:} The average prediction accuracy must be in the top 25\% of parcels that satisfy (1).

Since the data quality varies slightly across subjects, we used percentiles rather than numerical cutoffs. The selection process for parcels in subject 1's left hemisphere is shown in Figure~\ref{fig:parcel_selection} in Appendix~\ref{app:parcel-selection-process}. Cutoffs are determined separately for each hemisphere to maintain comparable parcel counts.

Among the 500 Schaefer parcels in each hemisphere for every subject, $409\pm 6$ parcels satisfy condition (1). After filtering out parcels with low mean noise ceiling or low mean model prediction accuracy, $179 \pm 12$ parcels per subject are chosen for further experimentation.

\subsection{Superstimulus Generation Process} 

We choose images that maximally activate (mean z-scored beta values) a parcel of interest using three different methods:
\begin{enumerate}
    \item \textbf{Natural Scenes Dataset (NSD) Ground Truth Images:} Images from the held-out NSD \citep{allen_massive_2022} test split, ranked based on ground truth data.
    \item \textbf{Diffusion-generated superstimuli:} BrainDIVE \cite{luo_brain_2023} uses a generative backbone guided by gradients from a brain encoder to generate images that can maximally activate specified brain parcels. We generated 400 images per parcel and reranked them with the brain encoder.
    \item \textbf{Encoder-selected ImageNet superstimuli:} ImageNet \cite{deng_imagenet_2009} images that maximally activate the parcel, according to the encoder.
\end{enumerate}

\section{Experiments}
\label{sec:experiments}

\subsection{Setup}

We used the NSD \citep{allen_massive_2022}, the largest fMRI dataset to date, with 7T fMRI responses from 8 subjects who each viewed up to 10,000 distinct natural scenes. Each image is presented up to three times, and our model is trained on the neural response averaged over the presentations. We report results for subjects who completed all NSD scan sessions (1, 2, 5, and 7), though we observed comparable results in all subjects. fMRI responses were preprocessed according to \citep{allen_massive_2022}. The resulting beta estimates were centered to zero mean and scaled to unit variance before training and experiments. ROI labels were obtained from NSD. V1--hV4 ROIs are derived from a pRF experiment; body-, face-, place-, word-selective ROIs are derived from a fLoc experiment \citep{allen_massive_2022}.

\subsection{Brain encoder results}
\label{subsec:model-comparison}

\begin{figure}[htbp]
    \centering
    \includegraphics[width=0.5\linewidth]{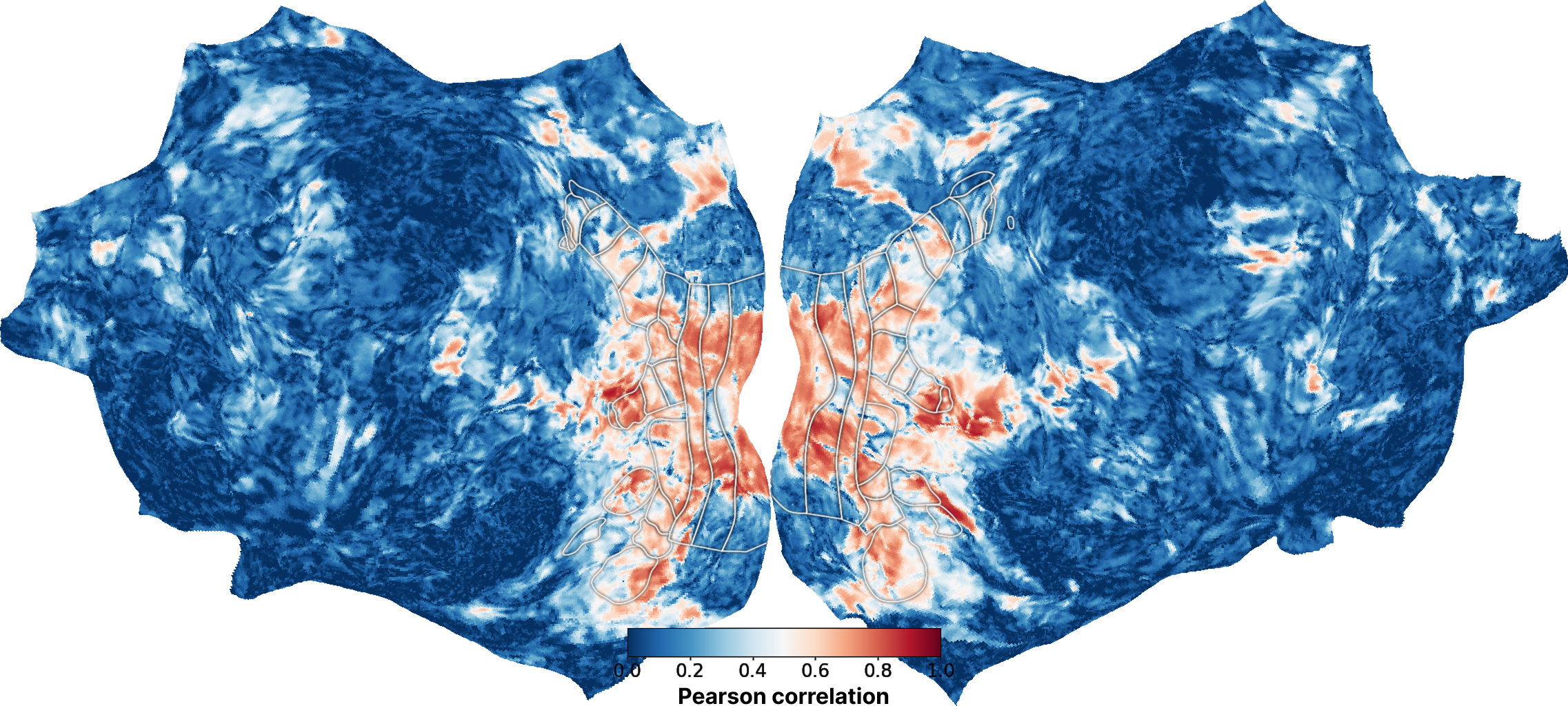}
    \caption{\textbf{Brain encoder prediction accuracy.} Pearson correlation between model predictions and ground truth data for subject 1 on the held-out test set.}
    \label{fig:test_corr}
\end{figure}

\paragraph{Model prediction accuracy.} Figure~\ref{fig:test_corr} shows the performance of the ensemble model with DINOv2 (ViT-B) backbone and the transformer cross-attention mapping function for subject 1 projected onto the cortical surface using Pycortex \cite{gao_pycortex_2015}. As expected, the model performs well on predicting the activity in the visual cortex (the area in the center of the flatmap), but also on several regions beyond the typical visual pathways.

\paragraph{Model comparison.} We evaluate several encoding models with different backbones and mapping functions to find the most suitable model for whole-brain voxel-wise prediction. The Algonauts 2023 challenge \citep{gifford2023algonauts} leaderboard showed that transformer-based backbones generally outperformed other model families such as convolutional networks (CNNs) at predicting fMRI activity. Therefore, we focused on evaluating features from several transformer backbones paired with either transformer attention-based mapping to neural data  \citep{adeli_predicting_2023} or a parameter-matched model that linearly maps the CLS token to vertex values. For each architecture, we used an ensemble of 16 models for each subject (4 feature backbone layers x 2 hemispheres x 2 random seeds). We compared the encoding accuracy, which is Pearson's correlation on the held-out set corrected for noise ceilings (see \cite{allen_massive_2022} Methods, Noise ceiling estimation).

\begin{table}[h]
  \centering
  \caption{Brain encoder encoding accuracy using different architectures}
  \label{tab:backbone_enc}
  \begin{tabular}{lcccccc}
    \toprule
    Architecture & S1 & S2 & S5 & S7 & Backbone + Mapping fn size (M) \\
    \midrule
    DINOv2 (ViT-B) + Linear & 0.33 & 0.34 & 0.39 & 0.33 & $\sim 87 + 252$ \\
    CLIP vision + Transf. & 0.44 & 0.41 & 0.46 & 0.43 & $\sim304 + 258$ \\
    RADIOv2.5-H + Transf. & 0.26 & 0.35 & 0.34 & 0.30 & $\sim 652 + 258$  \\
    DINOv2 (ViT-G w/reg) \\+ Transf. & \textbf{0.45} & \textbf{0.43} & \textbf{0.48} & \textbf{0.45} & $\sim 1136 + 258$  \\
    DINOv2 (ViT-B) + Transf. & \textbf{0.45} & \textbf{0.43} & \textbf{0.48} & 0.43 & $\sim 87 + 258$ \\
    \bottomrule
  \end{tabular}
\end{table}

\begin{figure}[htbp]
    \centering
    \includegraphics[width=1\linewidth]{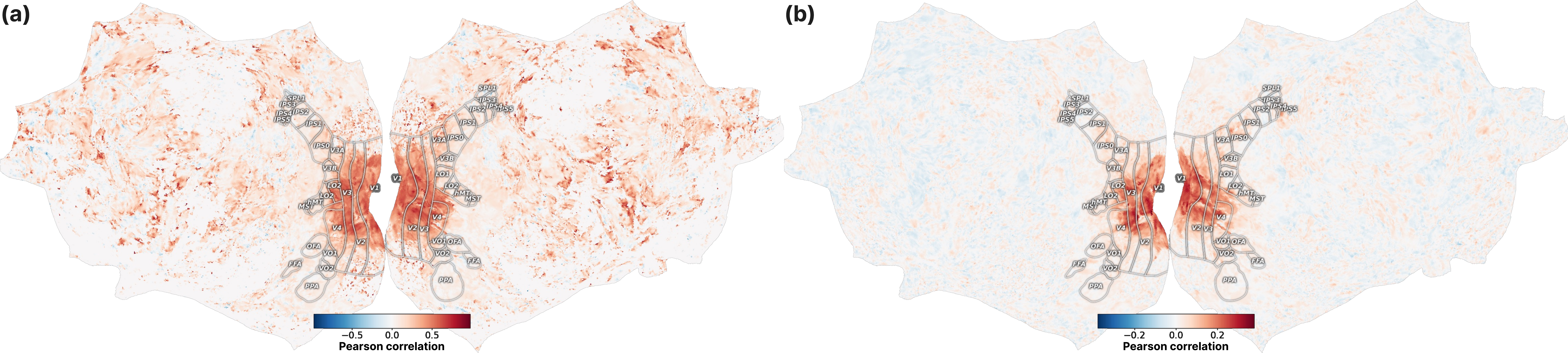}
    \caption{\textbf{(a) DINOv2 Transformer vs. Linear head-to-head comparison.} Difference in prediction accuracy for subject 1 between DINOv2 (ViT-B) with transformer vs. linear mapping functions ($>0$ or red means transformer is better). \textbf{(b) DINOv2 vs. CLIP head-to-head comparison.} Difference in prediction accuracy for subject 1 between DINOv2 (ViT-B) and CLIP vision ($>0$ or red means DINOv2 is better).}
    \label{fig:model_h2h_comp}
\end{figure}

Across the subjects that completed all NSD scan sessions (Table \ref{tab:backbone_enc}), the transformer models with the DINOv2 backbone \cite{oquab_dinov2_2024} outperform the other non-DINOv2 backbones. The two DINOv2 backbones perform similarly, which is consistent with past work \citep{adeli2023affinitybasedattentionselfsupervisedtransformers} that showed diminishing or worse performance for DINOv2 models larger than the base (ViT-B). The transformer-based encoder significantly outperformed the linear baseline, leveraging the attention mechanism to flexibly route information \citep{adeli_predicting_2023, azabou2023unified, pierzchlewicz2023energy}. Figure~\ref{fig:model_h2h_comp}a plots the difference in prediction accuracy (Pearson's correlation) between the two.

Figure~\ref{fig:model_h2h_comp}b plots the difference in prediction accuracy between DINOv2 and CLIP vision. While DINOv2 performs far better in the visual cortex, they perform similarly in many regions outside the visual area. For the rest of the paper, we performed our experiments on the brain encoder with DINOv2 (ViT-B) + Transformer, since it offered the best overall speed-performance tradeoff. 

\subsection{Sanity check on known regions}

\begin{figure}[h]
    \centering
    \includegraphics[width=1\linewidth]{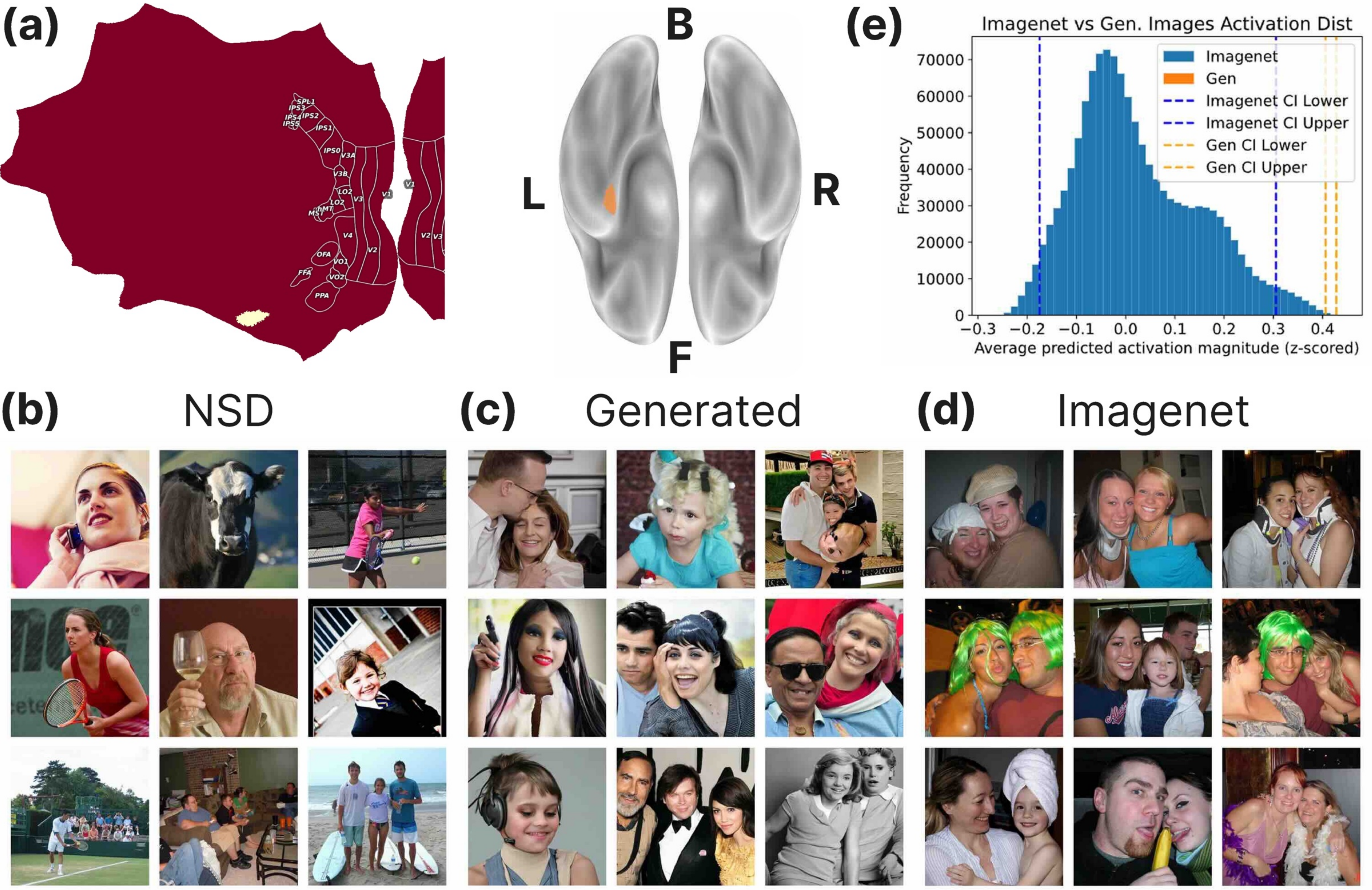}
    \caption{\textbf{Verifying the selectivity of aTL-faces.} \textbf{(a)} The location of the parcel. The inflated cortical surface (ventral view), with the left hemisphere on the left. \textbf{(b)} Held-out NSD images that maximally activate the parcel (based on ground-truth fMRI). \textbf{(c)} BrainDIVE generated and re-ranked images optimized for the parcel. \textbf{(d)} Maximally-activating images from ImageNet according to the encoder. \textbf{(e)} Distribution of predicted parcel activation for all of ImageNet images compared to images in \textbf{(c)}.}
    \label{fig:s1_aTL-faces_0}
\end{figure}

We first validate our paradigm by replicating the well-documented category selectivity of ventral pathway categorical areas. We show the results for a sample parcel from the labeled area, one which significantly overlaps with aTL-faces (47.6\% of the vertices in the parcel overlap with aTL-faces).
As shown in Figure~\ref{fig:s1_aTL-faces_0}, the images maximally activating the parcel overlapping with aTL-faces prominently feature faces, which agrees with previous work on the selectivity of this area \citep{sergent_functional_1992}. In Appendix \ref{app:superstimuli-known-areas}, we reproduce the selectivity of body-, place-, and word-selective areas. In Appendix \ref{app:reproducing-visual-hierarchy}, we show that stimuli generated by BrainDIVE reproduce the fine-to-coarse visual hierarchy progressing from V1–V4 to FFA.

\subsection{Extension to unlabeled areas}

We now extend our analyses to parcels beyond the visual cortex. We show images selected by the encoder to maximally activate each parcel for qualitative evaluation, then verify that our labels indeed explain parcel activations using formal statistical tests.

\paragraph{Labeling unlabeled parcels.} We apply the same image-selection method to generate selectivity hypotheses for unlabeled parcels. Figure~\ref{fig:teaser}c displays ImageNet images predicted by the encoder to maximally activate two unlabeled subject-specific parcels: one appears to depict skateboarding, the other a child eating food. Full results, including parcel location and the corresponding NSD and BrainDIVE images, are shown in Figures~\ref{fig:skateboarding_full} and \ref{fig:corn_eating_full} (Appendix~\ref{app:teaser-locations}). Many parcels with selectivity that appeared to be consistent and complex were found across the whole brain, suggesting that there exist regions in the brain that respond to more complex concepts than the basic categories labeled in visual cortex.

\label{subsec:evaluating-our-labels}
\paragraph{Evaluating our labels.} Our sanity checks demonstrated that our brain encoder retrieves images that align with the category selectivity of the labeled parcels. However, since the parcels we are interested in are outside the visual area and therefore unlabeled, we would like to quantitatively show that the images selected by our encoder faithfully reflect underlying neural selectivity.

We conducted two tests against the NSD test set—the only held-out fMRI data not used during training. If the selected images (hereafter “labels”) genuinely reflect parcel selectivity, other images with greater semantic similarity to the label should result in higher parcel activation. Prior studies have shown that categorical selectivity is not binary but graded across diverse stimuli \citep{BURNS2019209,Mur8649}. For example, fMRI responses to 1,705 object and action categories revealed that brain-response similarity strongly correlates with semantic similarity \citep{huth_continuous_2012}. Accordingly, we expect a parcel’s mean activity to scale with how semantically close an image is to that parcel’s preferred concept, as defined by our encoder’s maximally activating examples.

In each test, we compared two predictors: (1) labels derived from our encoder, and (2) a baseline that forms its hypothesis from each parcel’s most activating NSD-training images, selected using their measured responses. This baseline mirrors conventional fMRI studies that rely solely on stimuli shown in the scanner. If our encoder performs better than the baseline training set, it shows we can use brain encoders to discover novel concepts not explicitly shown to the subject.

For Test 1, we asked whether semantic similarity to a label predicted the activation rank order of NSD test images better than chance. Our pipeline represents a label as the mean CLIP embedding of the top 32 ImageNet or BrainDIVE images (both shown separately); the baseline uses the top 32 NSD-training images. Appendix \ref{app:stats-tests} details the procedure. Table \ref{tab:rank_corr_vs_null} reports the number of parcels for which each label significantly outperforms a random ranking. To ensure that the results reflect an expansion of the stimulus space and not merely dense sampling near peak activation, Appendix \ref{app:topk} reports the same results with a varying number of top images used to create the label. Each parcel whose model-derived label outperforms the null is deemed successfully labeled. Across tests, the encoder-selected stimuli generally outperformed the baseline, labeling a greater number of parcels.

\begin{table}[h]
  \centering
  \caption{Fraction of parcels whose model-derived label predicts parcel activation rankings significantly better than chance ($p<0.05$, FDR corrected)}
  \label{tab:rank_corr_vs_null}
  \begin{tabular}{lccccc}
    \toprule
    Model type & S1 & S2 & S5 & S7 \\
    \midrule
    NSD train & \bm{$150/181$} & $163/192$ & $136/175$ & $155/196$ \\
    Our encoder w/ImageNet & $139/181$ & $167/192$ & $130/175$ & $150/196$ \\
    Our encoder w/BrainDIVE & $135/181$ & \bm{$170/192$} & \bm{$139/175$} & \bm{$156/196$} \\
    \bottomrule
  \end{tabular}
\end{table}

For Test 2, we compare Spearman's rank correlations between the ground-truth activation ordering and each model’s predicted ordering, quantifying which model’s selectivity hypothesis best explains parcel activations. Table \ref{tab:spearman_rank} summarizes these coefficients across all parcels. A head-to-head comparison between our encoder and the baseline appears in Table \ref{tab:h2h_nsd_train_baseline} (Appendix \ref{app:h2d-comp}).

\begin{table}[h]
  \centering
  \small
  \setlength{\tabcolsep}{4pt}
  \caption{Spearman's $\rho$ (mean $\pm$ std) between the model-predicted and ground-truth activation rankings on the NSD test set, averaged across parcels.}
  \label{tab:spearman_rank}
  \begin{tabular}{lcccc}
    \toprule
    Model type & S1 & S2 & S5 & S7 \\
    \midrule
    Null & $0.000 \pm 0.045$ & $0.000 \!\pm\! 0.045$ & $0.000 \!\pm\! 0.045$ & $0.000 \!\pm\! 0.045$ \\
    NSD train & $0.162 \pm  0.098$ & $0.164 \pm  0.091$ & $0.150 \pm  0.094$ & \bm{$0.148 \pm  0.086$} \\
    Our encoder w/INet & \bm{$0.168 \pm  0.106$} & $0.163 \pm  0.082$ & $0.142 \pm  0.092$ & $0.133 \pm  0.075$ \\
    Our encoder w/BD & $0.163 \pm  0.121$ & \bm{$0.190 \pm  0.099$} & \bm{$0.154 \pm  0.094$} & $0.133 \pm  0.083$ \\
    \bottomrule
    \addlinespace[0.5em]
    \multicolumn{5}{@{}p{\linewidth}@{}}{\footnotesize
    \emph{Notes.} INet = ImageNet. BD = BrainDIVE.}
  \end{tabular}
\end{table}

The rank ordering derived from the encoder-selected stimuli generally outperforms the baseline ordering based on the NSD test set. To contextualize the rank correlation magnitude, we report in Appendix \ref{app:intreting-statistical-test} the results for known parcels in the visual area as a benchmark for what constitutes meaningful selectivity. The average rank correlation is in the range of areas like PPA and RSC—both widely studied and accepted in the literature. Because these encoder-driven selectivity hypotheses better explain activation patterns in the NSD test images, our results indicate that the brain-encoder pipeline can generate finer-grained categorical hypotheses than those afforded by the stimuli actually shown to the subject in the scanner.

\begin{figure}[h]
    \centering
    \includegraphics[width=\linewidth]{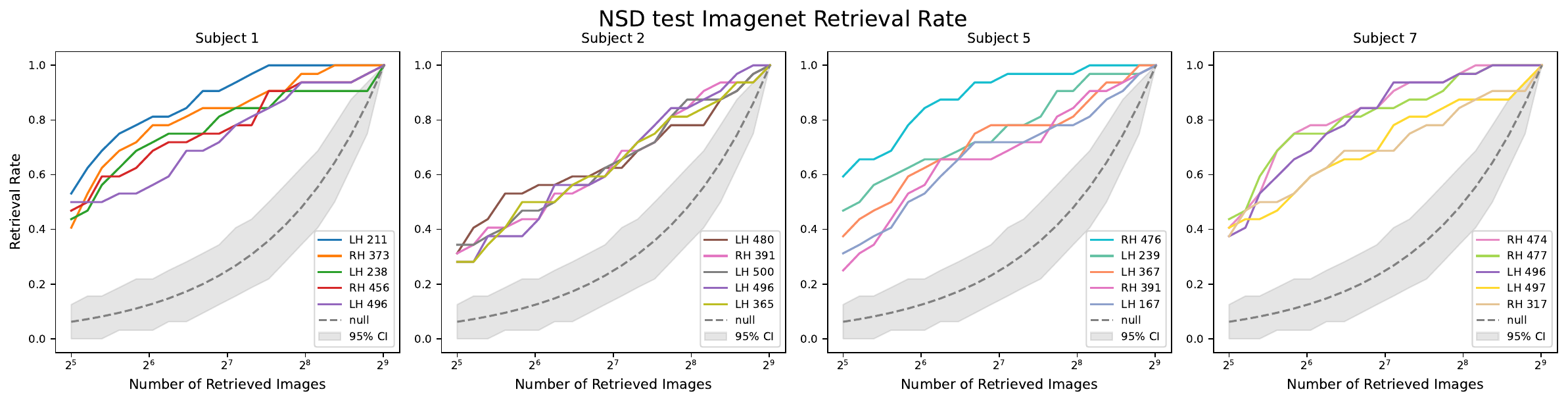}
    \caption{\textbf{Retrieval accuracy of parcel labels.} Using the concept vector as the label for each parcel, we retrieve a varying number of images from the test set based on cosine similarity, and calculate the fraction of overlap with 32 maximally-activating NSD test images (i.e. recall, on the y-axis). Curves that rise quickly and plateau high indicate concept vectors whose semantics closely match the parcel’s true selectivity.} 
    \label{fig:roc_within_subj}
\end{figure}

\paragraph{Choosing parcels for future fMRI experimentation.} For an fMRI study aimed at uncovering the selectivity of parcels outside visual cortex, an experimenter may be interested in parcels whose activity can be well explained by a semantic label. To prioritize such parcels, we define a metric that ranks the quality of the hypotheses generated by our pipeline. In Figure \ref{fig:roc_within_subj}, we plot the proportion of top 32 maximally-activating images successfully retrieved against the number of images retrieved by the pipeline. The top five parcels per subject (ranked by area under the curve) are highlighted.

From an experimenter’s perspective, the parcels shown in Figure \ref{fig:roc_within_subj} may be promising targets for follow-up fMRI studies, since semantic labels seem to capture the parcel selectivity exceptionally well. Our pipeline identifies images whose semantic representations appear to maximally activate these parcels—at least insofar as this can be evaluated with the modest NSD test set.

\subsection{Mapping Cross-subject Selectivity}

\begin{figure}[h]
    \centering
    \includegraphics[width=\linewidth]{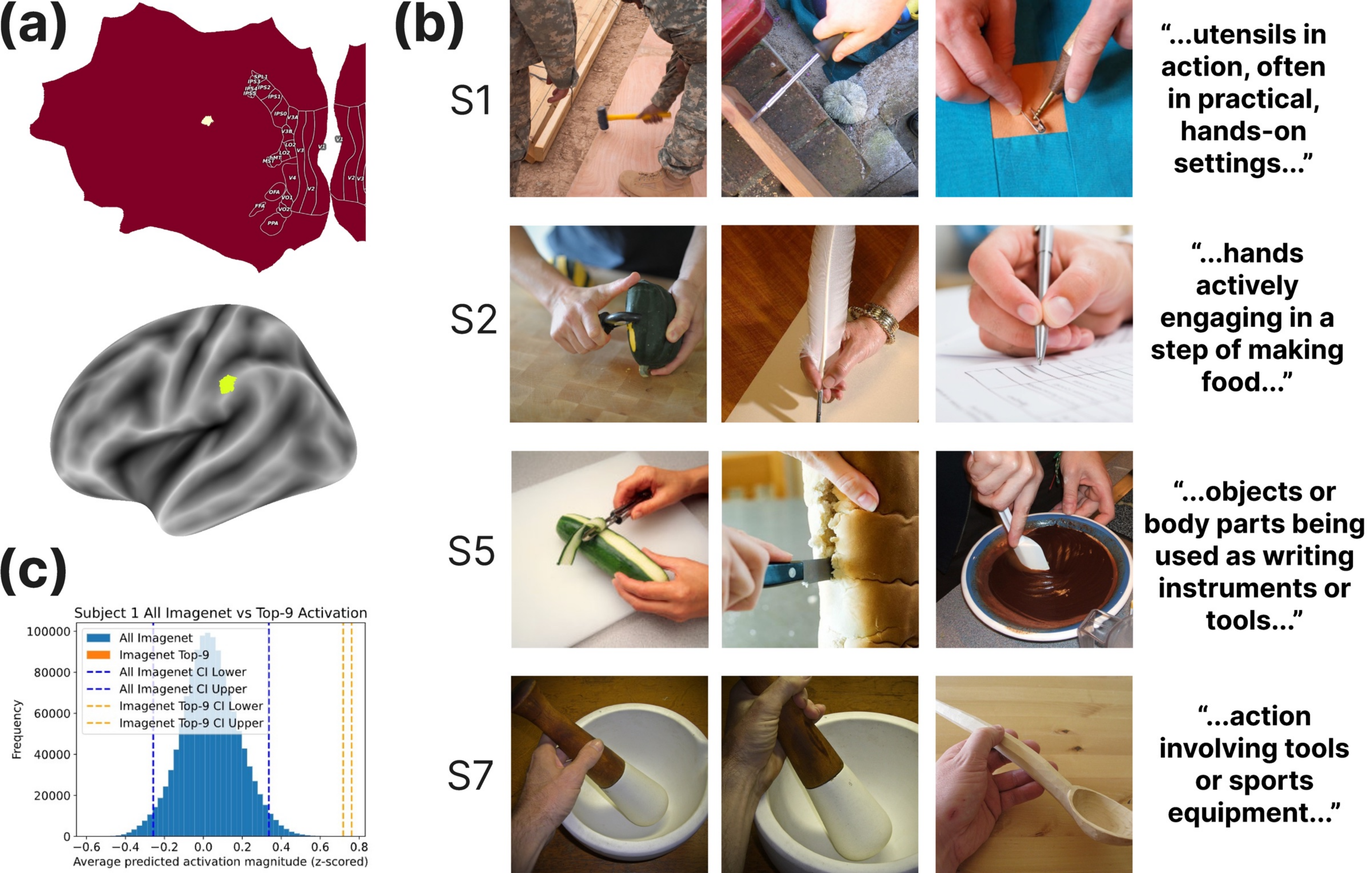}
    \caption{\textbf{(a)} Parcel location.  \textbf{(b)} Selected from the 9 images in ImageNet that maximally activate the parcel of interest, across subjects 1, 2, 5, and 7. ChatGPT 4o generated labels for top-25 ImageNet images, shown on the right (see Appendix~\ref{app:prompts} for details). \textbf{(c)} A comparison of the activation magnitude of the parcel from all ImageNet images and the top 9 images, with 95\% confidence intervals.}
    \label{fig:cross_subj_217}
\end{figure}

We now explore the selectivity of a parcel outside visual cortex (Figure~\ref{fig:cross_subj_217}a) by examining ImageNet images that the encoder predicts will maximally activate it. From the top 9 images per subject, we display 3 from each participant who completed all NSD scanning sessions. These maximally-activating images lie substantially outside the parcel's activation distribution for the remainder of ImageNet (Figure~\ref{fig:cross_subj_217}c). The selected images consistently depict hands manipulating tools—such as writing instruments or cooking utensils (see full collages in Appendix \ref{app:tool-full-collage}). When asked to identify a common theme among the top 25 ImageNet images, ChatGPT likewise highlights hands with objects. Recent work has reported tool-use representations in nearby cortical regions \citep{cortinovis_tool_2025}.

\paragraph{Evaluating our labels.} We qualitatively showed that our pipeline can uncover a parcel exhibiting consistent semantic selectivity across subjects. We now seek to quantitatively verify these hypotheses on ground-truth data. We ran statistical test similar to that in Section~\ref{subsec:evaluating-our-labels}, with two key differences because we are evaluating selectivity across subjects.

\begin{enumerate}
    \item \textbf{Combining across subjects:} For each parcel, we form the optimal image set by combining the top-32 maximally-activating images chosen by encoder models trained on the other three subjects.
    \item \textbf{Retrieval set:} Because each subject’s NSD training set contains distinct images, the ranking retrieval is performed on that subject’s own NSD training set.
\end{enumerate}

We again compare the selectivity hypotheses generated by our brain-encoder pipeline with the NSD-training-set baseline (described in the within-subject statistical tests). Table \ref{tab:rank_corr_vs_null_cross_subj} reports the results of this modified test (cf. Table \ref{tab:rank_corr_vs_null}); additional details are provided in Appendix \ref{app:cross-subj-stats}.

\begin{table}[h]
  \centering
  \caption{Fraction of parcels shared across subjects whose model-derived label predicts parcel activation rankings significantly better than chance ($p<0.05$, FDR corrected)}
  \label{tab:rank_corr_vs_null_cross_subj}
  \begin{tabular}{lccccc}
    \toprule
    Model type & S1 & S2 & S5 & S7 \\
    \midrule
    NSD train & $37/49$ & $44/49$ & $40/49$ & \bm{$40/49$} & \\
    Our encoder w/ImageNet & \bm{$38/49$} & \bm{$45/49$} & \bm{$42/49$} & $39/49$ \\
    Our encoder w/BrainDIVE & $37/49$ & $40/49$ & $38/49$ & $38/49$ \\
    \bottomrule
  \end{tabular}
\end{table}

Comparing the number of parcels for which each hypothesized label ranks NSD-training images better than chance, our pipeline generally outperforms the NSD-train baseline. These findings indicate that for parcels with high visual responsiveness shared across subjects, our pipeline can generate better hypotheses that describe the shared selectivity.

Table \ref{tab:spearman_rank_cross_subj} presents results from the same statistical test used in Table \ref{tab:spearman_rank}, reporting the average Spearman rank correlations for each model across parcels. A head-to-head comparison appears in Table \ref{tab:h2h_nsd_train_baseline_cross_subj} (Appendix \ref{app:h2d-comp}), and Figure \ref{fig:roc_cross_subj} (Appendix \ref{app:choosing-shared-parcels}) highlights the most promising cross-subject parcels for future fMRI experiments.

\begin{table}[h]
  \centering
  \small
  \setlength{\tabcolsep}{4pt}
  \caption{Spearman's $\rho$ (mean $\pm$ std) between the model-predicted (from all subjects other than heldout) and ground-truth activation rankings on the NSD training set, averaged across parcels.}
  \label{tab:spearman_rank_cross_subj}
  \begin{tabular}{lccccc}
    \toprule
    Model type & S1 & S2 & S5 & S7 \\
    \midrule
    Null & $0.000 \!\pm\! 0.011$ & $0.000 \!\pm\! 0.011$ & $0.000 \!\pm\! 0.011$ & $0.000 \!\pm\! 0.011$ \\
    NSD train & $0.068 \pm  0.068$ & $0.094 \pm  0.065$ & $0.090 \pm  0.074$ & $0.091 \pm  0.070$ \\
    Our encoder w/INet & $0.072 \pm  0.071$ & \bm{$0.105 \pm  0.073$} & $0.105 \pm  0.078$ & \bm{$0.093 \pm  0.075$} \\
    Our encoder w/BD & \bm{$0.078 \pm  0.077$} & $0.103 \pm  0.087$ & \bm{$0.106 \pm  0.091$} & $0.089 \pm  0.079$ \\
    \bottomrule
    \addlinespace[0.5em]
    \multicolumn{5}{@{}p{\linewidth}@{}}{\footnotesize
    \emph{Notes.} INet = ImageNet. BD = BrainDIVE.}
  \end{tabular}
\end{table}

\section{Discussion}
\label{sec:discussion}

Leveraging recent advances in AI and the availability of large-scale datasets, we introduce a data-driven paradigm for discovering parcel selectivity beyond visual cortex, paving the way toward systematic whole-brain labeling of higher-order visual representations. In particular, our transformer-based encoder, with its cross-attention mechanism and nonlinear mappings, routes visual information more effectively for parcels in and outside classical visual areas.

\paragraph{Limitations.} Although we predict images that should maximally activate a parcel, we have not yet validated these “superstimuli” in new fMRI experiments. On a held-out NSD split we show that our semantic labels predict parcel activations well, but whether the synthesized superstimuli elicit even stronger responses remains unknown. Moreover, because the encoder was trained solely on NSD \citep{allen_massive_2022}, it may inherit dataset biases. For example, one parcel in subject 2’s left hemisphere appears zebra-selective (Figure \ref{fig:s2_top25_imgnet_p309}, Appendix~\ref{app:nsd-quirks}), however, this could be due to the over-representation of giraffe and zebra images in the dataset. We did not quantify how such biases affect our results; future work should train on multiple datasets to address this concern.

More fundamentally, category selectivity is limited as a theoretical construct for understanding brain computation. We hope our approach will help the field go beyond the technical limitations of prior work and help reveal the fundamental limits of understanding brain computation through the lens of category selectivity.   

\paragraph{Optimizing future fMRI experiments.} Our framework paves the way for future fMRI studies that mitigate the effect of small datasets and experimenter bias in image selection by demonstrating the promise of in silico mapping to superstimuli. These superstimuli can then be tested in follow-up scans to verify the selectivity of newly discovered areas. By letting the encoder drive hypothesis generation, researchers can discover optimal stimuli empirically---even for concepts never presented in the scanner---maximizing data-collection efficiency. Because our encoder architecture is modality-agnostic, the same semantic-mapping approach could be extended to multi-modal backbones, enabling superstimuli generation across sensory domains.

\clearpage
\section*{Acknowledgments}
\label{Acknowledgment}
Research reported in this publication was supported in part by the National Institute of Neurological Disorders and Stroke of the National Institutes of Health under award numbers 1RF1NS128897 and 4R01NS128897. The content is solely the responsibility of the authors and does not necessarily represent the official views of the National Institutes of Health.
\bibliography{references}

\begin{thebibliography}{74}
\providecommand{\natexlab}[1]{#1}
\providecommand{\url}[1]{\texttt{#1}}
\expandafter\ifx\csname urlstyle\endcsname\relax
  \providecommand{\doi}[1]{doi: #1}\else
  \providecommand{\doi}{doi: \begingroup \urlstyle{rm}\Url}\fi

\bibitem[Adeli et~al.(2023{\natexlab{a}})Adeli, Ahn, Kriegeskorte, and Zelinsky]{adeli2023affinitybasedattentionselfsupervisedtransformers}
Hossein Adeli, Seoyoung Ahn, Nikolaus Kriegeskorte, and Gregory Zelinsky.
\newblock Affinity-based attention in self-supervised transformers predicts dynamics of object grouping in humans, 2023{\natexlab{a}}.
\newblock URL \url{https://arxiv.org/abs/2306.00294}.

\bibitem[Adeli et~al.(2023{\natexlab{b}})Adeli, Minni, and Kriegeskorte]{adeli_predicting_2023}
Hossein Adeli, Sun Minni, and Nikolaus Kriegeskorte.
\newblock Predicting brain activity using {Transformers}, August 2023{\natexlab{b}}.
\newblock URL \url{http://biorxiv.org/lookup/doi/10.1101/2023.08.02.551743}.

\bibitem[Adeli et~al.(2025)Adeli, Sun, and Kriegeskorte]{adeli_transformer_2025}
Hossein Adeli, Minni Sun, and Nikolaus Kriegeskorte.
\newblock Transformer brain encoders explain human high-level visual responses, May 2025.
\newblock URL \url{http://arxiv.org/abs/2505.17329}.
\newblock arXiv:2505.17329 [q-bio].

\bibitem[Allen et~al.(2022)Allen, St-Yves, Wu, Breedlove, Prince, Dowdle, Nau, Caron, Pestilli, Charest, Hutchinson, Naselaris, and Kay]{allen_massive_2022}
Emily~J. Allen, Ghislain St-Yves, Yihan Wu, Jesse~L. Breedlove, Jacob~S. Prince, Logan~T. Dowdle, Matthias Nau, Brad Caron, Franco Pestilli, Ian Charest, J.~Benjamin Hutchinson, Thomas Naselaris, and Kendrick Kay.
\newblock A massive {7T} {fMRI} dataset to bridge cognitive neuroscience and artificial intelligence.
\newblock \emph{Nature Neuroscience}, 25\penalty0 (1):\penalty0 116--126, January 2022.
\newblock ISSN 1097-6256, 1546-1726.
\newblock \doi{10.1038/s41593-021-00962-x}.
\newblock URL \url{https://www.nature.com/articles/s41593-021-00962-x}.

\bibitem[Allison et~al.(1994)Allison, McCarthy, Nobre, Puce, and Belger]{allison1994human}
Truett Allison, Gregory McCarthy, Anna Nobre, Aina Puce, and Aysenil Belger.
\newblock Human extrastriate visual cortex and the perception of faces, words, numbers, and colors.
\newblock \emph{Cerebral cortex}, 4\penalty0 (5):\penalty0 544--554, 1994.

\bibitem[Azabou et~al.(2023{\natexlab{a}})Azabou, Arora, Ganesh, Mao, Nachimuthu, Mendelson, Richards, Perich, Lajoie, and Dyer]{azabou2023unified}
Mehdi Azabou, Vinam Arora, Venkataramana Ganesh, Ximeng Mao, Santosh Nachimuthu, Michael Mendelson, Blake Richards, Matthew Perich, Guillaume Lajoie, and Eva~L. Dyer.
\newblock A unified, scalable framework for neural population decoding.
\newblock In \emph{Thirty-seventh Conference on Neural Information Processing Systems}, 2023{\natexlab{a}}.

\bibitem[Azabou et~al.(2023{\natexlab{b}})Azabou, Arora, Ganesh, Mao, Nachimuthu, Mendelson, Richards, Perich, Lajoie, and Dyer]{azabou2023unifiedscalableframeworkneural}
Mehdi Azabou, Vinam Arora, Venkataramana Ganesh, Ximeng Mao, Santosh Nachimuthu, Michael~J. Mendelson, Blake Richards, Matthew~G. Perich, Guillaume Lajoie, and Eva~L. Dyer.
\newblock A unified, scalable framework for neural population decoding, 2023{\natexlab{b}}.
\newblock URL \url{https://arxiv.org/abs/2310.16046}.

\bibitem[Baker et~al.(2007)Baker, Liu, Wald, Kwong, Benner, and Kanwisher]{baker_visual_2007}
Chris~I. Baker, Jia Liu, Lawrence~L. Wald, Kenneth~K. Kwong, Thomas Benner, and Nancy Kanwisher.
\newblock Visual word processing and experiential origins of functional selectivity in human extrastriate cortex.
\newblock \emph{Proceedings of the National Academy of Sciences of the United States of America}, 104\penalty0 (21):\penalty0 9087--9092, May 2007.
\newblock ISSN 0027-8424 1091-6490.
\newblock \doi{10.1073/pnas.0703300104}.
\newblock Place: United States.

\bibitem[Barrós-Loscertales et~al.(2012)Barrós-Loscertales, González, Pulvermüller, Ventura-Campos, Bustamante, Costumero, Parcet, and Ávila]{barros-loscertales_reading_2012}
Alfonso Barrós-Loscertales, Julio González, Friedemann Pulvermüller, Noelia Ventura-Campos, Juan~Carlos Bustamante, Víctor Costumero, María~Antonia Parcet, and César Ávila.
\newblock Reading salt activates gustatory brain regions: {fMRI} evidence for semantic grounding in a novel sensory modality.
\newblock \emph{Cerebral cortex (New York, N.Y. : 1991)}, 22\penalty0 (11):\penalty0 2554--2563, November 2012.
\newblock ISSN 1460-2199 1047-3211.
\newblock \doi{10.1093/cercor/bhr324}.
\newblock Place: United States.

\bibitem[Bashivan et~al.(2019)Bashivan, Kar, and DiCarlo]{bashivan2019neural}
Pouya Bashivan, Kohitij Kar, and James~J DiCarlo.
\newblock Neural population control via deep image synthesis.
\newblock \emph{Science}, 364\penalty0 (6439):\penalty0 eaav9436, 2019.

\bibitem[Beliy et~al.(2025)Beliy, Wasserman, Zalcher, and Irani]{beliy2025wisdomcrowdbrainsuniversal}
Roman Beliy, Navve Wasserman, Amit Zalcher, and Michal Irani.
\newblock The wisdom of a crowd of brains: A universal brain encoder, 2025.
\newblock URL \url{https://arxiv.org/abs/2406.12179}.

\bibitem[Burns et~al.(2019)Burns, Arnold, and Bukach]{BURNS2019209}
Edwin~J. Burns, Taylor Arnold, and Cindy~M. Bukach.
\newblock P-curving the fusiform face area: Meta-analyses support the expertise hypothesis.
\newblock \emph{Neuroscience \& Biobehavioral Reviews}, 104:\penalty0 209--221, 2019.
\newblock ISSN 0149-7634.
\newblock \doi{https://doi.org/10.1016/j.neubiorev.2019.07.003}.
\newblock URL \url{https://www.sciencedirect.com/science/article/pii/S014976341830798X}.

\bibitem[Cerdas et~al.(2024)Cerdas, Sartzetaki, Petersen, Roig, Mettes, and Groen]{Cerdas2024.10.29.620889}
Diego~Garc{\'\i}a Cerdas, Christina Sartzetaki, Magnus Petersen, Gemma Roig, Pascal Mettes, and Iris Groen.
\newblock Brainactiv: Identifying visuo-semantic properties driving cortical selectivity using diffusion-based image manipulation.
\newblock \emph{bioRxiv}, 2024.
\newblock \doi{10.1101/2024.10.29.620889}.
\newblock URL \url{https://www.biorxiv.org/content/early/2024/10/31/2024.10.29.620889}.

\bibitem[Chen et~al.(2022)Chen, Qing, Xiang, Yue, and Zhou]{chen2022seeing}
Zijiao Chen, Jiaxin Qing, Tiange Xiang, Wan~Lin Yue, and Juan~Helen Zhou.
\newblock Seeing beyond the brain: Conditional diffusion model with sparse masked modeling for vision decoding.
\newblock \emph{arXiv preprint arXiv:2211.06956}, 1\penalty0 (2):\penalty0 4, 2022.

\bibitem[Cohen et~al.(2000)Cohen, Dehaene, Naccache, Lehéricy, Dehaene-Lambertz, Hénaff, and Michel]{cohen_visual_2000}
L.~Cohen, S.~Dehaene, L.~Naccache, S.~Lehéricy, G.~Dehaene-Lambertz, M.~A. Hénaff, and F.~Michel.
\newblock The visual word form area: spatial and temporal characterization of an initial stage of reading in normal subjects and posterior split-brain patients.
\newblock \emph{Brain : a journal of neurology}, 123 ( Pt 2):\penalty0 291--307, February 2000.
\newblock ISSN 0006-8950.
\newblock \doi{10.1093/brain/123.2.291}.
\newblock Place: England.

\bibitem[Cortinovis et~al.(2025)Cortinovis, Peelen, and Bracci]{cortinovis_tool_2025}
Davide Cortinovis, Marius~V. Peelen, and Stefania Bracci.
\newblock Tool {Representations} in {Human} {Visual} {Cortex}.
\newblock \emph{Journal of Cognitive Neuroscience}, 37\penalty0 (3):\penalty0 515--531, March 2025.
\newblock ISSN 0898-929X.
\newblock \doi{10.1162/jocn_a_02281}.
\newblock URL \url{https://doi.org/10.1162/jocn_a_02281}.

\bibitem[Deng et~al.(2009)Deng, Dong, Socher, Li, {Kai Li}, and {Li Fei-Fei}]{deng_imagenet_2009}
Jia Deng, Wei Dong, Richard Socher, Li-Jia Li, {Kai Li}, and {Li Fei-Fei}.
\newblock {ImageNet}: {A} large-scale hierarchical image database.
\newblock In \emph{2009 {IEEE} {Conference} on {Computer} {Vision} and {Pattern} {Recognition}}, pages 248--255, Miami, FL, June 2009. IEEE.
\newblock ISBN 978-1-4244-3992-8.
\newblock \doi{10.1109/CVPR.2009.5206848}.
\newblock URL \url{https://ieeexplore.ieee.org/document/5206848/}.

\bibitem[Desimone and Schein(1987)]{desimone_visual_1987}
R.~Desimone and S.~J. Schein.
\newblock Visual properties of neurons in area {V4} of the macaque: sensitivity to stimulus form.
\newblock \emph{Journal of neurophysiology}, 57\penalty0 (3):\penalty0 835--868, March 1987.
\newblock ISSN 0022-3077.
\newblock \doi{10.1152/jn.1987.57.3.835}.
\newblock Place: United States.

\bibitem[Doerig et~al.(2022)Doerig, Kietzmann, Allen, Wu, Naselaris, Kay, and Charest]{doerig2022semantic}
Adrien Doerig, Tim~C Kietzmann, Emily Allen, Yihan Wu, Thomas Naselaris, Kendrick Kay, and Ian Charest.
\newblock Semantic scene descriptions as an objective of human vision.
\newblock \emph{arXiv preprint arXiv:2209.11737}, 2022.

\bibitem[Downing et~al.(2001)Downing, Jiang, Shuman, and Kanwisher]{downing2001cortical}
Paul~E Downing, Yuhong Jiang, Miles Shuman, and Nancy Kanwisher.
\newblock A cortical area selective for visual processing of the human body.
\newblock \emph{Science}, 293\penalty0 (5539):\penalty0 2470--2473, 2001.

\bibitem[Epstein and Kanwisher(1998)]{epstein1998cortical}
Russell Epstein and Nancy Kanwisher.
\newblock A cortical representation of the local visual environment.
\newblock \emph{Nature}, 392\penalty0 (6676):\penalty0 598--601, 1998.

\bibitem[Ferrante et~al.(2023)Ferrante, Ozcelik, Boccato, VanRullen, and Toschi]{ferrante2023brain}
Matteo Ferrante, Furkan Ozcelik, Tommaso Boccato, Rufin VanRullen, and Nicola Toschi.
\newblock Brain captioning: Decoding human brain activity into images and text.
\newblock \emph{arXiv preprint arXiv:2305.11560}, 2023.

\bibitem[Franke et~al.(2025)Franke, Karantzas, Willeke, Diamantaki, Ramakrishnan, Elumalai, Restivo, Fahey, Nealley, Shinn, et~al.]{franke2025dual}
Katrin Franke, Nikos Karantzas, Konstantin Willeke, Maria Diamantaki, Kandan Ramakrishnan, Pavithra Elumalai, Kelli Restivo, Paul Fahey, Cate Nealley, Tori Shinn, et~al.
\newblock Dual-feature selectivity enables bidirectional coding in visual cortical neurons.
\newblock \emph{bioRxiv}, pages 2025--07, 2025.

\bibitem[Freteault et~al.(2025)Freteault, Le~Clei, Tetrel, Bellec, and Farrugia]{Freteault_alignment_2025}
Maëlle Freteault, Maximilien Le~Clei, Loic Tetrel, Lune Bellec, and Nicolas Farrugia.
\newblock Alignment of auditory artificial networks with massive individual fmri brain data leads to generalisable improvements in brain encoding and downstream tasks.
\newblock \emph{Imaging Neuroscience}, 3, 04 2025.
\newblock ISSN 2837-6056.
\newblock \doi{10.1162/imag\_a\_00525}.
\newblock URL \url{https://doi.org/10.1162/imag\_a\_00525}.

\bibitem[Gao et~al.(2015)Gao, Huth, Lescroart, and Gallant]{gao_pycortex_2015}
James~S. Gao, Alexander~G. Huth, Mark~D. Lescroart, and Jack~L. Gallant.
\newblock Pycortex: an interactive surface visualizer for {fMRI}.
\newblock \emph{Frontiers in Neuroinformatics}, 9, September 2015.
\newblock ISSN 1662-5196.
\newblock \doi{10.3389/fninf.2015.00023}.
\newblock URL \url{http://journal.frontiersin.org/Article/10.3389/fninf.2015.00023/abstract}.

\bibitem[Gifford et~al.(2023)Gifford, Lahner, Saba-Sadiya, Vilas, Lascelles, Oliva, Kay, Roig, and Cichy]{gifford2023algonauts}
Alessandro~T Gifford, Benjamin Lahner, Sari Saba-Sadiya, Martina~G Vilas, Alex Lascelles, Aude Oliva, Kendrick Kay, Gemma Roig, and Radoslaw~M Cichy.
\newblock The algonauts project 2023 challenge: How the human brain makes sense of natural scenes.
\newblock \emph{arXiv preprint arXiv:2301.03198}, 2023.

\bibitem[Gifford et~al.(2025)Gifford, Jastrzębowska, Singer, and Cichy]{gifford2025silicodiscoveryrepresentationalrelationships}
Alessandro~T. Gifford, Maya~A. Jastrzębowska, Johannes J.~D. Singer, and Radoslaw~M. Cichy.
\newblock In silico discovery of representational relationships across visual cortex, 2025.
\newblock URL \url{https://arxiv.org/abs/2411.10872}.

\bibitem[Gu et~al.(2022)Gu, Jamison, Khosla, Allen, Wu, St-Yves, Naselaris, Kay, Sabuncu, and Kuceyeski]{gu_neurogen_2022}
Zijin Gu, Keith~Wakefield Jamison, Meenakshi Khosla, Emily~J. Allen, Yihan Wu, Ghislain St-Yves, Thomas Naselaris, Kendrick Kay, Mert~R. Sabuncu, and Amy Kuceyeski.
\newblock {NeuroGen}: {Activation} optimized image synthesis for discovery neuroscience.
\newblock \emph{NeuroImage}, 247:\penalty0 118812, February 2022.
\newblock ISSN 10538119.
\newblock \doi{10.1016/j.neuroimage.2021.118812}.
\newblock URL \url{https://linkinghub.elsevier.com/retrieve/pii/S1053811921010831}.

\bibitem[G{\"u}{\c c}l{\"u} and van Gerven(2015)]{Guclu10005}
Umut G{\"u}{\c c}l{\"u} and Marcel A.~J. van Gerven.
\newblock Deep neural networks reveal a gradient in the complexity of neural representations across the ventral stream.
\newblock \emph{Journal of Neuroscience}, 35\penalty0 (27):\penalty0 10005--10014, 2015.
\newblock ISSN 0270-6474.
\newblock \doi{10.1523/JNEUROSCI.5023-14.2015}.
\newblock URL \url{https://www.jneurosci.org/content/35/27/10005}.

\bibitem[Horikawa et~al.(2020)Horikawa, Cowen, Keltner, and Kamitani]{horikawa_neural_2020}
Tomoyasu Horikawa, Alan~S. Cowen, Dacher Keltner, and Yukiyasu Kamitani.
\newblock The {Neural} {Representation} of {Visually} {Evoked} {Emotion} {Is} {High}-{Dimensional}, {Categorical}, and {Distributed} across {Transmodal} {Brain} {Regions}.
\newblock \emph{iScience}, 23\penalty0 (5), May 2020.
\newblock ISSN 2589-0042.
\newblock \doi{10.1016/j.isci.2020.101060}.
\newblock URL \url{https://doi.org/10.1016/j.isci.2020.101060}.
\newblock Publisher: Elsevier.

\bibitem[Hubel and Wiesel(1962)]{hubel_receptive_1962}
D.~H. Hubel and T.~N. Wiesel.
\newblock Receptive fields, binocular interaction and functional architecture in the cat's visual cortex.
\newblock \emph{The Journal of physiology}, 160\penalty0 (1):\penalty0 106--154, January 1962.
\newblock ISSN 0022-3751 1469-7793.
\newblock \doi{10.1113/jphysiol.1962.sp006837}.
\newblock Place: England.

\bibitem[Huth et~al.(2016)Huth, De~Heer, Griffiths, Theunissen, and Gallant]{huth_natural_2016}
Alexander~G. Huth, Wendy~A. De~Heer, Thomas~L. Griffiths, Frédéric~E. Theunissen, and Jack~L. Gallant.
\newblock Natural speech reveals the semantic maps that tile human cerebral cortex.
\newblock \emph{Nature}, 532\penalty0 (7600):\penalty0 453--458, April 2016.
\newblock ISSN 0028-0836, 1476-4687.
\newblock \doi{10.1038/nature17637}.
\newblock URL \url{https://www.nature.com/articles/nature17637}.

\bibitem[Huth et~al.(2012)Huth, Nishimoto, Vu, and Gallant]{huth_continuous_2012}
Alexander G. Huth, Shinji Nishimoto, An T. Vu, and Jack L. Gallant.
\newblock A {Continuous} {Semantic} {Space} {Describes} the {Representation} of {Thousands} of {Object} and {Action} {Categories} across the {Human} {Brain}.
\newblock \emph{Neuron}, 76\penalty0 (6):\penalty0 1210--1224, December 2012.
\newblock ISSN 08966273.
\newblock \doi{10.1016/j.neuron.2012.10.014}.
\newblock URL \url{https://linkinghub.elsevier.com/retrieve/pii/S0896627312009348}.

\bibitem[Jain et~al.(2023)Jain, Wang, Henderson, Lin, Prince, Tarr, and Wehbe]{Jain2023}
Nidhi Jain, Aria Wang, Margaret~M. Henderson, Ruogu Lin, Jacob~S. Prince, Michael~J. Tarr, and Leila Wehbe.
\newblock Selectivity for food in human ventral visual cortex.
\newblock \emph{Communications Biology 2023 6:1}, 6:\penalty0 1--14, 2 2023.
\newblock ISSN 2399-3642.
\newblock \doi{10.1038/s42003-023-04546-2}.

\bibitem[Kanwisher et~al.(1997)Kanwisher, McDermott, and Chun]{kanwisher1997fusiform}
Nancy Kanwisher, Josh McDermott, and Marvin~M Chun.
\newblock The fusiform face area: a module in human extrastriate cortex specialized for face perception.
\newblock \emph{Journal of neuroscience}, 17\penalty0 (11):\penalty0 4302--4311, 1997.

\bibitem[Kay et~al.(2008)Kay, Naselaris, Prenger, and Gallant]{kay2008identifying}
Kendrick~N Kay, Thomas Naselaris, Ryan~J Prenger, and Jack~L Gallant.
\newblock Identifying natural images from human brain activity.
\newblock \emph{Nature}, 452\penalty0 (7185):\penalty0 352--355, 2008.

\bibitem[Kell et~al.(2018)Kell, Yamins, Shook, Norman-Haignere, and McDermott]{kell_task-optimized_2018}
Alexander J.~E. Kell, Daniel L.~K. Yamins, Erica~N. Shook, Sam~V. Norman-Haignere, and Josh~H. McDermott.
\newblock A {Task}-{Optimized} {Neural} {Network} {Replicates} {Human} {Auditory} {Behavior}, {Predicts} {Brain} {Responses}, and {Reveals} a {Cortical} {Processing} {Hierarchy}.
\newblock \emph{Neuron}, 98\penalty0 (3):\penalty0 630--644.e16, May 2018.
\newblock ISSN 1097-4199 0896-6273.
\newblock \doi{10.1016/j.neuron.2018.03.044}.
\newblock Place: United States.

\bibitem[Khosla et~al.(2022)Khosla, Apurva Ratan~Murty, and Kanwisher]{khosla2022}
Meenakshi Khosla, N.~Apurva Ratan~Murty, and Nancy Kanwisher.
\newblock A highly selective response to food in human visual cortex revealed by hypothesis-free voxel decomposition.
\newblock \emph{Current Biology}, 32:\penalty0 1--13, 2022.

\bibitem[Kingma and Ba(2014)]{kingma2014adam}
Diederik~P Kingma and Jimmy Ba.
\newblock Adam: A method for stochastic optimization.
\newblock \emph{arXiv preprint arXiv:1412.6980}, 2014.

\bibitem[Kong et~al.(2023)Kong, Tan, Wulan, Ooi, Farahibozorg, Harrison, Bijsterbosch, Bernhardt, Eickhoff, and Thomas~Yeo]{kong_comparison_2023}
Ru~Kong, Yan~Rui Tan, Naren Wulan, Leon Qi~Rong Ooi, Seyedeh-Rezvan Farahibozorg, Samuel Harrison, Janine~D. Bijsterbosch, Boris~C. Bernhardt, Simon Eickhoff, and B.~T. Thomas~Yeo.
\newblock Comparison between gradients and parcellations for functional connectivity prediction of behavior.
\newblock \emph{NeuroImage}, 273:\penalty0 120044, June 2023.
\newblock ISSN 1095-9572 1053-8119.
\newblock \doi{10.1016/j.neuroimage.2023.120044}.
\newblock Place: United States.

\bibitem[Kriegeskorte(2015)]{kriegeskorte2015deep}
Nikolaus Kriegeskorte.
\newblock Deep neural networks: a new framework for modeling biological vision and brain information processing.
\newblock \emph{Annual review of vision science}, 1:\penalty0 417--446, 2015.

\bibitem[Kriegeskorte et~al.(2008)Kriegeskorte, Mur, Ruff, Kiani, Bodurka, Esteky, Tanaka, and Bandettini]{Kriegeskorte2008matching}
Nikolaus Kriegeskorte, Marieke Mur, Douglas~A. Ruff, Roozbeh Kiani, Jerzy Bodurka, Hossein Esteky, Keiji Tanaka, and Peter~A. Bandettini.
\newblock Matching categorical object representations in inferior temporal cortex of man and monkey.
\newblock \emph{Neuron}, 60\penalty0 (6):\penalty0 1126--1141, Dec 2008.
\newblock ISSN 0896-6273.
\newblock \doi{10.1016/j.neuron.2008.10.043}.
\newblock URL \url{https://doi.org/10.1016/j.neuron.2008.10.043}.

\bibitem[Li et~al.(2022)Li, Yang, and Gu]{Li2022.11.06.515387}
Yuanning Li, Huzheng Yang, and Shi Gu.
\newblock Upgrading voxel-wise encoding model via integrated integration over features and brain networks.
\newblock \emph{bioRxiv}, 2022.
\newblock \doi{10.1101/2022.11.06.515387}.
\newblock URL \url{https://www.biorxiv.org/content/early/2022/11/07/2022.11.06.515387}.

\bibitem[Liu et~al.(2023)Liu, Ma, Zhou, Zhu, and Zheng]{liu2023brainclip}
Yulong Liu, Yongqiang Ma, Wei Zhou, Guibo Zhu, and Nanning Zheng.
\newblock Brainclip: Bridging brain and visual-linguistic representation via clip for generic natural visual stimulus decoding from fmri.
\newblock \emph{arXiv preprint arXiv:2302.12971}, 2023.

\bibitem[Luo et~al.(2024)Luo, Henderson, Tarr, and Wehbe]{luo2024brainscuba}
Andrew Luo, Margaret~Marie Henderson, Michael~J. Tarr, and Leila Wehbe.
\newblock Brainscuba: Fine-grained natural language captions of visual cortex selectivity.
\newblock In \emph{The Twelfth International Conference on Learning Representations}, 2024.
\newblock URL \url{https://openreview.net/forum?id=mQYHXUUTkU}.

\bibitem[Luo et~al.(2023)Luo, Henderson, Wehbe, and Tarr]{luo_brain_2023}
Andrew~F. Luo, Margaret~M. Henderson, Leila Wehbe, and Michael~J. Tarr.
\newblock Brain {Diffusion} for {Visual} {Exploration}: {Cortical} {Discovery} using {Large} {Scale} {Generative} {Models}, November 2023.
\newblock URL \url{http://arxiv.org/abs/2306.03089}.
\newblock arXiv:2306.03089 [cs].

\bibitem[Matsuyama et~al.(2025)Matsuyama, Nishimoto, and Takagi]{matsuyama_lavca_2025}
Takuya Matsuyama, Shinji Nishimoto, and Yu~Takagi.
\newblock {LaVCa}: {LLM}-assisted {Visual} {Cortex} {Captioning}, February 2025.
\newblock URL \url{http://arxiv.org/abs/2502.13606}.
\newblock arXiv:2502.13606 [q-bio].

\bibitem[McCarthy et~al.(1997)McCarthy, Puce, Gore, and Allison]{mccarthy1997face}
Gregory McCarthy, Aina Puce, John~C Gore, and Truett Allison.
\newblock Face-specific processing in the human fusiform gyrus.
\newblock \emph{Journal of cognitive neuroscience}, 9\penalty0 (5):\penalty0 605--610, 1997.

\bibitem[Mur et~al.(2012)Mur, Ruff, Bodurka, De~Weerd, Bandettini, and Kriegeskorte]{Mur8649}
Marieke Mur, Douglas~A. Ruff, Jerzy Bodurka, Peter De~Weerd, Peter~A. Bandettini, and Nikolaus Kriegeskorte.
\newblock Categorical, yet graded {\textendash} single-image activation profiles of human category-selective cortical regions.
\newblock \emph{Journal of Neuroscience}, 32\penalty0 (25):\penalty0 8649--8662, 2012.
\newblock ISSN 0270-6474.
\newblock \doi{10.1523/JNEUROSCI.2334-11.2012}.
\newblock URL \url{https://www.jneurosci.org/content/32/25/8649}.

\bibitem[Nierhaus et~al.(2023)Nierhaus, Wesolek, Pach, Witt, Blankenburg, and Schmidt]{nierhaus_content_2023}
Till Nierhaus, Sara Wesolek, Daniel Pach, Claudia~M. Witt, Felix Blankenburg, and Timo~T. Schmidt.
\newblock Content {Representation} of {Tactile} {Mental} {Imagery} in {Primary} {Somatosensory} {Cortex}.
\newblock \emph{eNeuro}, 10\penalty0 (6):\penalty0 ENEURO.0408--22.2023, June 2023.
\newblock ISSN 2373-2822.
\newblock \doi{10.1523/ENEURO.0408-22.2023}.
\newblock Place: United States.

\bibitem[Okumura et~al.(2024)Okumura, Kida, Yokoi, Nakai, Nishimoto, Touhara, and Okamoto]{okumura_semantic_2024}
Toshiki Okumura, Ikuhiro Kida, Atsushi Yokoi, Tomoya Nakai, Shinji Nishimoto, Kazushige Touhara, and Masako Okamoto.
\newblock Semantic context-dependent neural representations of odors in the human piriform cortex revealed by {7T} {MRI}.
\newblock \emph{Human brain mapping}, 45\penalty0 (6):\penalty0 e26681, April 2024.
\newblock ISSN 1097-0193 1065-9471.
\newblock \doi{10.1002/hbm.26681}.
\newblock Place: United States.

\bibitem[OpenAI et~al.(2024)OpenAI, Achiam, Adler, Agarwal, Ahmad, Akkaya, Aleman, Almeida, Altenschmidt, Altman, Anadkat, Avila, Babuschkin, Balaji, Balcom, Baltescu, Bao, Bavarian, Belgum, Bello, Berdine, Bernadett-Shapiro, Berner, Bogdonoff, Boiko, Boyd, Brakman, Brockman, Brooks, Brundage, Button, Cai, Campbell, Cann, Carey, Carlson, Carmichael, Chan, Chang, Chantzis, Chen, Chen, Chen, Chen, Chen, Chess, Cho, Chu, Chung, Cummings, Currier, Dai, Decareaux, Degry, Deutsch, Deville, Dhar, Dohan, Dowling, Dunning, Ecoffet, Eleti, Eloundou, Farhi, Fedus, Felix, Fishman, Forte, Fulford, Gao, Georges, Gibson, Goel, Gogineni, Goh, Gontijo-Lopes, Gordon, Grafstein, Gray, Greene, Gross, Gu, Guo, Hallacy, Han, Harris, He, Heaton, Heidecke, Hesse, Hickey, Hickey, Hoeschele, Houghton, Hsu, Hu, Hu, Huizinga, Jain, Jain, Jang, Jiang, Jiang, Jin, Jin, Jomoto, Jonn, Jun, Kaftan, Łukasz Kaiser, Kamali, Kanitscheider, Keskar, Khan, Kilpatrick, Kim, Kim, Kim, Kirchner, Kiros, Knight, Kokotajlo, Łukasz Kondraciuk, Kondrich,
  Konstantinidis, Kosic, Krueger, Kuo, Lampe, Lan, Lee, Leike, Leung, Levy, Li, Lim, Lin, Lin, Litwin, Lopez, Lowe, Lue, Makanju, Malfacini, Manning, Markov, Markovski, Martin, Mayer, Mayne, McGrew, McKinney, McLeavey, McMillan, McNeil, Medina, Mehta, Menick, Metz, Mishchenko, Mishkin, Monaco, Morikawa, Mossing, Mu, Murati, Murk, Mély, Nair, Nakano, Nayak, Neelakantan, Ngo, Noh, Ouyang, O'Keefe, Pachocki, Paino, Palermo, Pantuliano, Parascandolo, Parish, Parparita, Passos, Pavlov, Peng, Perelman, de~Avila Belbute~Peres, Petrov, de~Oliveira~Pinto, Michael, Pokorny, Pokrass, Pong, Powell, Power, Power, Proehl, Puri, Radford, Rae, Ramesh, Raymond, Real, Rimbach, Ross, Rotsted, Roussez, Ryder, Saltarelli, Sanders, Santurkar, Sastry, Schmidt, Schnurr, Schulman, Selsam, Sheppard, Sherbakov, Shieh, Shoker, Shyam, Sidor, Sigler, Simens, Sitkin, Slama, Sohl, Sokolowsky, Song, Staudacher, Such, Summers, Sutskever, Tang, Tezak, Thompson, Tillet, Tootoonchian, Tseng, Tuggle, Turley, Tworek, Uribe, Vallone, Vijayvergiya,
  Voss, Wainwright, Wang, Wang, Wang, Ward, Wei, Weinmann, Welihinda, Welinder, Weng, Weng, Wiethoff, Willner, Winter, Wolrich, Wong, Workman, Wu, Wu, Wu, Xiao, Xu, Yoo, Yu, Yuan, Zaremba, Zellers, Zhang, Zhang, Zhao, Zheng, Zhuang, Zhuk, and Zoph]{openai2024gpt4technicalreport}
OpenAI, Josh Achiam, Steven Adler, Sandhini Agarwal, Lama Ahmad, Ilge Akkaya, Florencia~Leoni Aleman, Diogo Almeida, Janko Altenschmidt, Sam Altman, Shyamal Anadkat, Red Avila, Igor Babuschkin, Suchir Balaji, Valerie Balcom, Paul Baltescu, Haiming Bao, Mohammad Bavarian, Jeff Belgum, Irwan Bello, Jake Berdine, Gabriel Bernadett-Shapiro, Christopher Berner, Lenny Bogdonoff, Oleg Boiko, Madelaine Boyd, Anna-Luisa Brakman, Greg Brockman, Tim Brooks, Miles Brundage, Kevin Button, Trevor Cai, Rosie Campbell, Andrew Cann, Brittany Carey, Chelsea Carlson, Rory Carmichael, Brooke Chan, Che Chang, Fotis Chantzis, Derek Chen, Sully Chen, Ruby Chen, Jason Chen, Mark Chen, Ben Chess, Chester Cho, Casey Chu, Hyung~Won Chung, Dave Cummings, Jeremiah Currier, Yunxing Dai, Cory Decareaux, Thomas Degry, Noah Deutsch, Damien Deville, Arka Dhar, David Dohan, Steve Dowling, Sheila Dunning, Adrien Ecoffet, Atty Eleti, Tyna Eloundou, David Farhi, Liam Fedus, Niko Felix, Simón~Posada Fishman, Juston Forte, Isabella Fulford, Leo
  Gao, Elie Georges, Christian Gibson, Vik Goel, Tarun Gogineni, Gabriel Goh, Rapha Gontijo-Lopes, Jonathan Gordon, Morgan Grafstein, Scott Gray, Ryan Greene, Joshua Gross, Shixiang~Shane Gu, Yufei Guo, Chris Hallacy, Jesse Han, Jeff Harris, Yuchen He, Mike Heaton, Johannes Heidecke, Chris Hesse, Alan Hickey, Wade Hickey, Peter Hoeschele, Brandon Houghton, Kenny Hsu, Shengli Hu, Xin Hu, Joost Huizinga, Shantanu Jain, Shawn Jain, Joanne Jang, Angela Jiang, Roger Jiang, Haozhun Jin, Denny Jin, Shino Jomoto, Billie Jonn, Heewoo Jun, Tomer Kaftan, Łukasz Kaiser, Ali Kamali, Ingmar Kanitscheider, Nitish~Shirish Keskar, Tabarak Khan, Logan Kilpatrick, Jong~Wook Kim, Christina Kim, Yongjik Kim, Jan~Hendrik Kirchner, Jamie Kiros, Matt Knight, Daniel Kokotajlo, Łukasz Kondraciuk, Andrew Kondrich, Aris Konstantinidis, Kyle Kosic, Gretchen Krueger, Vishal Kuo, Michael Lampe, Ikai Lan, Teddy Lee, Jan Leike, Jade Leung, Daniel Levy, Chak~Ming Li, Rachel Lim, Molly Lin, Stephanie Lin, Mateusz Litwin, Theresa Lopez, Ryan
  Lowe, Patricia Lue, Anna Makanju, Kim Malfacini, Sam Manning, Todor Markov, Yaniv Markovski, Bianca Martin, Katie Mayer, Andrew Mayne, Bob McGrew, Scott~Mayer McKinney, Christine McLeavey, Paul McMillan, Jake McNeil, David Medina, Aalok Mehta, Jacob Menick, Luke Metz, Andrey Mishchenko, Pamela Mishkin, Vinnie Monaco, Evan Morikawa, Daniel Mossing, Tong Mu, Mira Murati, Oleg Murk, David Mély, Ashvin Nair, Reiichiro Nakano, Rajeev Nayak, Arvind Neelakantan, Richard Ngo, Hyeonwoo Noh, Long Ouyang, Cullen O'Keefe, Jakub Pachocki, Alex Paino, Joe Palermo, Ashley Pantuliano, Giambattista Parascandolo, Joel Parish, Emy Parparita, Alex Passos, Mikhail Pavlov, Andrew Peng, Adam Perelman, Filipe de~Avila Belbute~Peres, Michael Petrov, Henrique~Ponde de~Oliveira~Pinto, Michael, Pokorny, Michelle Pokrass, Vitchyr~H. Pong, Tolly Powell, Alethea Power, Boris Power, Elizabeth Proehl, Raul Puri, Alec Radford, Jack Rae, Aditya Ramesh, Cameron Raymond, Francis Real, Kendra Rimbach, Carl Ross, Bob Rotsted, Henri Roussez,
  Nick Ryder, Mario Saltarelli, Ted Sanders, Shibani Santurkar, Girish Sastry, Heather Schmidt, David Schnurr, John Schulman, Daniel Selsam, Kyla Sheppard, Toki Sherbakov, Jessica Shieh, Sarah Shoker, Pranav Shyam, Szymon Sidor, Eric Sigler, Maddie Simens, Jordan Sitkin, Katarina Slama, Ian Sohl, Benjamin Sokolowsky, Yang Song, Natalie Staudacher, Felipe~Petroski Such, Natalie Summers, Ilya Sutskever, Jie Tang, Nikolas Tezak, Madeleine~B. Thompson, Phil Tillet, Amin Tootoonchian, Elizabeth Tseng, Preston Tuggle, Nick Turley, Jerry Tworek, Juan Felipe~Cerón Uribe, Andrea Vallone, Arun Vijayvergiya, Chelsea Voss, Carroll Wainwright, Justin~Jay Wang, Alvin Wang, Ben Wang, Jonathan Ward, Jason Wei, CJ~Weinmann, Akila Welihinda, Peter Welinder, Jiayi Weng, Lilian Weng, Matt Wiethoff, Dave Willner, Clemens Winter, Samuel Wolrich, Hannah Wong, Lauren Workman, Sherwin Wu, Jeff Wu, Michael Wu, Kai Xiao, Tao Xu, Sarah Yoo, Kevin Yu, Qiming Yuan, Wojciech Zaremba, Rowan Zellers, Chong Zhang, Marvin Zhang, Shengjia
  Zhao, Tianhao Zheng, Juntang Zhuang, William Zhuk, and Barret Zoph.
\newblock Gpt-4 technical report, 2024.
\newblock URL \url{https://arxiv.org/abs/2303.08774}.

\bibitem[Oquab et~al.(2024)Oquab, Darcet, Moutakanni, Vo, Szafraniec, Khalidov, Fernandez, Haziza, Massa, El-Nouby, Assran, Ballas, Galuba, Howes, Huang, Li, Misra, Rabbat, Sharma, Synnaeve, Xu, Jegou, Mairal, Labatut, Joulin, and Bojanowski]{oquab_dinov2_2024}
Maxime Oquab, Timothée Darcet, Théo Moutakanni, Huy Vo, Marc Szafraniec, Vasil Khalidov, Pierre Fernandez, Daniel Haziza, Francisco Massa, Alaaeldin El-Nouby, Mahmoud Assran, Nicolas Ballas, Wojciech Galuba, Russell Howes, Po-Yao Huang, Shang-Wen Li, Ishan Misra, Michael Rabbat, Vasu Sharma, Gabriel Synnaeve, Hu~Xu, Hervé Jegou, Julien Mairal, Patrick Labatut, Armand Joulin, and Piotr Bojanowski.
\newblock {DINOv2}: {Learning} {Robust} {Visual} {Features} without {Supervision}, February 2024.
\newblock URL \url{http://arxiv.org/abs/2304.07193}.
\newblock arXiv:2304.07193 [cs].

\bibitem[Peelen and Downing(2005)]{peelen_selectivity_2005}
Marius~V. Peelen and Paul~E. Downing.
\newblock Selectivity for the human body in the fusiform gyrus.
\newblock \emph{Journal of neurophysiology}, 93\penalty0 (1):\penalty0 603--608, January 2005.
\newblock ISSN 0022-3077.
\newblock \doi{10.1152/jn.00513.2004}.
\newblock Place: United States.

\bibitem[Pereira et~al.(2018)Pereira, Lou, Pritchett, Ritter, Gershman, Kanwisher, Botvinick, and Fedorenko]{pereira_toward_2018}
Francisco Pereira, Bin Lou, Brianna Pritchett, Samuel Ritter, Samuel~J. Gershman, Nancy Kanwisher, Matthew Botvinick, and Evelina Fedorenko.
\newblock Toward a universal decoder of linguistic meaning from brain activation.
\newblock \emph{Nature Communications}, 9\penalty0 (1):\penalty0 963, March 2018.
\newblock ISSN 2041-1723.
\newblock \doi{10.1038/s41467-018-03068-4}.
\newblock URL \url{https://doi.org/10.1038/s41467-018-03068-4}.

\bibitem[Pierzchlewicz et~al.(2023)Pierzchlewicz, Willeke, Nix, Elumalai, Restivo, Shinn, Nealley, Rodriguez, Patel, Franke, et~al.]{pierzchlewicz2023energy}
Pawe{\l}~A Pierzchlewicz, Konstantin~F Willeke, Arne~F Nix, Pavithra Elumalai, Kelli Restivo, Tori Shinn, Cate Nealley, Gabrielle Rodriguez, Saumil Patel, Katrin Franke, et~al.
\newblock Energy guided diffusion for generating neurally exciting images.
\newblock In \emph{Proceedings of the 37th International Conference on Neural Information Processing Systems}, pages 32574--32601, 2023.

\bibitem[Qian et~al.(2024)Qian, Wang, Huo, Sun, Fu, and Feng]{qian2024fmripte}
Xuelin Qian, Yun Wang, Jingyang Huo, Xinwei Sun, Yanwei Fu, and Jianfeng Feng.
\newblock f{MRI}-{PTE}: A large-scale f{MRI} pretrained transformer encoder for multi-subject brain activity decoding, 2024.
\newblock URL \url{https://openreview.net/forum?id=BZkKMQ25Z7}.

\bibitem[Ratan~Murty et~al.(2021)Ratan~Murty, Bashivan, Abate, DiCarlo, and Kanwisher]{ratan2021computational}
N~Apurva Ratan~Murty, Pouya Bashivan, Alex Abate, James~J DiCarlo, and Nancy Kanwisher.
\newblock Computational models of category-selective brain regions enable high-throughput tests of selectivity.
\newblock \emph{Nature communications}, 12\penalty0 (1):\penalty0 5540, 2021.

\bibitem[Schaefer et~al.(2018)Schaefer, Kong, Gordon, Laumann, Zuo, Holmes, Eickhoff, and Yeo]{schaefer_local-global_2018}
Alexander Schaefer, Ru~Kong, Evan~M Gordon, Timothy~O Laumann, Xi-Nian Zuo, Avram~J Holmes, Simon~B Eickhoff, and B~T~Thomas Yeo.
\newblock Local-{Global} {Parcellation} of the {Human} {Cerebral} {Cortex} from {Intrinsic} {Functional} {Connectivity} {MRI}.
\newblock \emph{Cerebral Cortex}, 28\penalty0 (9):\penalty0 3095--3114, September 2018.
\newblock ISSN 1047-3211, 1460-2199.
\newblock \doi{10.1093/cercor/bhx179}.
\newblock URL \url{https://academic.oup.com/cercor/article/28/9/3095/3978804}.

\bibitem[Schrimpf et~al.(2018)Schrimpf, Kubilius, Hong, Majaj, Rajalingham, Issa, Kar, Bashivan, Prescott-Roy, Geiger, et~al.]{schrimpf2018brain}
Martin Schrimpf, Jonas Kubilius, Ha~Hong, Najib~J Majaj, Rishi Rajalingham, Elias~B Issa, Kohitij Kar, Pouya Bashivan, Jonathan Prescott-Roy, Franziska Geiger, et~al.
\newblock Brain-score: Which artificial neural network for object recognition is most brain-like?
\newblock \emph{BioRxiv}, page 407007, 2018.

\bibitem[Schwarzlose et~al.(2005)Schwarzlose, Baker, and Kanwisher]{schwarzlose_separate_2005}
Rebecca~F. Schwarzlose, Chris~I. Baker, and Nancy Kanwisher.
\newblock Separate face and body selectivity on the fusiform gyrus.
\newblock \emph{The Journal of neuroscience : the official journal of the Society for Neuroscience}, 25\penalty0 (47):\penalty0 11055--11059, November 2005.
\newblock ISSN 1529-2401 0270-6474.
\newblock \doi{10.1523/JNEUROSCI.2621-05.2005}.
\newblock Place: United States.

\bibitem[Scotti et~al.(2024)Scotti, Tripathy, Villanueva, Kneeland, Chen, Narang, Santhirasegaran, Xu, Naselaris, Norman, et~al.]{scotti2024mindeye2}
Paul~S Scotti, Mihir Tripathy, Cesar Kadir~Torrico Villanueva, Reese Kneeland, Tong Chen, Ashutosh Narang, Charan Santhirasegaran, Jonathan Xu, Thomas Naselaris, Kenneth~A Norman, et~al.
\newblock Mindeye2: Shared-subject models enable fmri-to-image with 1 hour of data.
\newblock \emph{arXiv preprint arXiv:2403.11207}, 2024.

\bibitem[Sergent et~al.(1992)Sergent, Ohta, and Macdonald]{sergent_functional_1992}
Justine Sergent, Shinsuke Ohta, and Brenna Macdonald.
\newblock Functional {Neuroanatomy} of {Face} and {Object} {Processing}: {A} {Positron} {Emission} {Tomography} {Study}.
\newblock \emph{Brain}, 115\penalty0 (1):\penalty0 15--36, February 1992.
\newblock ISSN 0006-8950.
\newblock \doi{10.1093/brain/115.1.15}.
\newblock URL \url{https://doi.org/10.1093/brain/115.1.15}.
\newblock \_eprint: https://academic.oup.com/brain/article-pdf/115/1/15/836448/115-1-15.pdf.

\bibitem[Simanova et~al.(2014)Simanova, Hagoort, Oostenveld, and van Gerven]{simanova_modality-independent_2014}
Irina Simanova, Peter Hagoort, Robert Oostenveld, and Marcel A.~J. van Gerven.
\newblock Modality-{Independent} {Decoding} of {Semantic} {Information} from the {Human} {Brain}.
\newblock \emph{Cerebral Cortex}, 24\penalty0 (2):\penalty0 426--434, February 2014.
\newblock ISSN 1047-3211.
\newblock \doi{10.1093/cercor/bhs324}.
\newblock URL \url{https://doi.org/10.1093/cercor/bhs324}.

\bibitem[Simmons et~al.(2005)Simmons, Martin, and Barsalou]{simmons_pictures_2005}
W.~Kyle Simmons, Alex Martin, and Lawrence~W. Barsalou.
\newblock Pictures of appetizing foods activate gustatory cortices for taste and reward.
\newblock \emph{Cerebral cortex (New York, N.Y. : 1991)}, 15\penalty0 (10):\penalty0 1602--1608, October 2005.
\newblock ISSN 1047-3211.
\newblock \doi{10.1093/cercor/bhi038}.
\newblock Place: United States.

\bibitem[Takagi and Nishimoto(2022)]{takagi2022high}
Yu~Takagi and Shinji Nishimoto.
\newblock High-resolution image reconstruction with latent diffusion models from human brain activity.
\newblock \emph{bioRxiv}, pages 2022--11, 2022.

\bibitem[Tang et~al.(2023)Tang, Du, Vo, Lal, and Huth]{tang_jerry_brain_encoding}
Jerry Tang, Meng Du, Vy~Vo, Vasudev Lal, and Alexander Huth.
\newblock Brain encoding models based on multimodal transformers can transfer across language and vision.
\newblock \emph{Advances in neural information processing systems}, 36:\penalty0 29654--29666, 12 2023.

\bibitem[Turishcheva et~al.(2024)Turishcheva, Fahey, Vystr{\v{c}}ilov{\'a}, Hansel, Froebe, Ponder, Qiu, Willeke, Bashiri, Wang, et~al.]{turishcheva2024dynamic}
Polina Turishcheva, Paul~G Fahey, Michaela Vystr{\v{c}}ilov{\'a}, Laura Hansel, Rachel Froebe, Kayla Ponder, Yongrong Qiu, Konstantin~F Willeke, Mohammad Bashiri, Eric Wang, et~al.
\newblock The dynamic sensorium competition for predicting large-scale mouse visual cortex activity from videos.
\newblock \emph{ArXiv}, pages arXiv--2305, 2024.

\bibitem[van~der Laan et~al.(2011)van~der Laan, de~Ridder, Viergever, and Smeets]{van_der_laan_first_2011}
L.~N. van~der Laan, D.~T.~D. de~Ridder, M.~A. Viergever, and P.~A.~M. Smeets.
\newblock The first taste is always with the eyes: a meta-analysis on the neural correlates of processing visual food cues.
\newblock \emph{NeuroImage}, 55\penalty0 (1):\penalty0 296--303, March 2011.
\newblock ISSN 1095-9572 1053-8119.
\newblock \doi{10.1016/j.neuroimage.2010.11.055}.
\newblock Place: United States.

\bibitem[{van der Schaaf} and {van Hateren}(1996)]{VANDERSCHAAF19962759}
A.~{van der Schaaf} and J.H. {van Hateren}.
\newblock Modelling the power spectra of natural images: Statistics and information.
\newblock \emph{Vision Research}, 36\penalty0 (17):\penalty0 2759--2770, 1996.
\newblock ISSN 0042-6989.
\newblock \doi{https://doi.org/10.1016/0042-6989(96)00002-8}.
\newblock URL \url{https://www.sciencedirect.com/science/article/pii/0042698996000028}.

\bibitem[Walker et~al.(2020)Walker, Cotton, Ma, and Tolias]{walker_neural_2020}
Edgar~Y. Walker, R.~James Cotton, Wei~Ji Ma, and Andreas~S. Tolias.
\newblock A neural basis of probabilistic computation in visual cortex.
\newblock \emph{Nature neuroscience}, 23\penalty0 (1):\penalty0 122--129, January 2020.
\newblock ISSN 1546-1726 1097-6256.
\newblock \doi{10.1038/s41593-019-0554-5}.
\newblock Place: United States.

\bibitem[Yamins and DiCarlo(2016)]{Yamins2016}
Daniel L.~K. Yamins and James~J. DiCarlo.
\newblock Using goal-driven deep learning models to understand sensory cortex.
\newblock \emph{Nature Neuroscience}, 19\penalty0 (3):\penalty0 356--365, Mar 2016.

\bibitem[Yang et~al.(2024)Yang, Gee, and Shi]{yang2024brain}
Huzheng Yang, James Gee, and Jianbo Shi.
\newblock Brain decodes deep nets.
\newblock In \emph{Proceedings of the IEEE/CVF Conference on Computer Vision and Pattern Recognition}, pages 23030--23040, 2024.

\bibitem[Yeung et~al.(2024)Yeung, Luo, Sarch, Henderson, Ramanan, and Tarr]{yeung2024neural}
Jacob Yeung, Andrew~F Luo, Gabriel Sarch, Margaret~M Henderson, Deva Ramanan, and Michael~J Tarr.
\newblock Neural representations of dynamic visual stimuli.
\newblock \emph{arXiv preprint arXiv:2406.02659}, 2024.

\end{thebibliography}

\clearpage
\appendix
\renewcommand{\thefigure}{S\arabic{figure}}
\setcounter{figure}{0}

\section{Technical Appendices and Supplementary Material}

\subsection{Social impacts}
\label{app:social-impacts}

In this work, we present a method for in silico mapping of semantic selectivity across cortical parcels in the whole brain. For cognitive neuroscience, our approach deepens the field’s understanding of visual processing well beyond classical visual cortex. Clinically, deviations from normative whole-brain semantic maps may act as early biomarkers for neurological and neurodegenerative disorders. Furthermore, these detailed semantic atlases lay foundational groundwork for next-generation brain--computer interfaces, where fine-grained category-level decoding could translate a user’s intended concepts—rather than low-level motor signals—into control commands for communication or assistive devices.

\clearpage
\subsection{Reproducing selectivity for known areas}
\label{app:superstimuli-known-areas}

We show superstimuli from parcels overlapping with body-, place-, and word-selective areas in order to show that our pipeline can reproduce the known selectivity.

\begin{figure}[h]
    \centering
    \includegraphics[width=\linewidth]{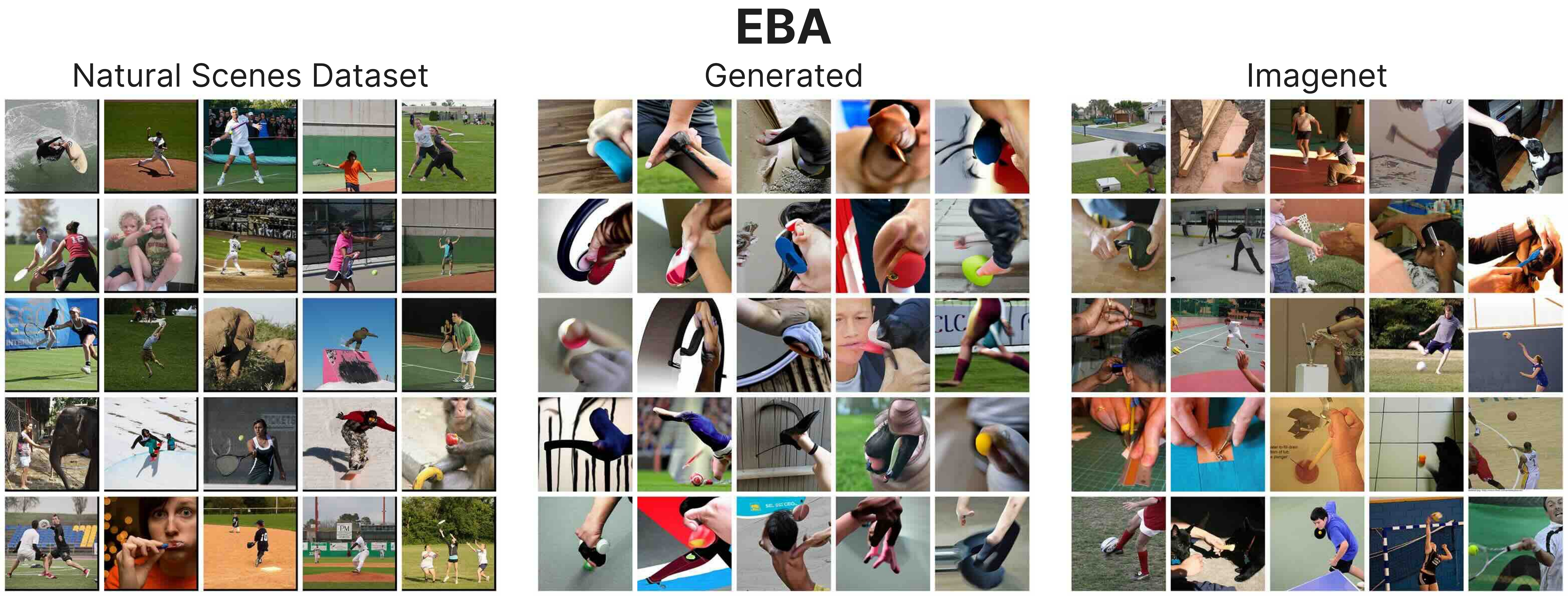}
\end{figure}

\begin{figure}[h]
    \centering
    \includegraphics[width=\linewidth]{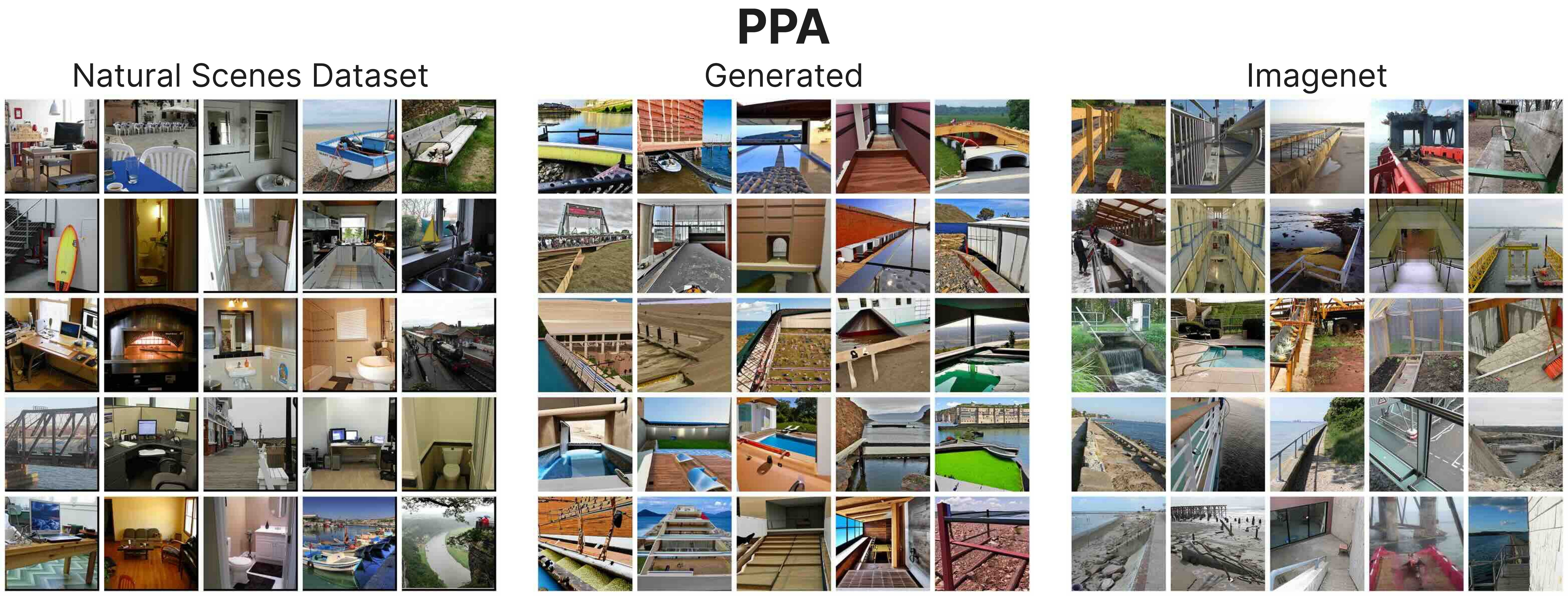}
\end{figure}

\begin{figure}[h]
    \centering
    \includegraphics[width=\linewidth]{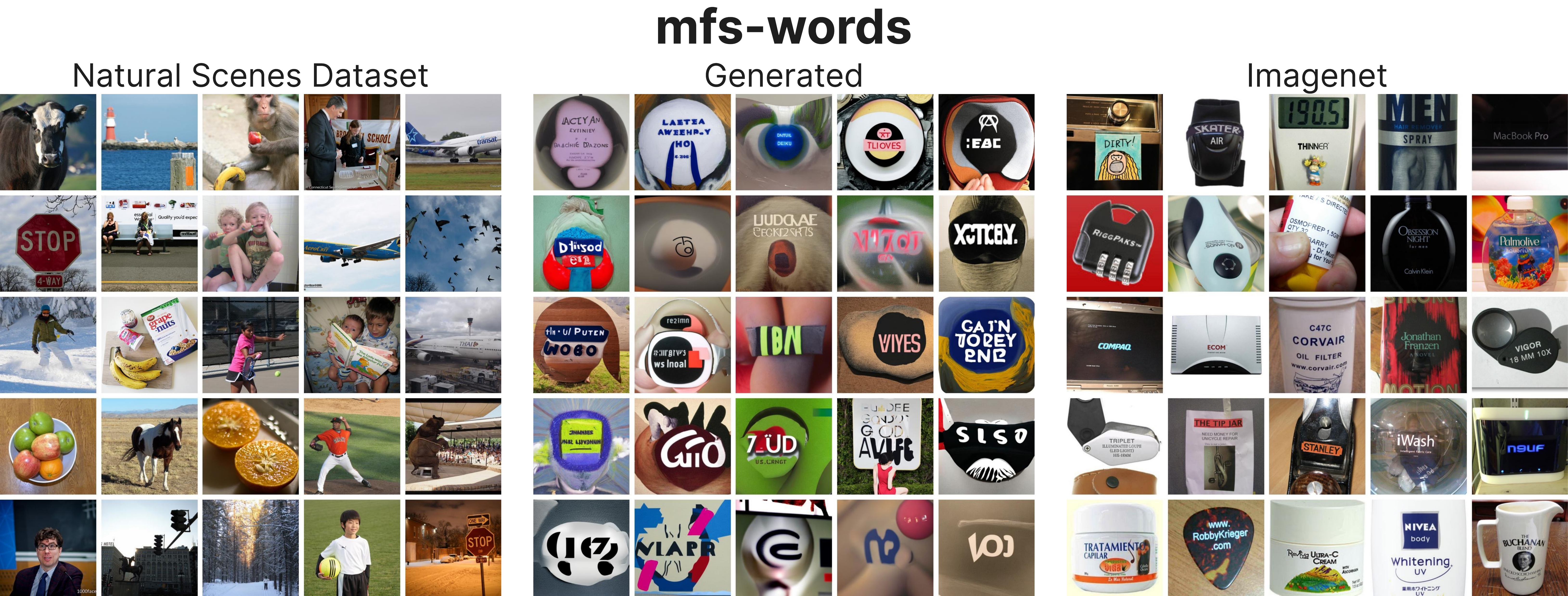}
\end{figure}

\clearpage
\subsection{Parcel selection process}
\label{app:parcel-selection-process}

\begin{figure}[h]
    \centering
    \includegraphics[width=\linewidth]{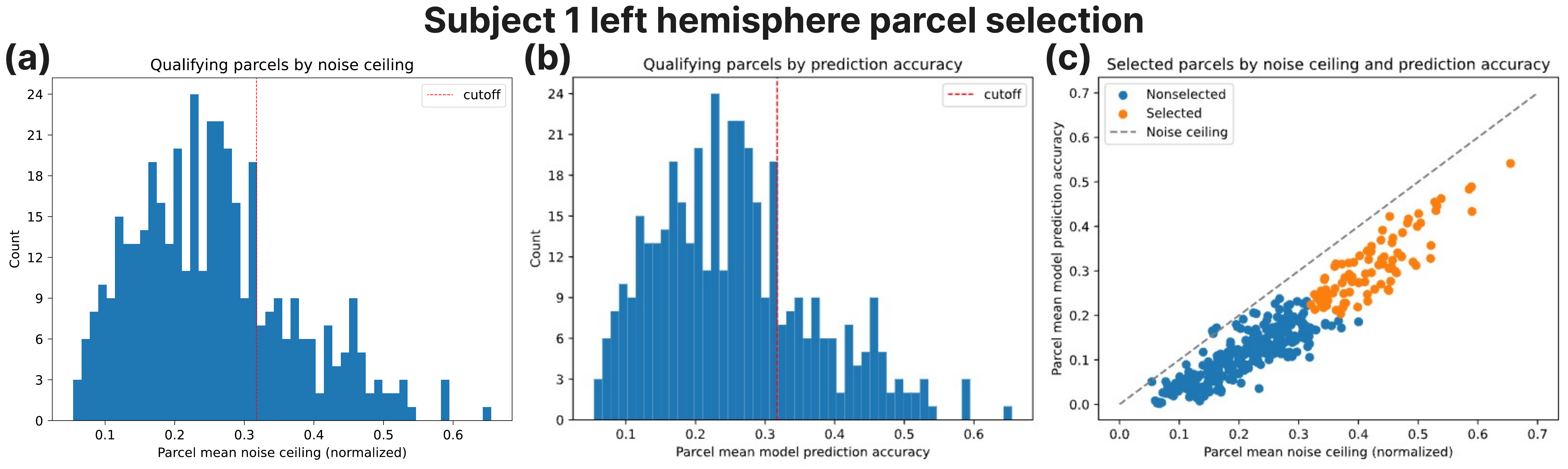}
    \caption{\textbf{Selecting parcels outside visual cortex for experimentation.} Parcels outside the visual area that have sufficiently high mean \textbf{(a)} noise ceiling and \textbf{(b)} model prediction accuracy are chosen for further experimentation. \textbf{(c)} Selected parcels by mean noise ceiling and model prediction accuracy.} 
    \label{fig:parcel_selection}
\end{figure}

\clearpage
\subsection{Reproducing known visual hierarchy properties}
\label{app:reproducing-visual-hierarchy}

We optimized 32 superstimuli (16 per hemisphere) that maximally activate parcels in V1, V2, and V4 using the BrainDIVE framework. We additionally include results on FFA as a representative IT-cortex ROI.

Generated superstimuli for early visual areas reproduce the classic fine-to-coarse hierarchy: V1 stimuli appear as cluttered scenes filled with dense, repetitive texture; V2 images add composite color patches and rudimentary objects; V4 stimuli reveal smoother, recognizable object forms; and FFA images almost exclusively depict close-up faces, often with clear emotional expressions.

To quantify hierarchical properties, we examined whether the spatial-frequency content of our stimuli mirrors classical physiological findings (e.g., \citep{hubel_receptive_1962, desimone_visual_1987}). We computed the radial average power spectrum for each image \citep{VANDERSCHAAF19962759} and calculated the proportion of spectral power above 10\%, 20\%, and 30\% of the maximum spatial frequency. At every threshold the high-frequency energy ratio decreases monotonically from V1 > V2 > V4 > FFA, indicating that the images that best drive higher-level areas (FFA) contain proportionally less fine-scale texture and relatively more coarse, low-frequency structure.

\begin{table}[h]
  \centering
  \caption{High-frequency energy ratio across ROIs at different thresholds, subject 1.}
  \label{tab:high_freq_energy}
  \begin{tabular}{lccc}
    \toprule
    Threshold & ROI & High-frequency energy ratio \\
    \midrule
    0.1 & V1  & 0.006547 \\
        & V2  & 0.004160 \\
        & V4  & 0.002500 \\
        & FFA & 0.001347 \\
    \midrule
    0.2 & V1  & 0.001838 \\
        & V2  & 0.000944 \\
        & V4  & 0.000699 \\
        & FFA & 0.000299 \\
    \midrule
    0.3 & V1  & 0.000762 \\
        & V2  & 0.000339 \\
        & V4  & 0.000285 \\
        & FFA & 0.000104 \\
    \bottomrule
  \end{tabular}
\end{table}

Below we include superstimuli optimized for each of the above ROIs to visualize the image frequency differences.

\begin{figure}[htbp]
  \centering

  \begin{subfigure}{0.48\textwidth}
    \centering
    \includegraphics[width=\linewidth]{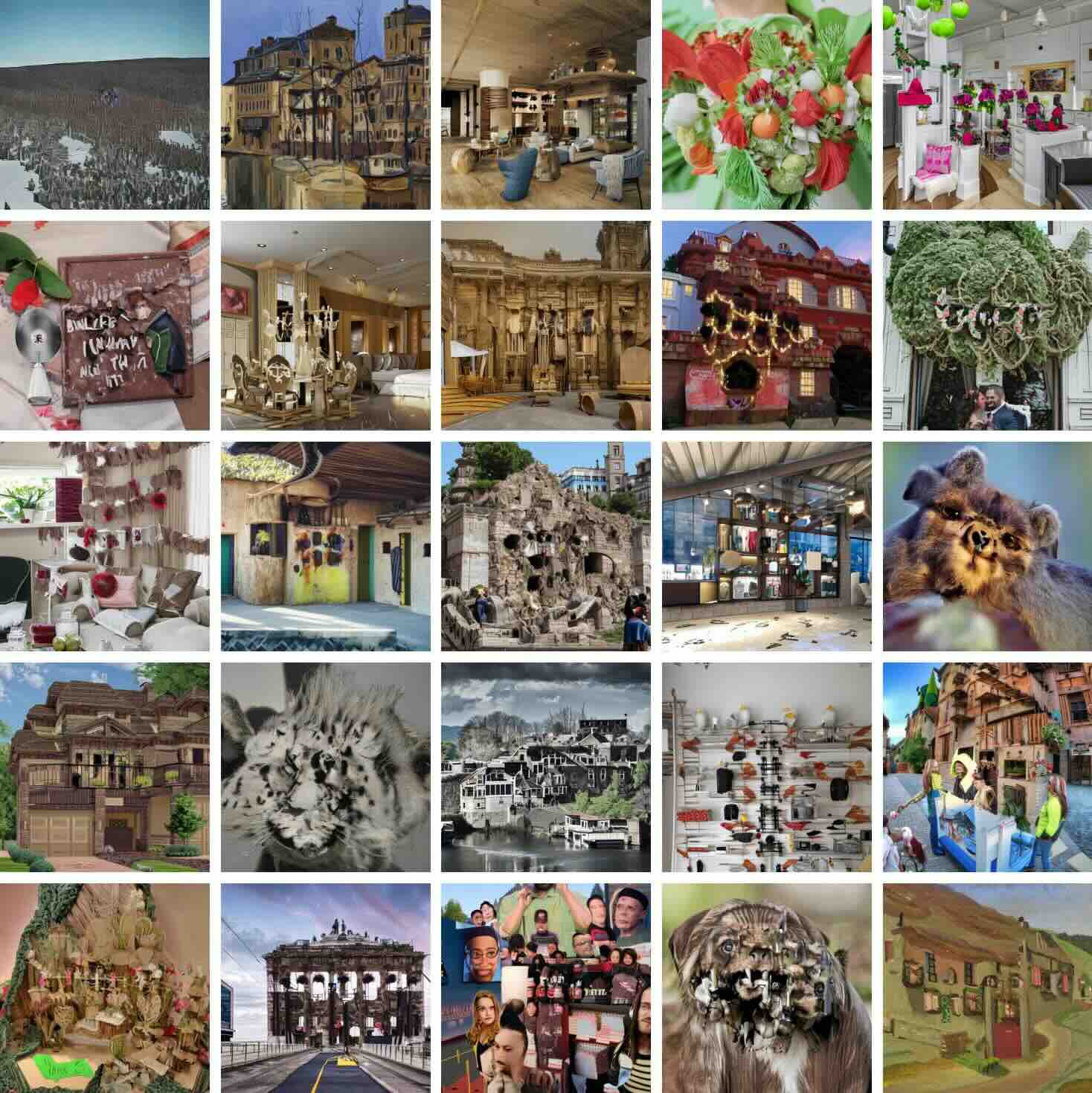}
    \caption{V1}
  \end{subfigure}\hfill
  \begin{subfigure}{0.48\textwidth}
    \centering
    \includegraphics[width=\linewidth]{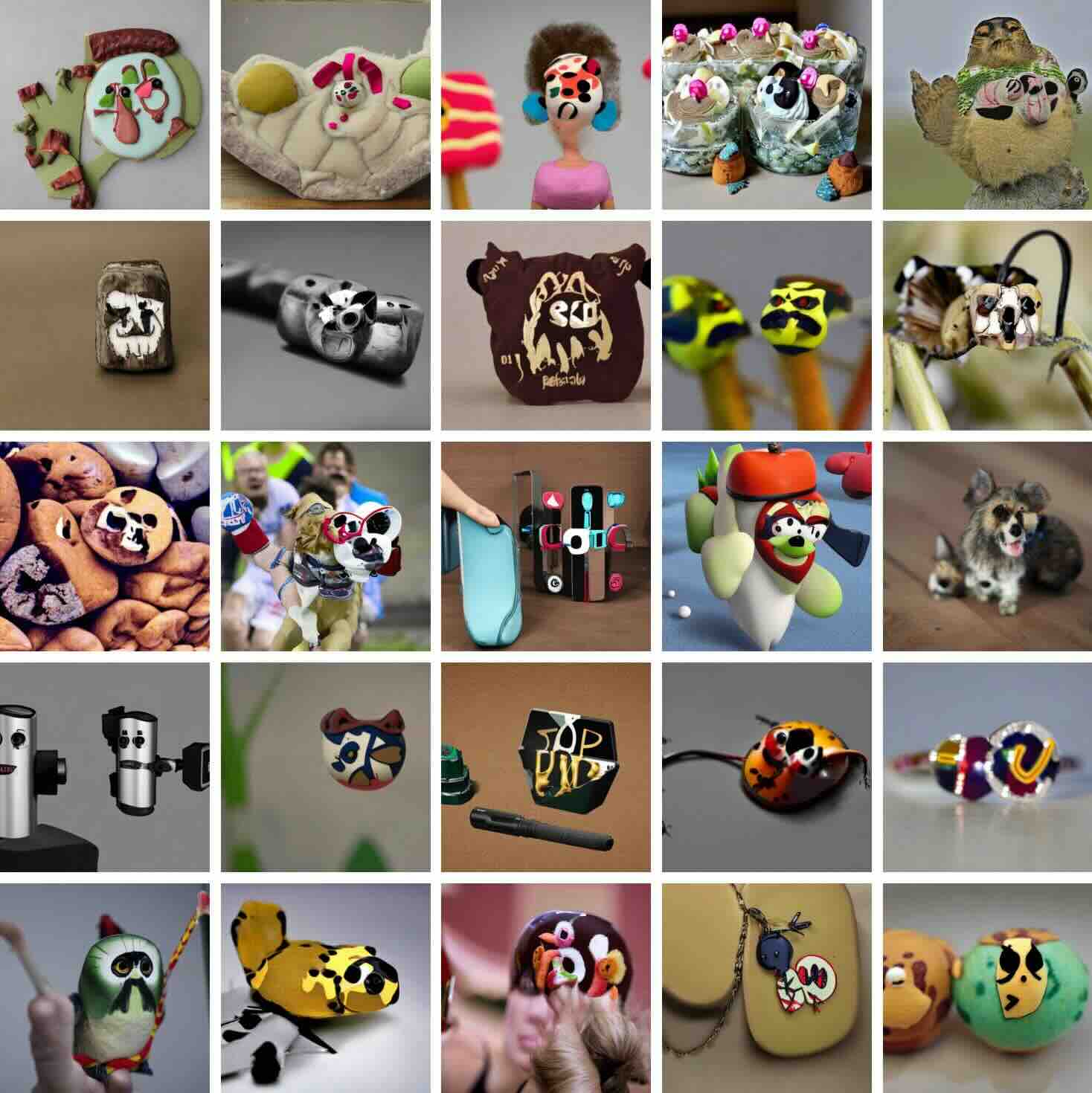}
    \caption{V2}
  \end{subfigure}

  \vspace{0.5em}

  \begin{subfigure}{0.48\textwidth}
    \centering
    \includegraphics[width=\linewidth]{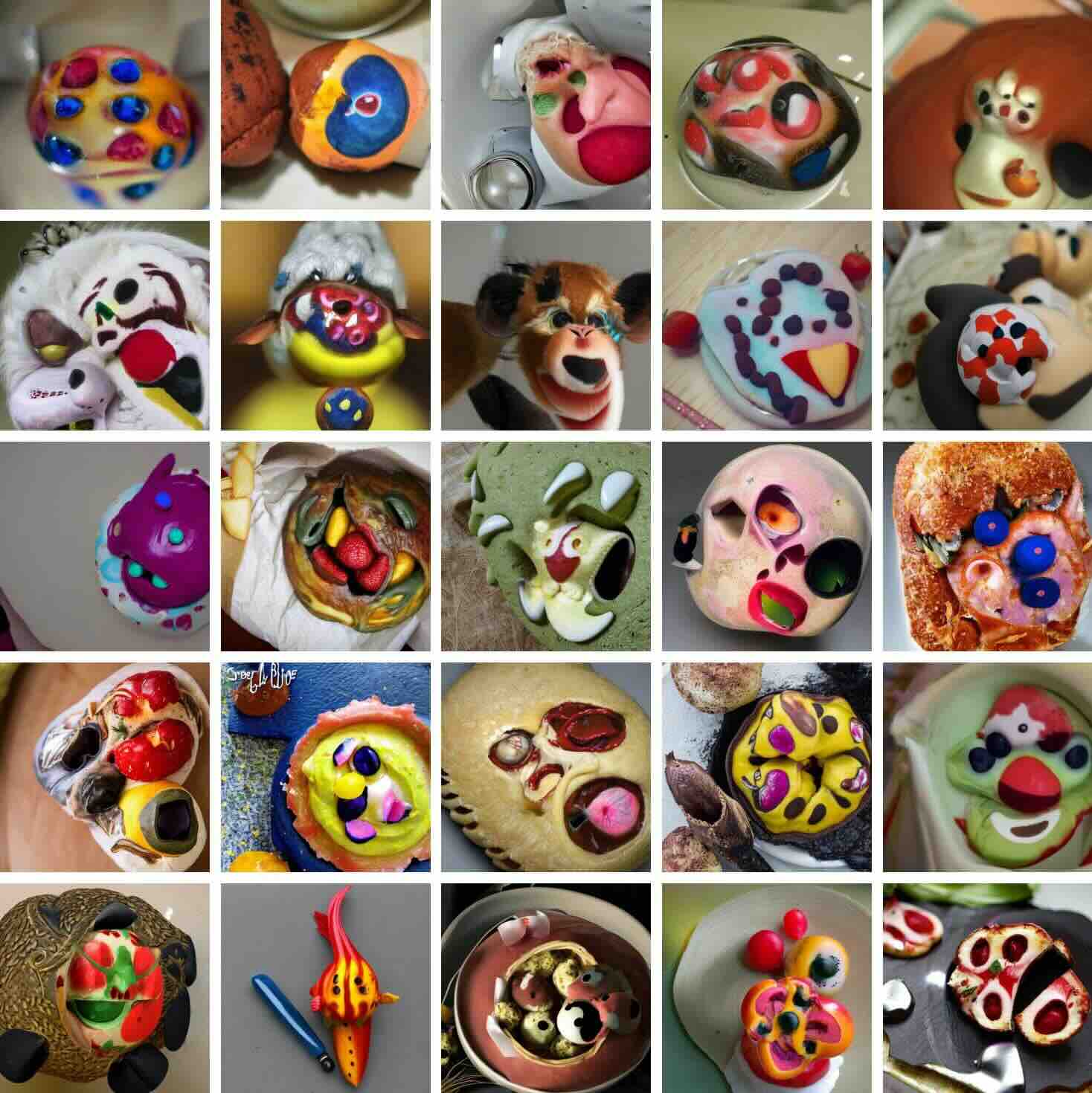}
    \caption{V4}
  \end{subfigure}\hfill
  \begin{subfigure}{0.48\textwidth}
    \centering
    \includegraphics[width=\linewidth]{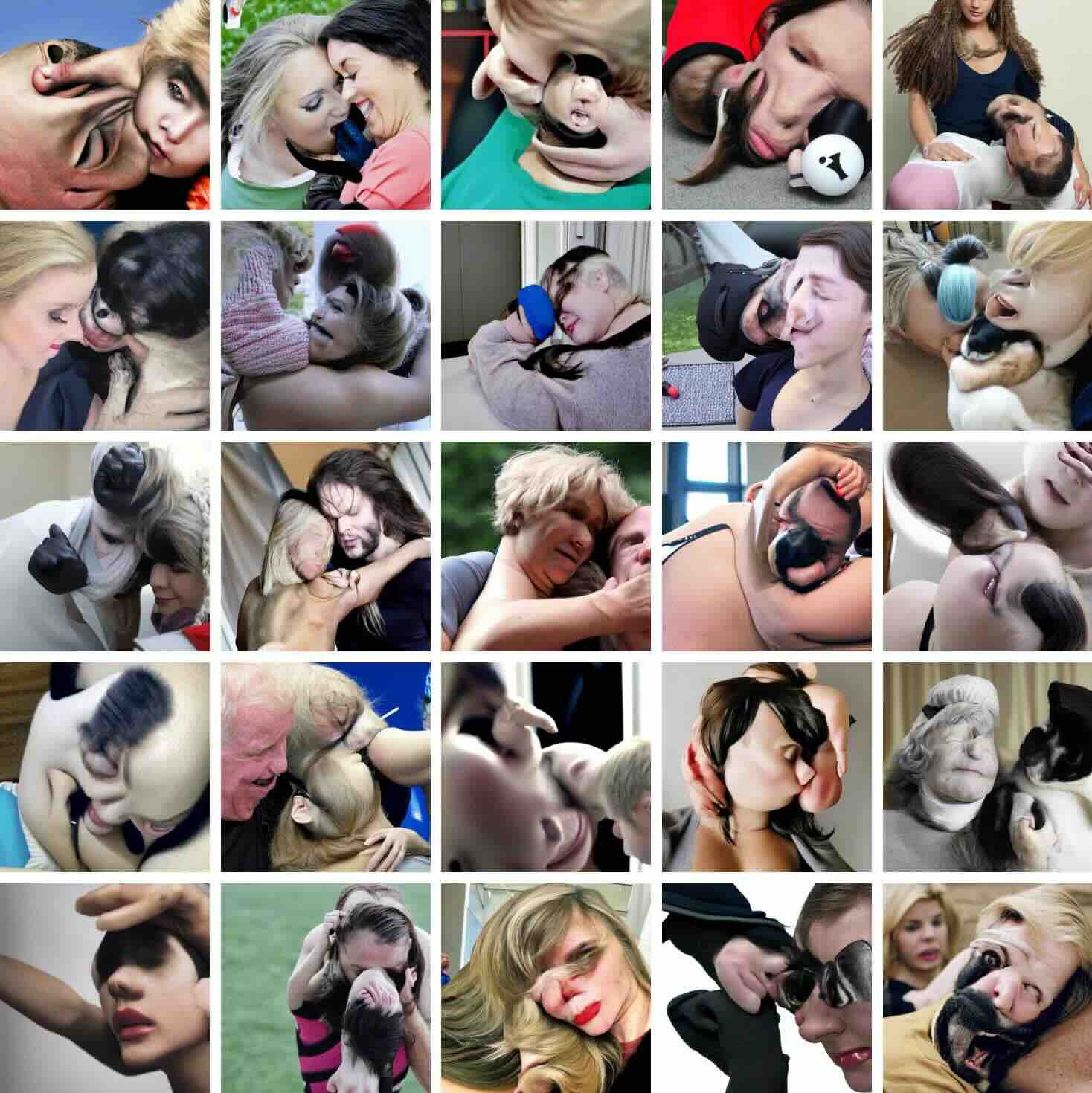}
    \caption{FFA-1}
  \end{subfigure}

  \caption{BrainDIVE stimuli optimized for parcels across the visual hierarchy, subject 1.}
\end{figure}

\clearpage
\subsection{Analyzing ROI queries to map parcel connectivity}

Taking advantage of our transformer-based approach, we map the connections between visual area parcels by examining the representational similarity of learned ROI queries, each of which corresponds to a single parcel. For both functional connectivity and ROI query similarity, we first take the full 1000 parcel x 1000 parcel matrix, then mask out the labeled area, and finally average across the top three parcels that overlap with each labeled area. (Thus, not every entry on the diagonal is 1.) High cosine similarity between ROI queries (queries averaged across ensemble models) suggests that two parcels are highly connected, as both attend to similar content in an image.

This similarity matrix for the visual areas closely replicates the functional connectivity matrix from the Schaefer parcellation (derived from resting-state correlated responses) \citep{kong_comparison_2023}; the Pearson correlation between the upper triangle of the two matrices is $\approx 0.5$.

Below we report the functional connectivity and ROI query cosine similarity matrices for subject 1.

\begin{figure}[htbp]
  \centering

  \begin{subfigure}{0.48\textwidth}
    \centering
    \includegraphics[width=\linewidth]{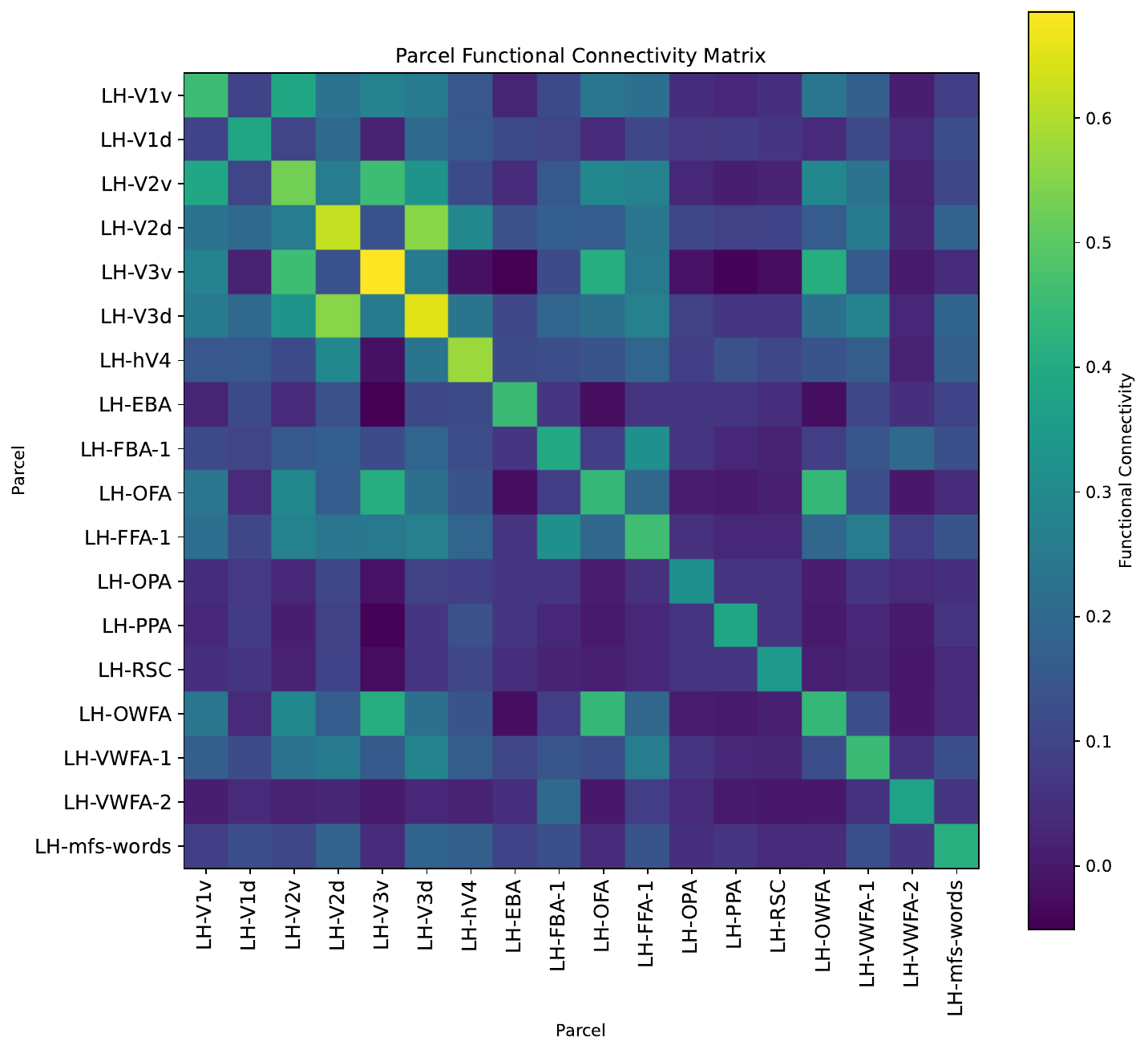}
    \caption{Functional connectivity, left hemisphere}
  \end{subfigure}\hfill
  \begin{subfigure}{0.48\textwidth}
    \centering
    \includegraphics[width=\linewidth]{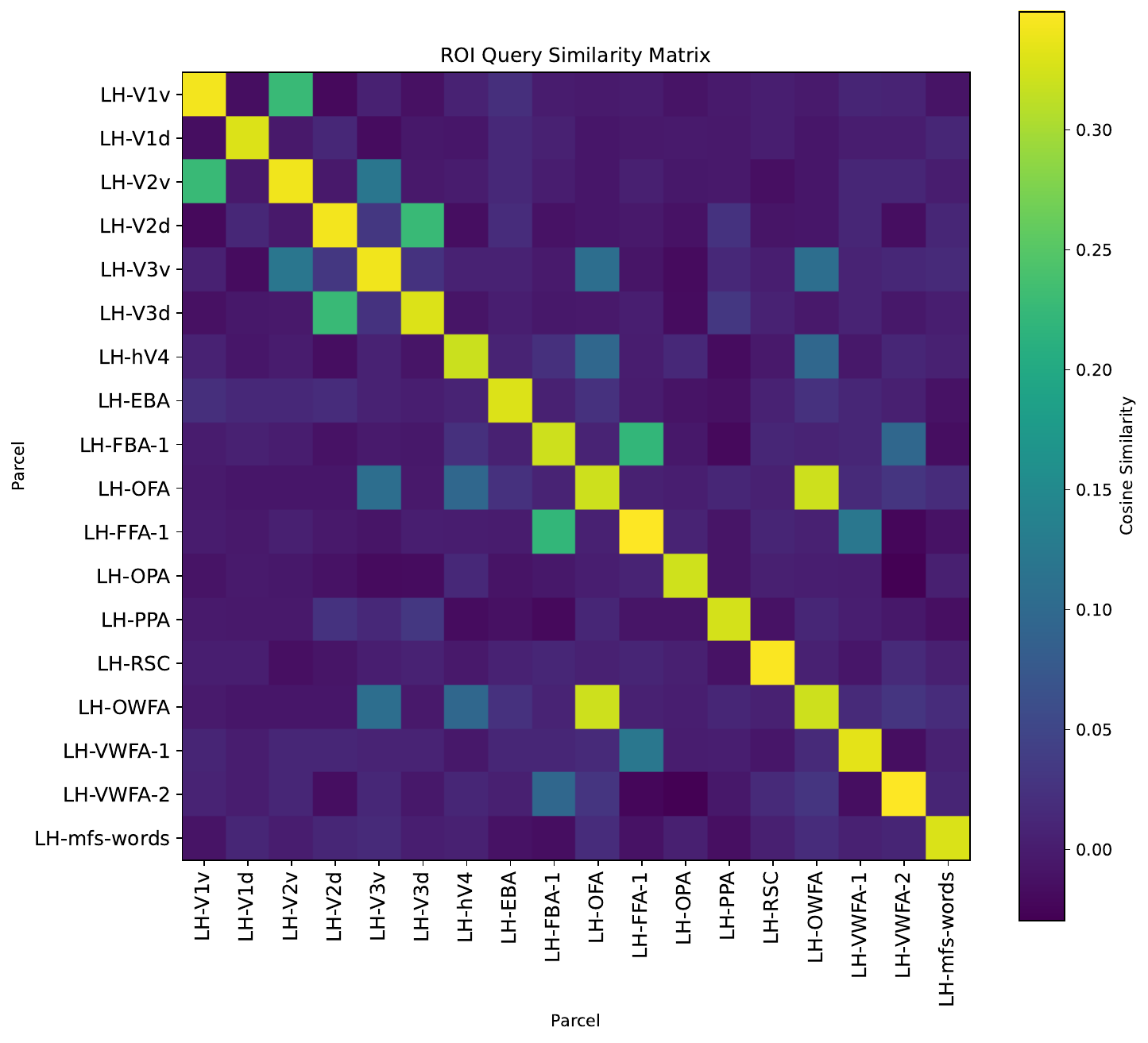}
    \caption{ROI query similarity, left hemisphere}
  \end{subfigure}

  \vspace{0.5em}

  \begin{subfigure}{0.48\textwidth}
    \centering
    \includegraphics[width=\linewidth]{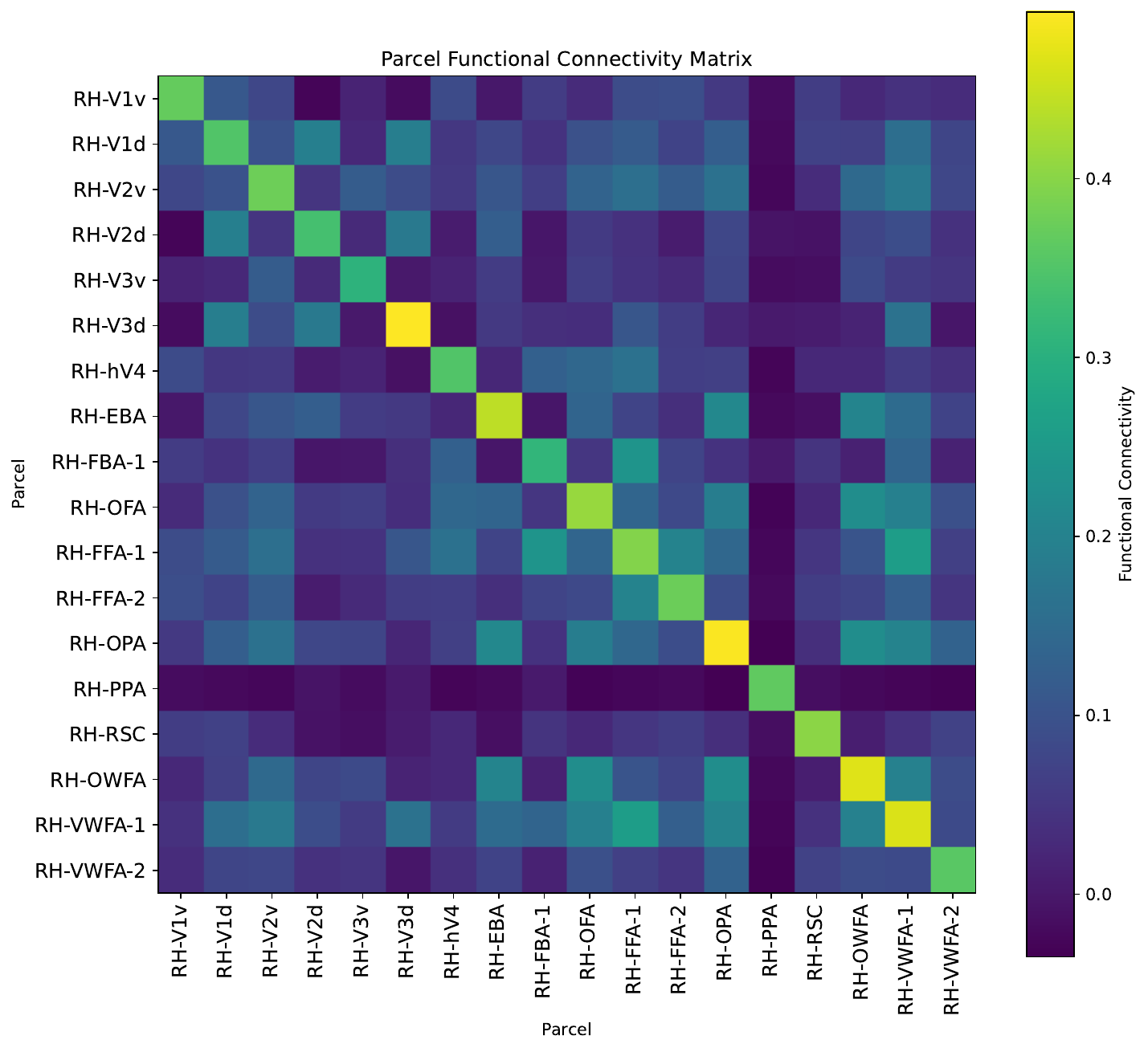}
    \caption{Functional connectivity, right hemisphere}
  \end{subfigure}\hfill
  \begin{subfigure}{0.48\textwidth}
    \centering
    \includegraphics[width=\linewidth]{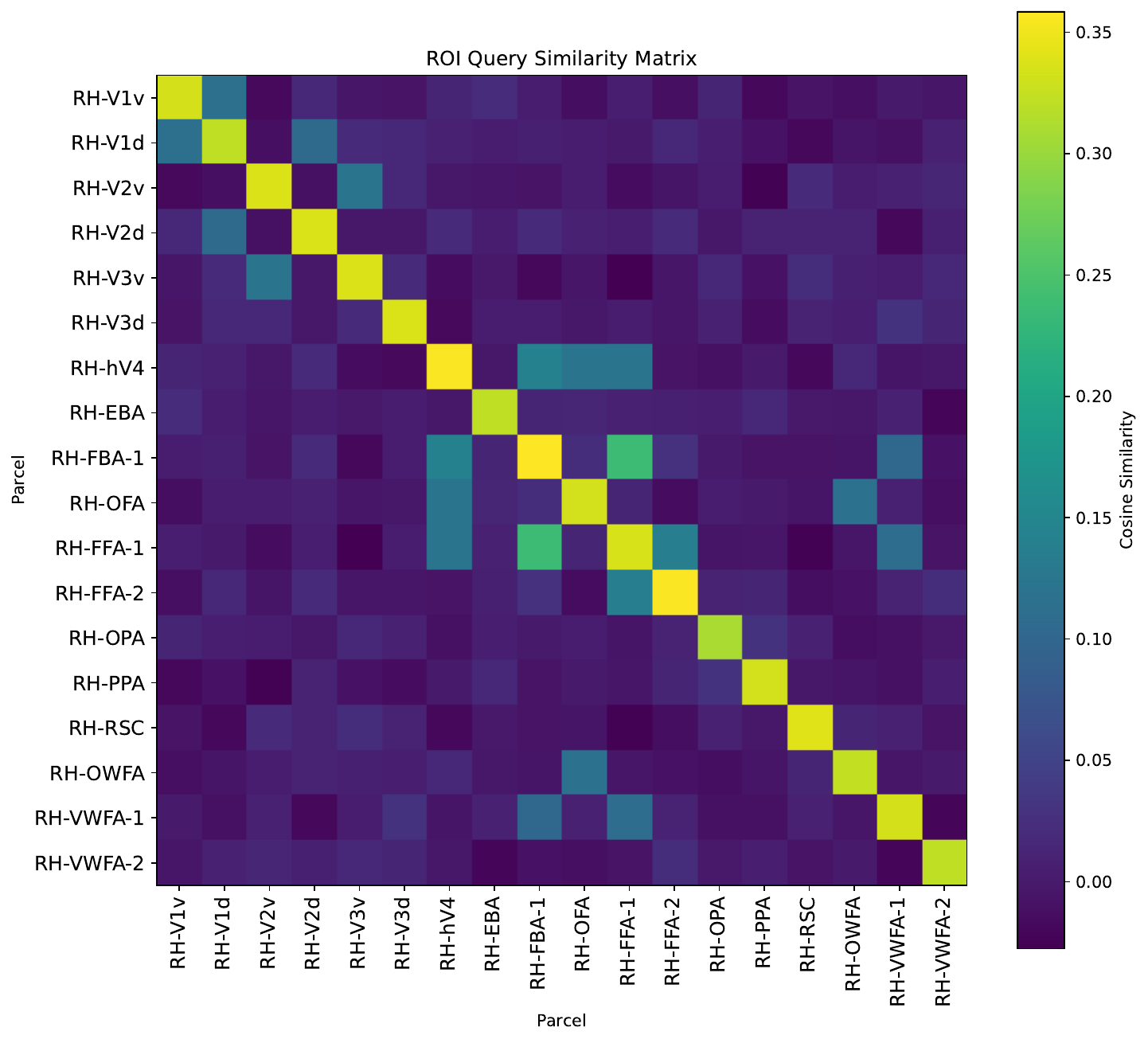}
    \caption{ROI query similarity, right hemisphere}
  \end{subfigure}

  \caption{Functional connectivity matrix vs ROI query similarity, subject 1.}
\end{figure}

\clearpage
\subsection{Attention maps}
\label{app:attention-maps}

\begin{figure}[h]
    \centering
    \includegraphics[width=1\linewidth]{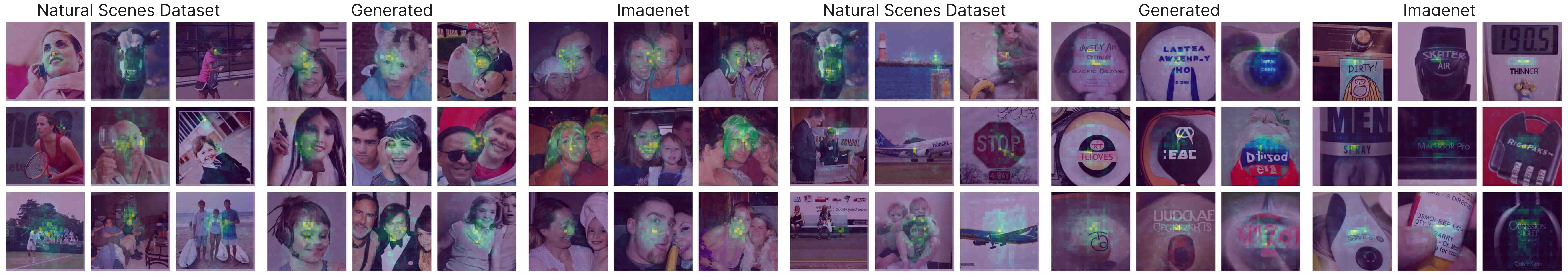}
    \caption{\textbf{Attention maps for aTL-faces and mfs-words.} Yellow areas represent the highest attention weights and purple represents the lowest. Held-out NSD images are selected based on ground-truth parcel activation, but attention maps are from the encoder since no ground-truth attention maps exist. See Figure~\ref{fig:s1_aTL-faces_0} for the original images.}
    \label{fig:s1_atlfaces_mfswords_att_map}
\end{figure}

While one can easily make guesses about parcel selectivity simply by looking at the prominent feature of the image, we can examine the attention scores from cross-attention  (Figure~\ref{fig:teaser}b) to interpret the selectivity of any parcel. The attention maps are visualized in Figure~\ref{fig:s1_atlfaces_mfswords_att_map} for the images from Figure~\ref{fig:s1_aTL-faces_0}.

As expected, the areas with highest attention weights for the aTL-faces parcel primarily overlap with faces, while the attention weights for the mfs-words parcel overlap with words (when present). Compared to previous work using a similar encoder model~\citep{adeli_transformer_2025}, our attention maps appear to follow the expected stimuli considerably less closely, possibly because our parcel boundaries were determined a priori and don't align well with the boundaries of the actual labeled area.

\clearpage
\subsection{Parcel locations and BrainDIVE/NSD images accompanying Figure~\ref{fig:teaser}}
\label{app:teaser-locations}

\begin{figure}[ht]
  \centering
  \begin{subfigure}[t]{0.3\textwidth}
    \centering
    \includegraphics[width=\linewidth]{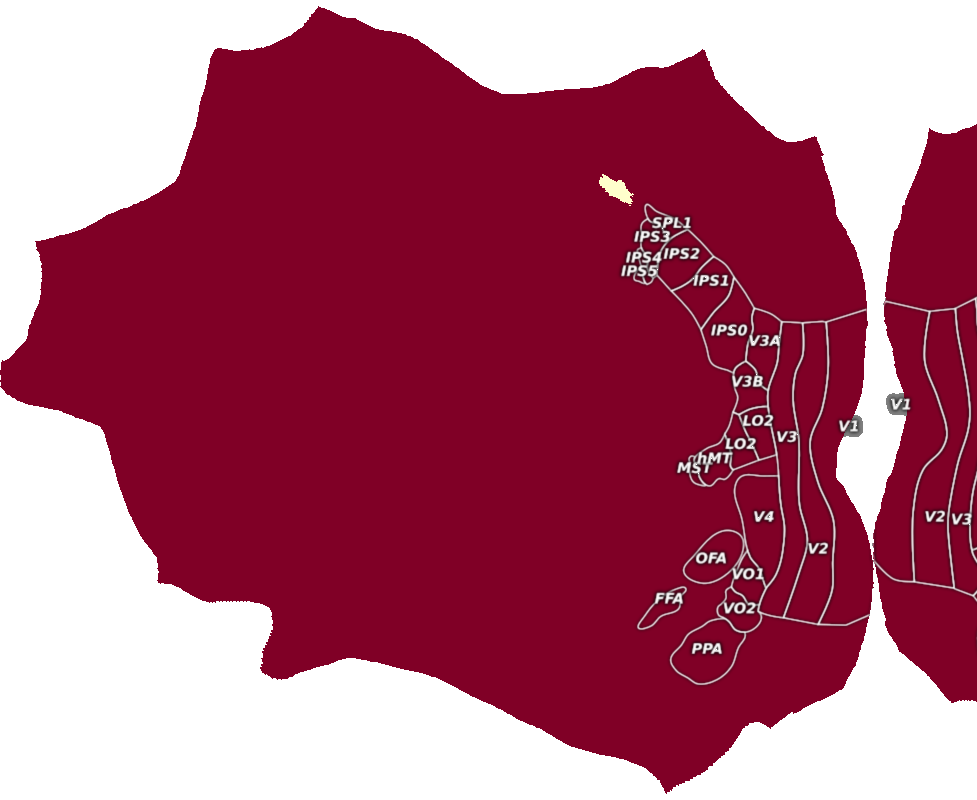}
    \caption{Parcel location in cortical flatmap.}
  \end{subfigure}
  \hfill
  \begin{subfigure}[t]{0.3\textwidth}
    \centering
    \includegraphics[width=\linewidth]{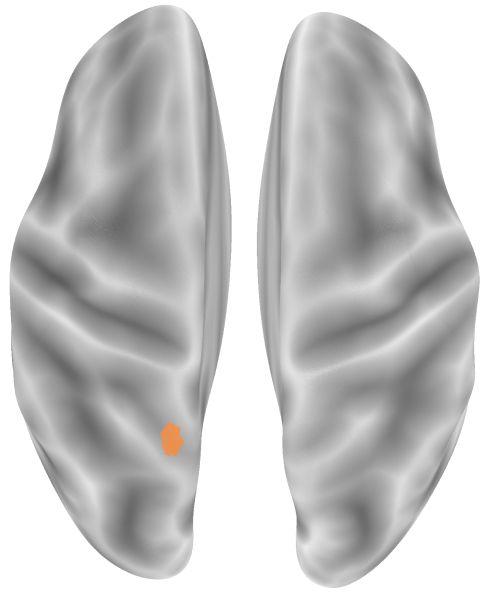}
    \caption{Parcel location on inflated map}
    \label{fig:sub2}
  \end{subfigure}

  \vspace{1em}

  \begin{subfigure}[t]{0.3\textwidth}
    \centering
    \includegraphics[width=\linewidth]{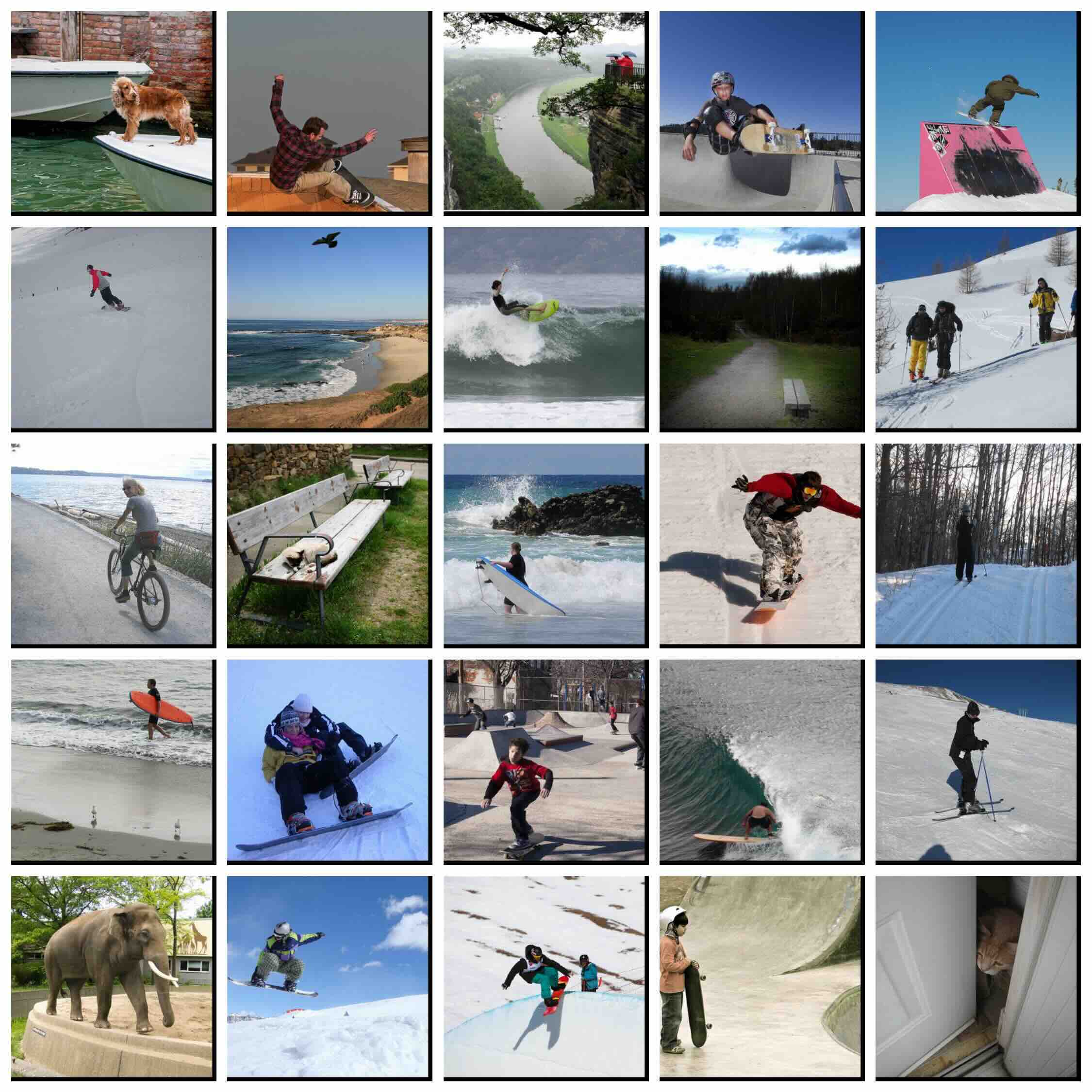}
    \caption{Held-out NSD that maximally activate the parcel.}
    \label{}
  \end{subfigure}
  \hfill
  \begin{subfigure}[t]{0.3\textwidth}
    \centering
    \includegraphics[width=\linewidth]{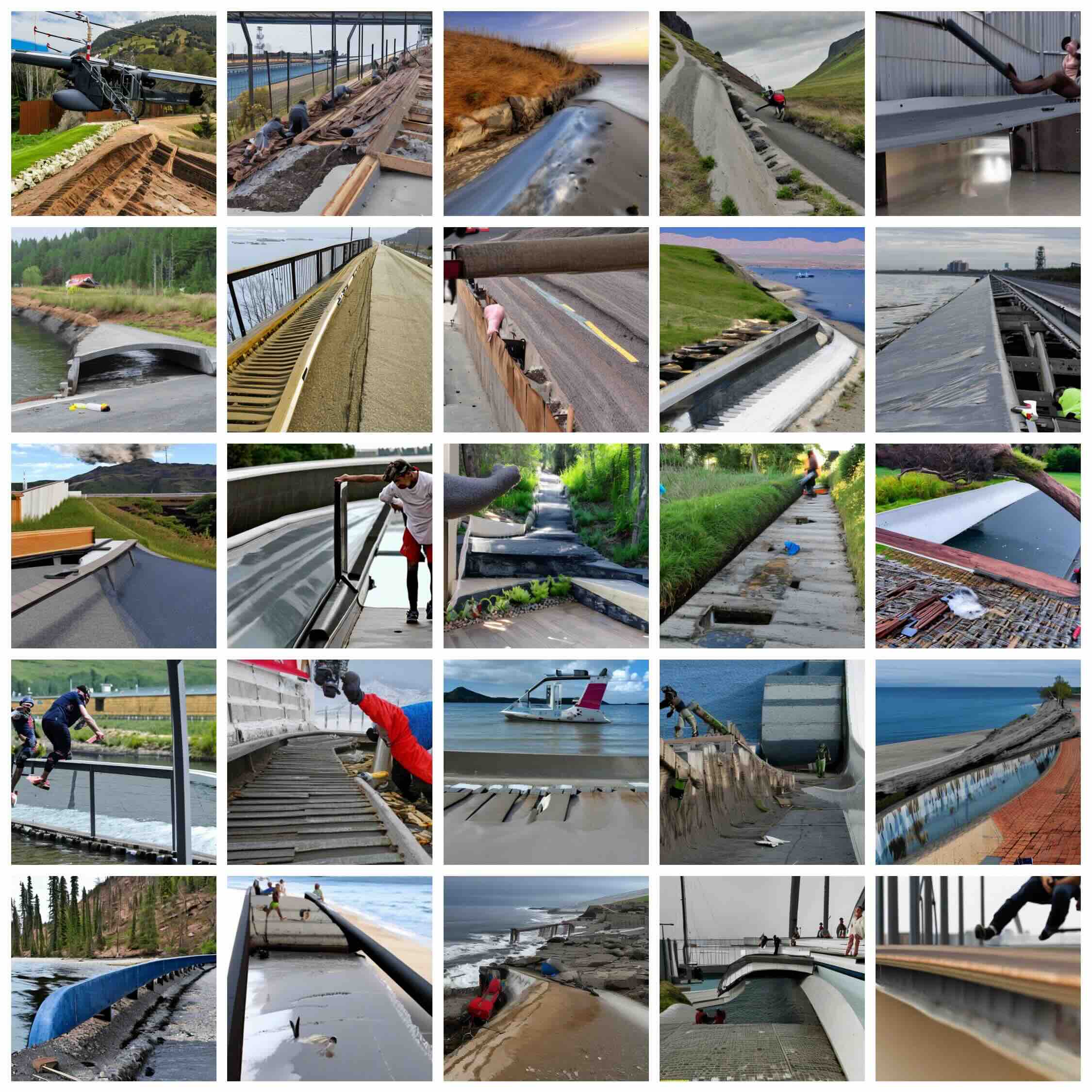}
    \caption{BrainDIVE generated images optimized for the parcel.}
    \label{}
  \end{subfigure}
  \hfill
  \begin{subfigure}[t]{0.3\textwidth}
    \centering
    \includegraphics[width=\linewidth]{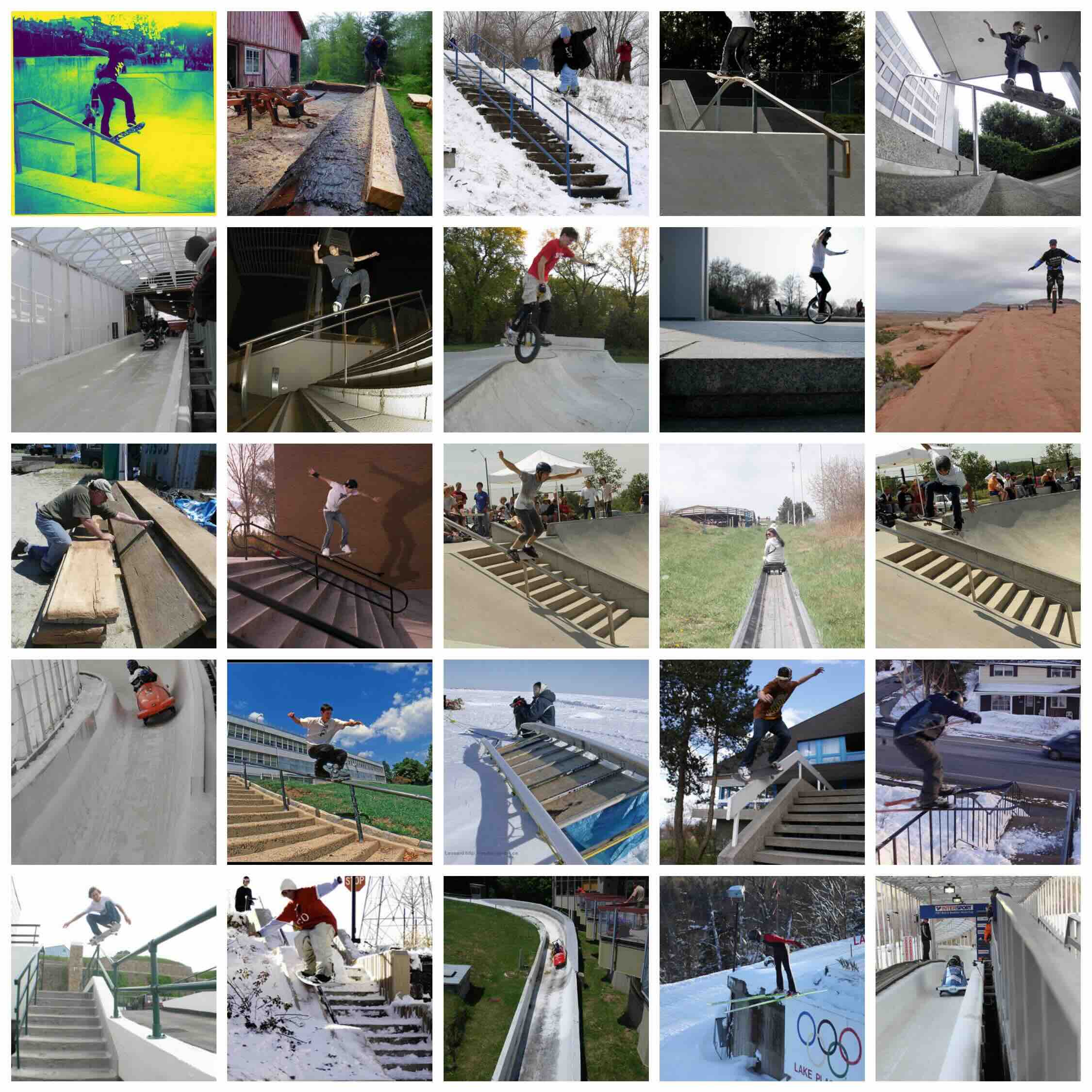}
    \caption{Maximally-activating ImageNet images.}
    \label{}
  \end{subfigure}

  \caption{Skateboarding parcel location and full collages, subject 1.}
  \label{fig:skateboarding_full}
\end{figure}

\begin{figure}[ht]
  \centering
  \begin{subfigure}[t]{0.3\textwidth}
    \centering
    \includegraphics[width=\linewidth]{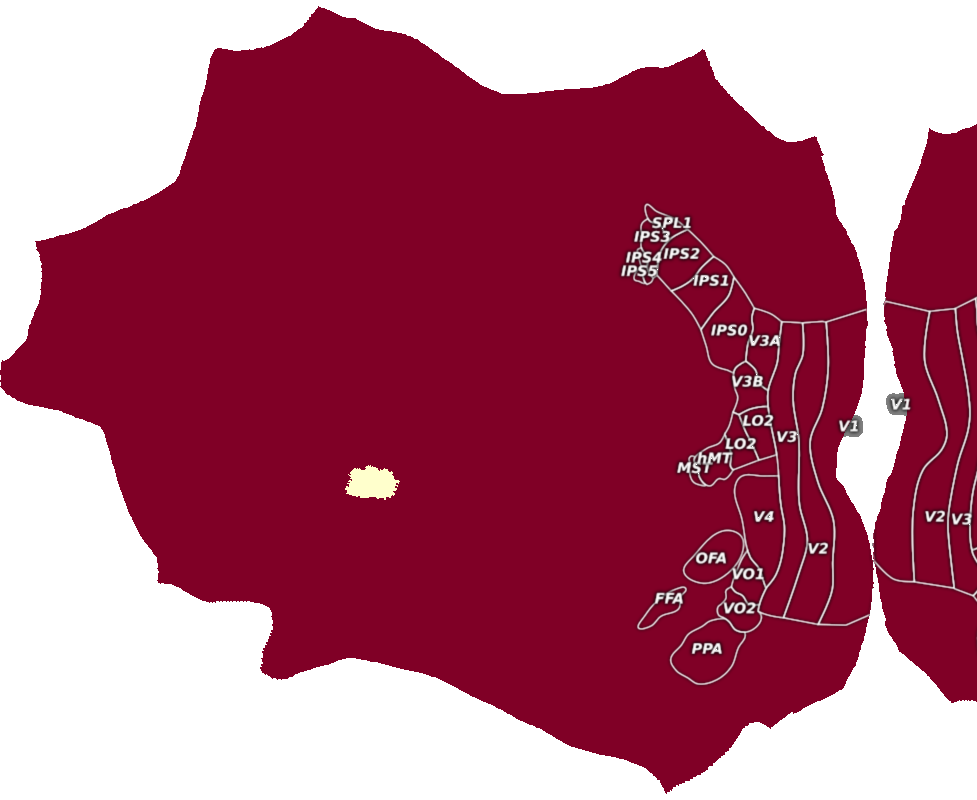}
    \caption{Parcel location in cortical flatmap.}
  \end{subfigure}
  \hfill
  \begin{subfigure}[t]{0.3\textwidth}
    \centering
    \includegraphics[width=\linewidth]{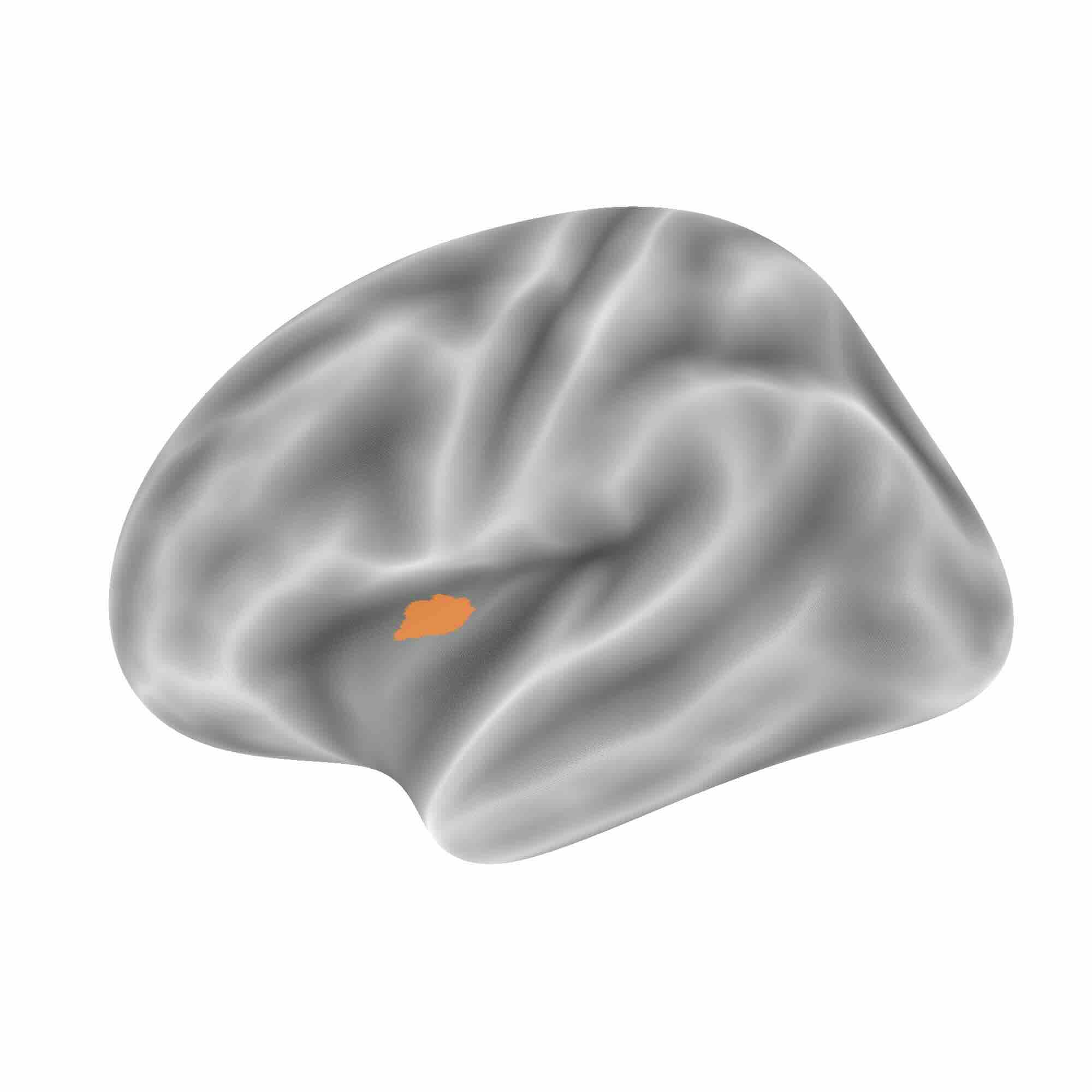}
    \caption{Parcel location on inflated map}
    \label{}
  \end{subfigure}

  \vspace{1em}

  \begin{subfigure}[t]{0.3\textwidth}
    \centering
    \includegraphics[width=\linewidth]{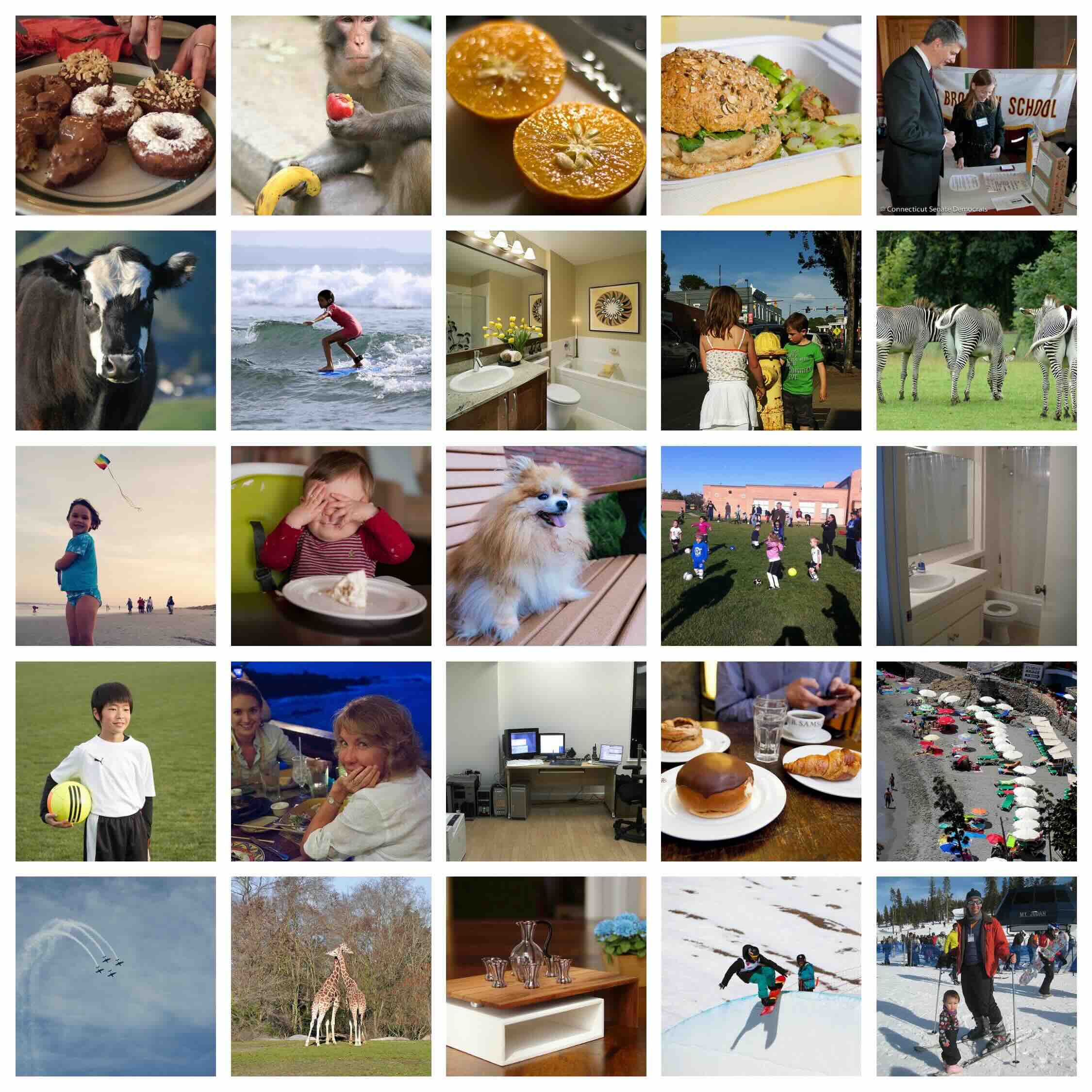}
    \caption{Held-out NSD that maximally activate the parcel.}
    \label{}
  \end{subfigure}
  \hfill
  \begin{subfigure}[t]{0.3\textwidth}
    \centering
    \includegraphics[width=\linewidth]{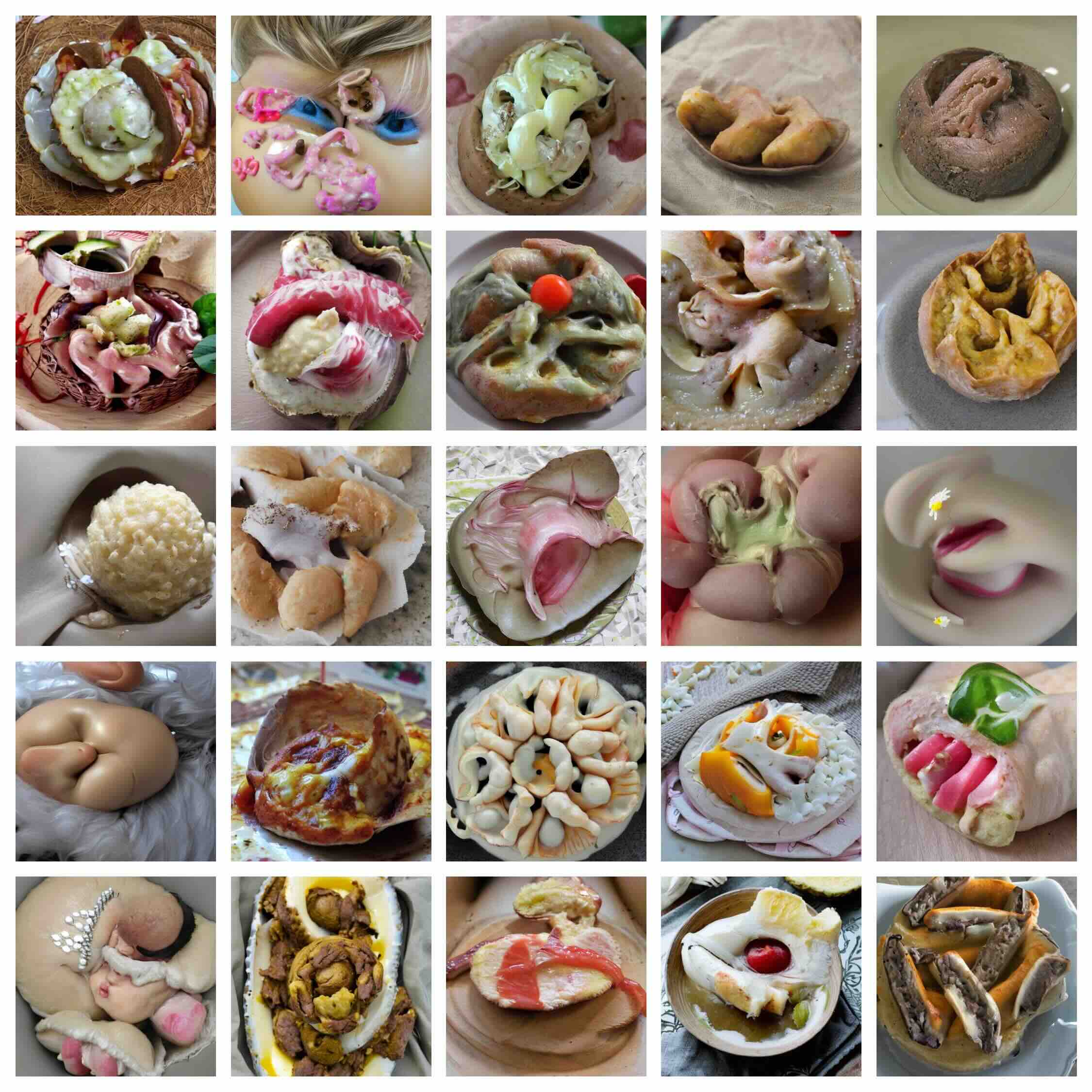}
    \caption{BrainDIVE generated images optimized for the parcel.}
    \label{}
  \end{subfigure}
  \hfill
  \begin{subfigure}[t]{0.3\textwidth}
    \centering
    \includegraphics[width=\linewidth]{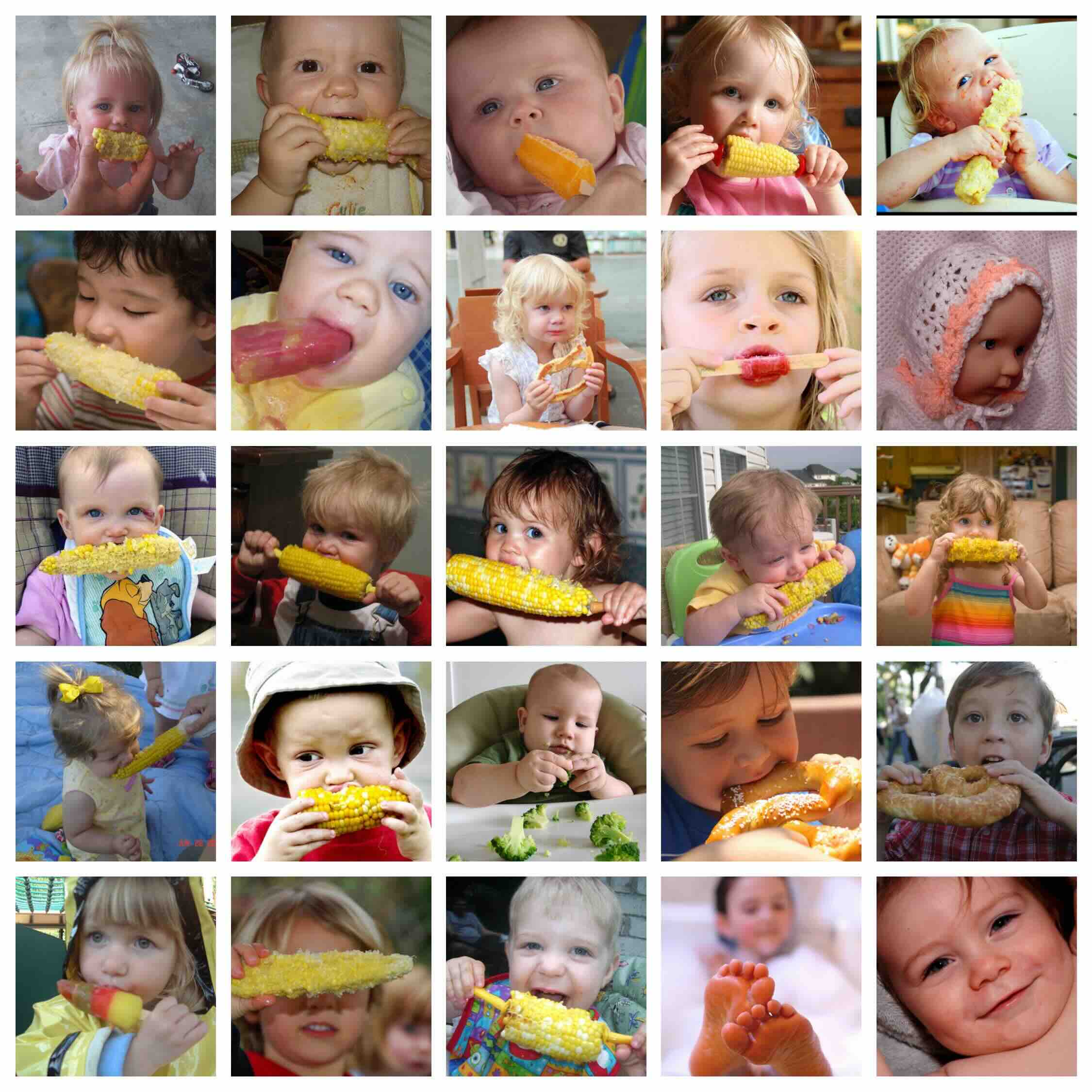}
    \caption{Maximally-activating ImageNet images.}
    \label{}
  \end{subfigure}

  \caption{Child eating parcel location and full collages, subject 1.}
  \label{fig:corn_eating_full}
\end{figure}

\clearpage
\subsection{Labeling top subject-specific parcels}
\label{app:labeling-top-subj-specifici-parcels}

For each of the parcels shown in Figure~\ref{fig:roc_within_subj}, we show the top ImageNet images and the caption generated by GPT-4o (see Appendix~\ref{app:prompts} for prompt details).

\begin{table}[ht]
  \centering
  \small
  \begin{subtable}[t]{0.45\textwidth}
    \centering
    \begin{tabular}{
    >{\raggedright\arraybackslash}p{0.22\textwidth}
    >{\centering\arraybackslash}m{0.25\textwidth}
    >{\raggedright\arraybackslash}m{0.38\textwidth}}
      \toprule
      Parcel & ImageNet top 4 & Caption \\
      \midrule
      S1 LH 211 & \includegraphics[width=\linewidth]{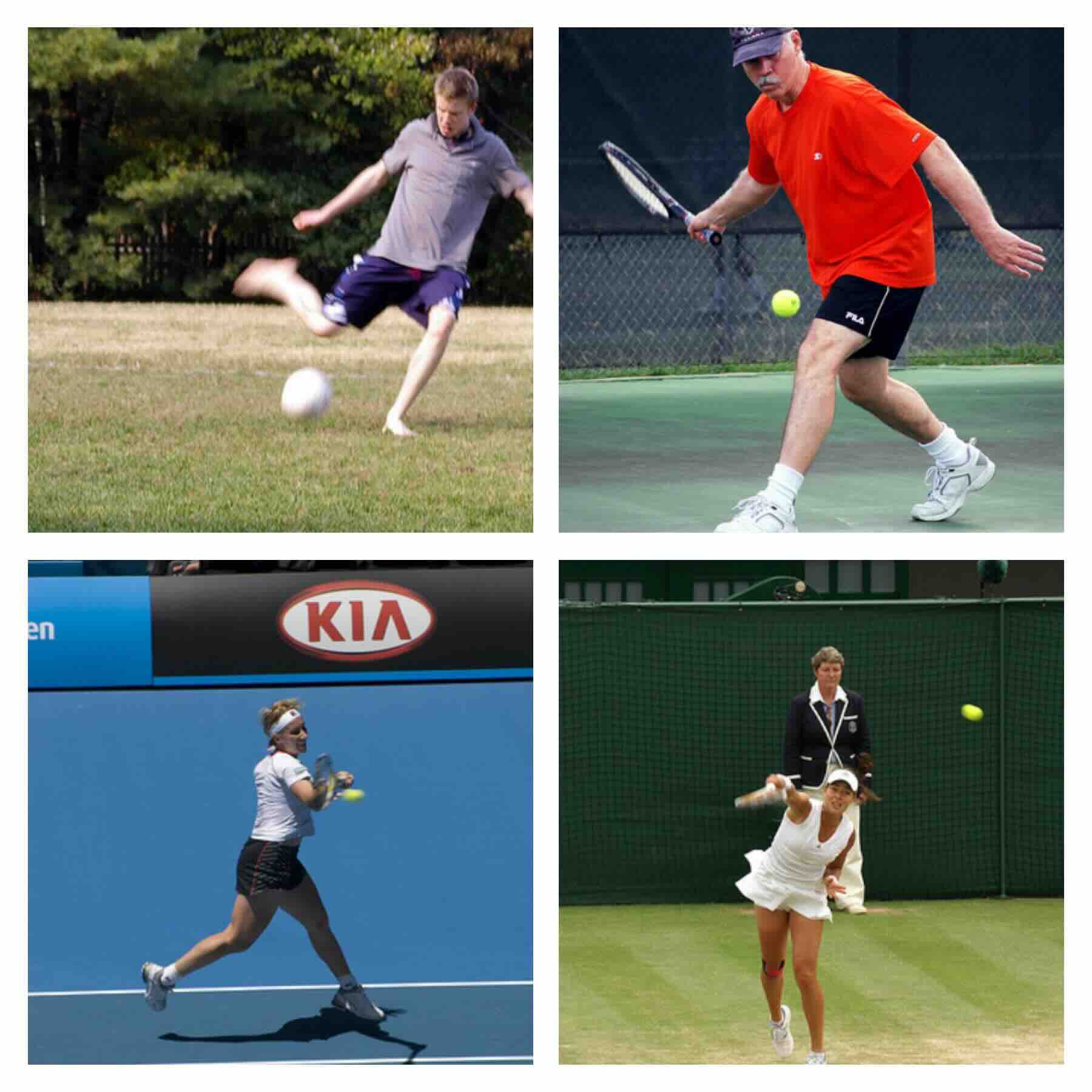} & Sports, Tennis, Soccer \\
      S1 RH 373 & \includegraphics[width=\linewidth]{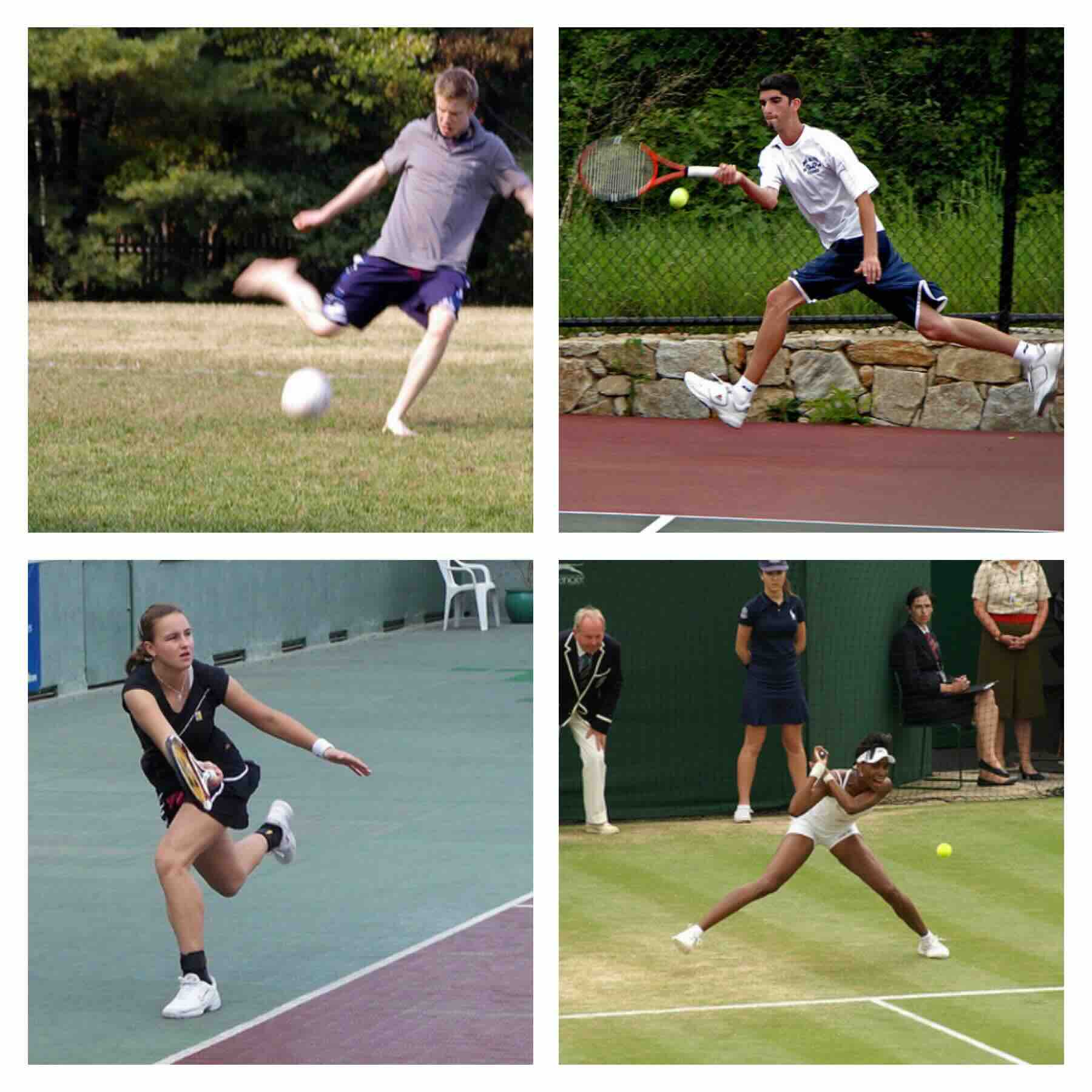} & Soccer, Tennis, Sports \\
        S1 LH 238 & \includegraphics[width=\linewidth]{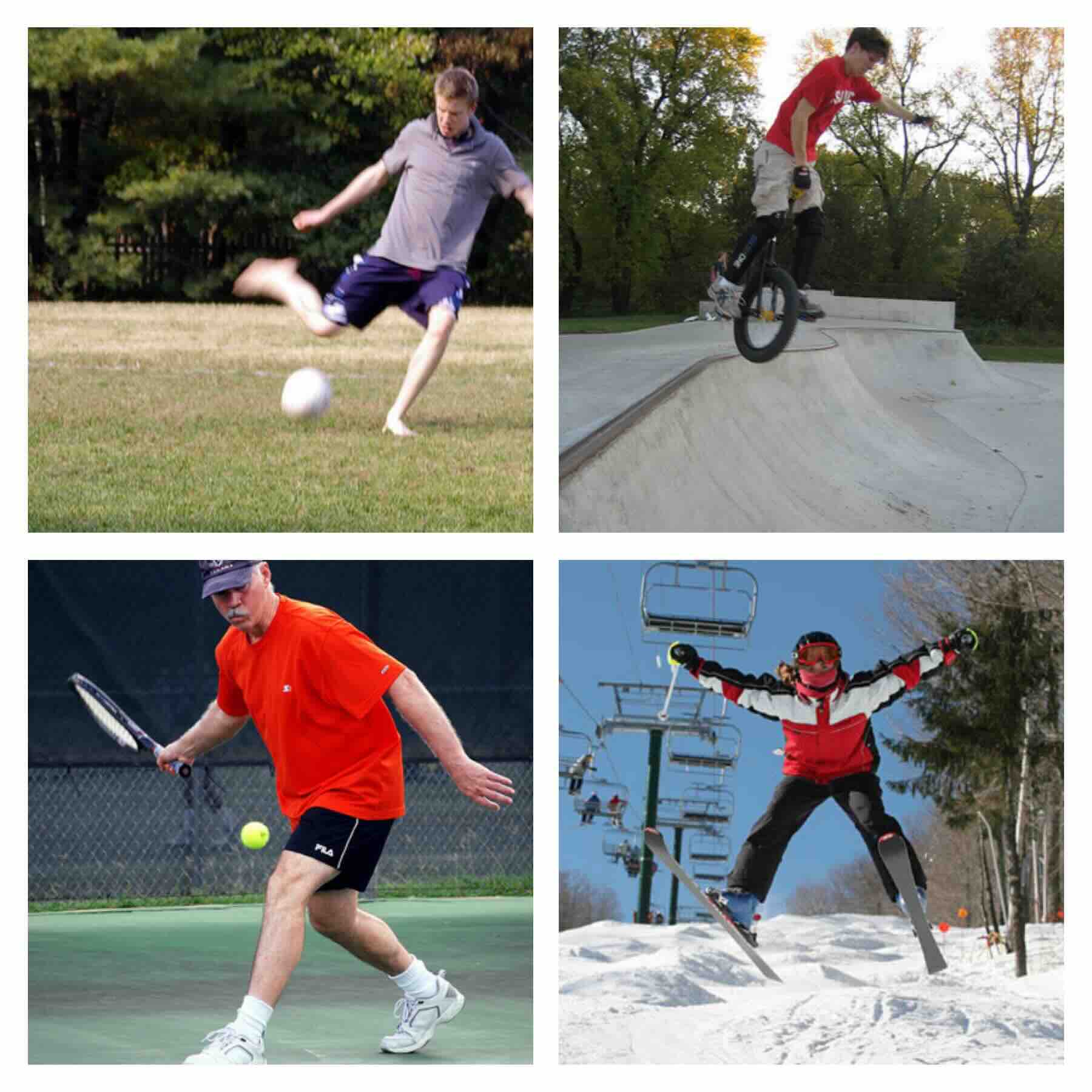} & Sports, Tennis, Soccer \\
      S1 RH 456 & \includegraphics[width=\linewidth]{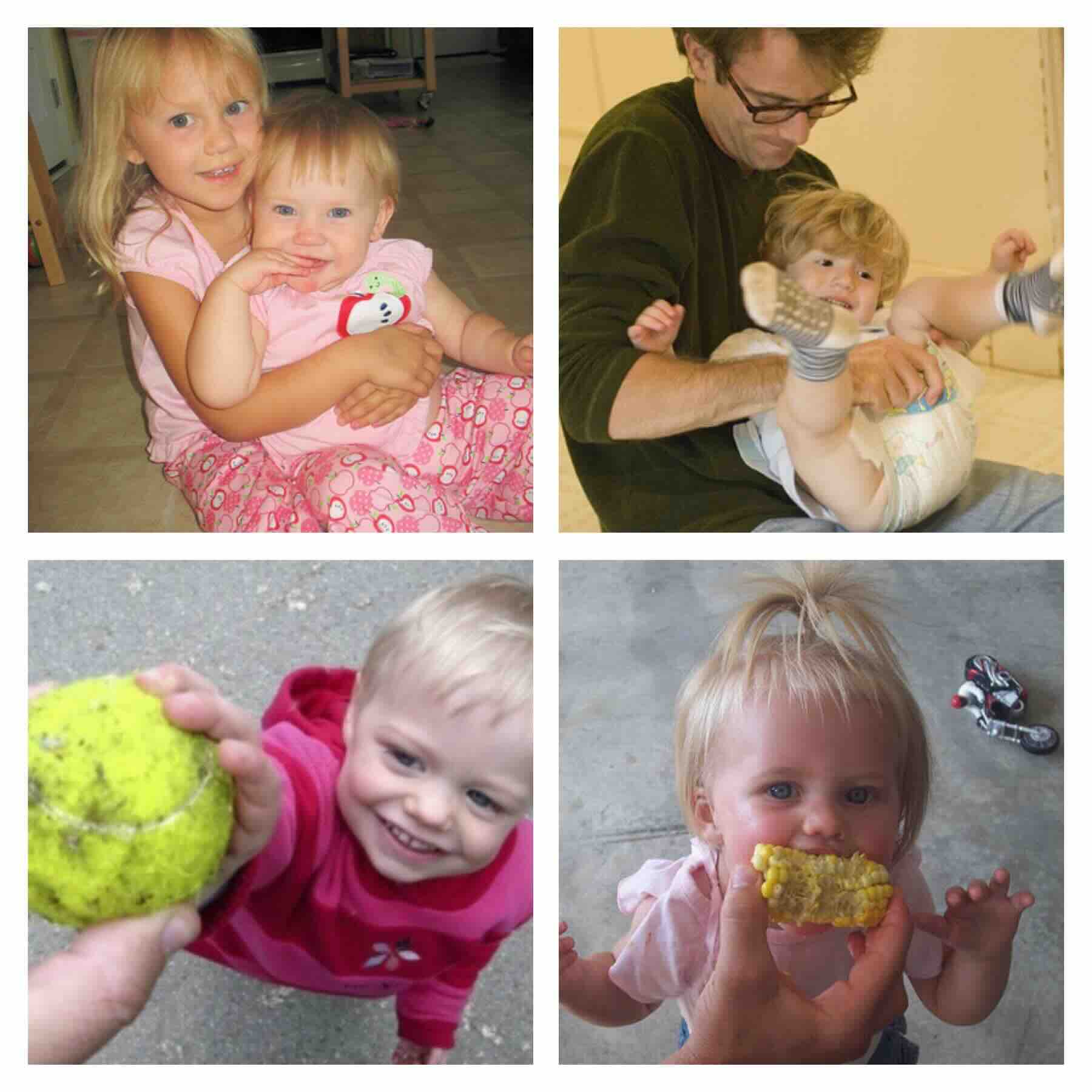} & Babies, Children, Family \\
      S1 LH 496 & \includegraphics[width=\linewidth]{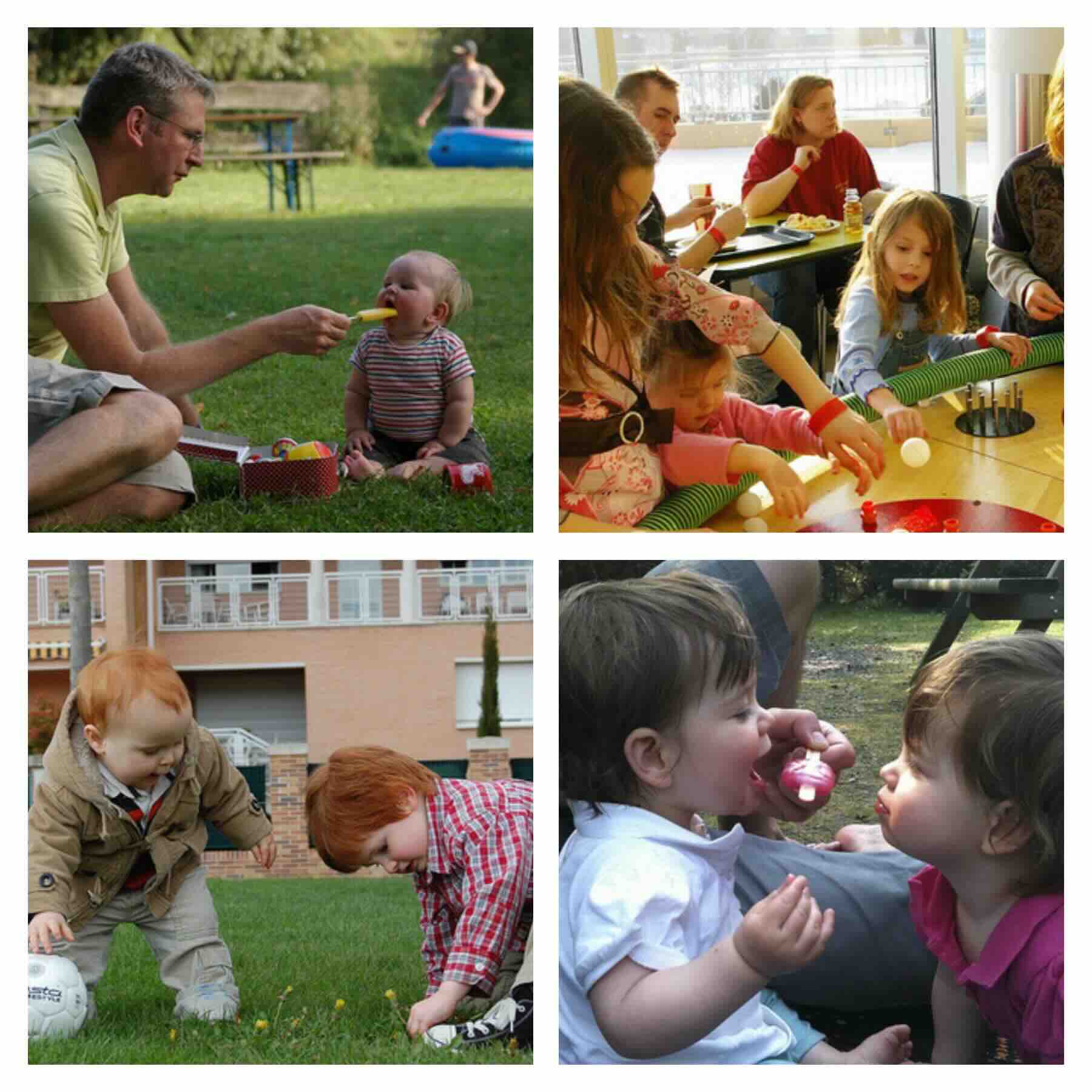} & Family, Parenting, Children \\
        \addlinespace[2pt]                     
        \cmidrule[0.4pt]{1-3}                  
        \addlinespace[2pt]
      S2 LH 480 & \includegraphics[width=\linewidth]{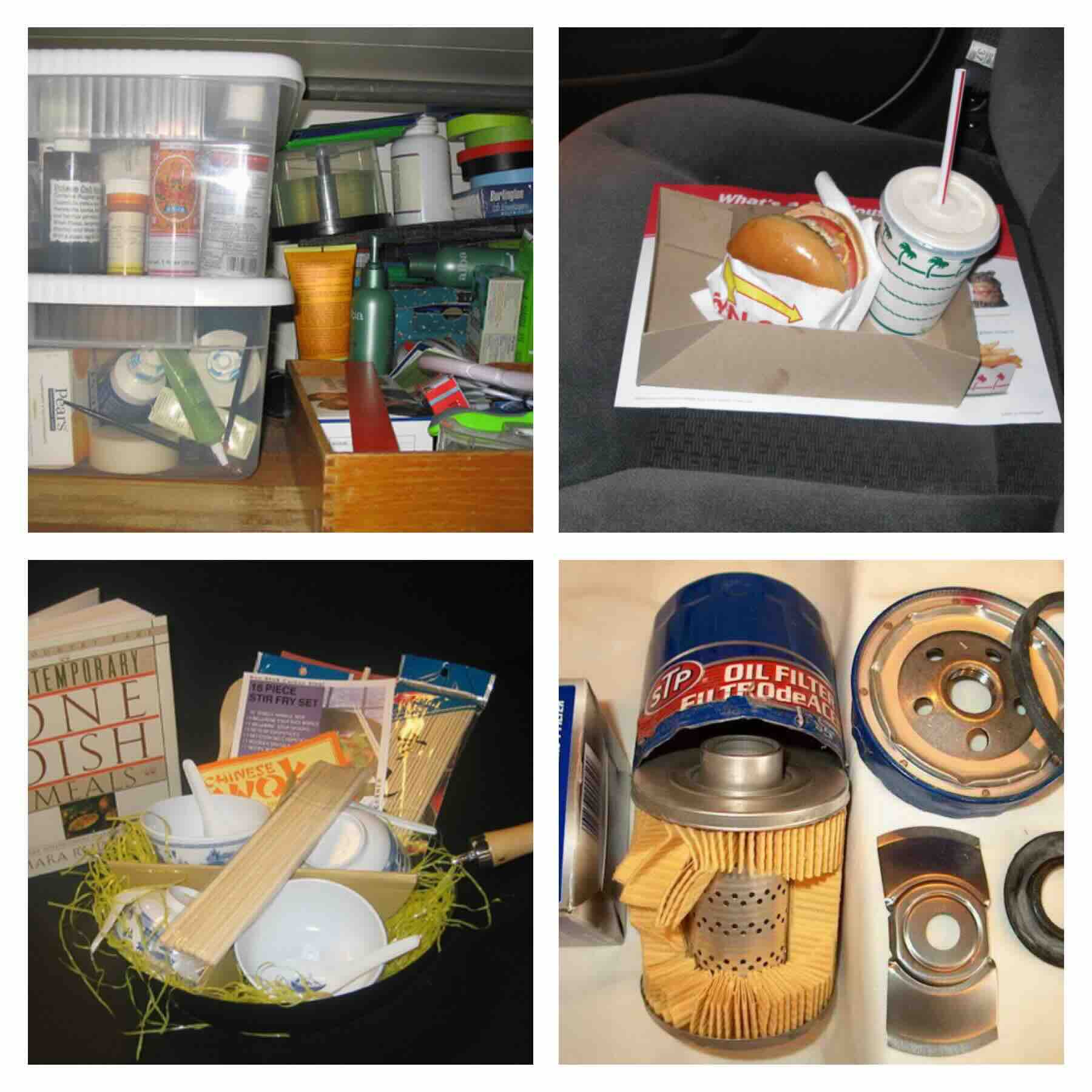}& Food storage, Office supplies, Kitchen appliances \\
      S2 RH 391 & \includegraphics[width=\linewidth]{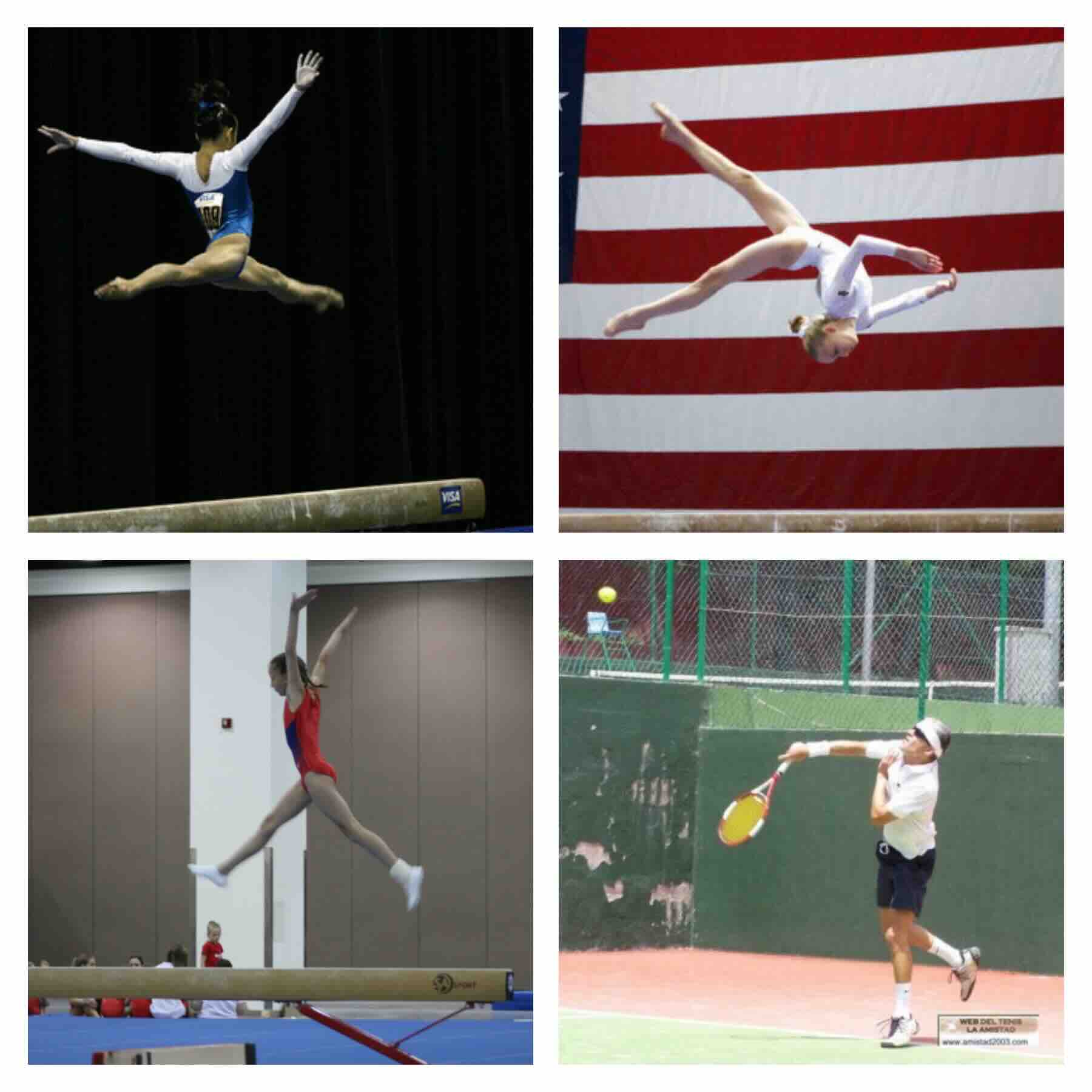} & Gymnastics, Dance, Tennis \\
      S2 LH 500 & \includegraphics[width=\linewidth]{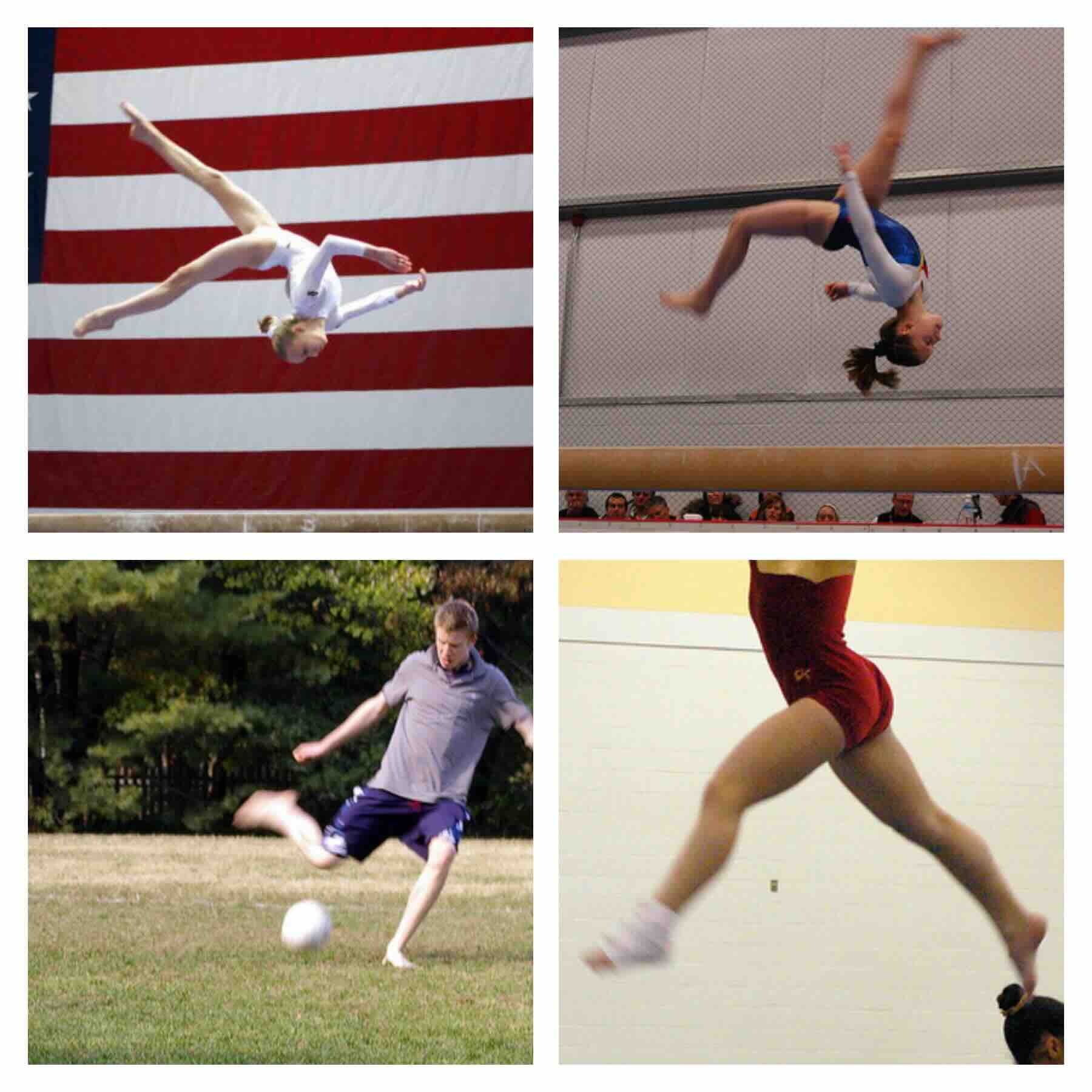} & Tennis, Gymnastics, Baseball \\
      S2 LH 496 & \includegraphics[width=\linewidth]{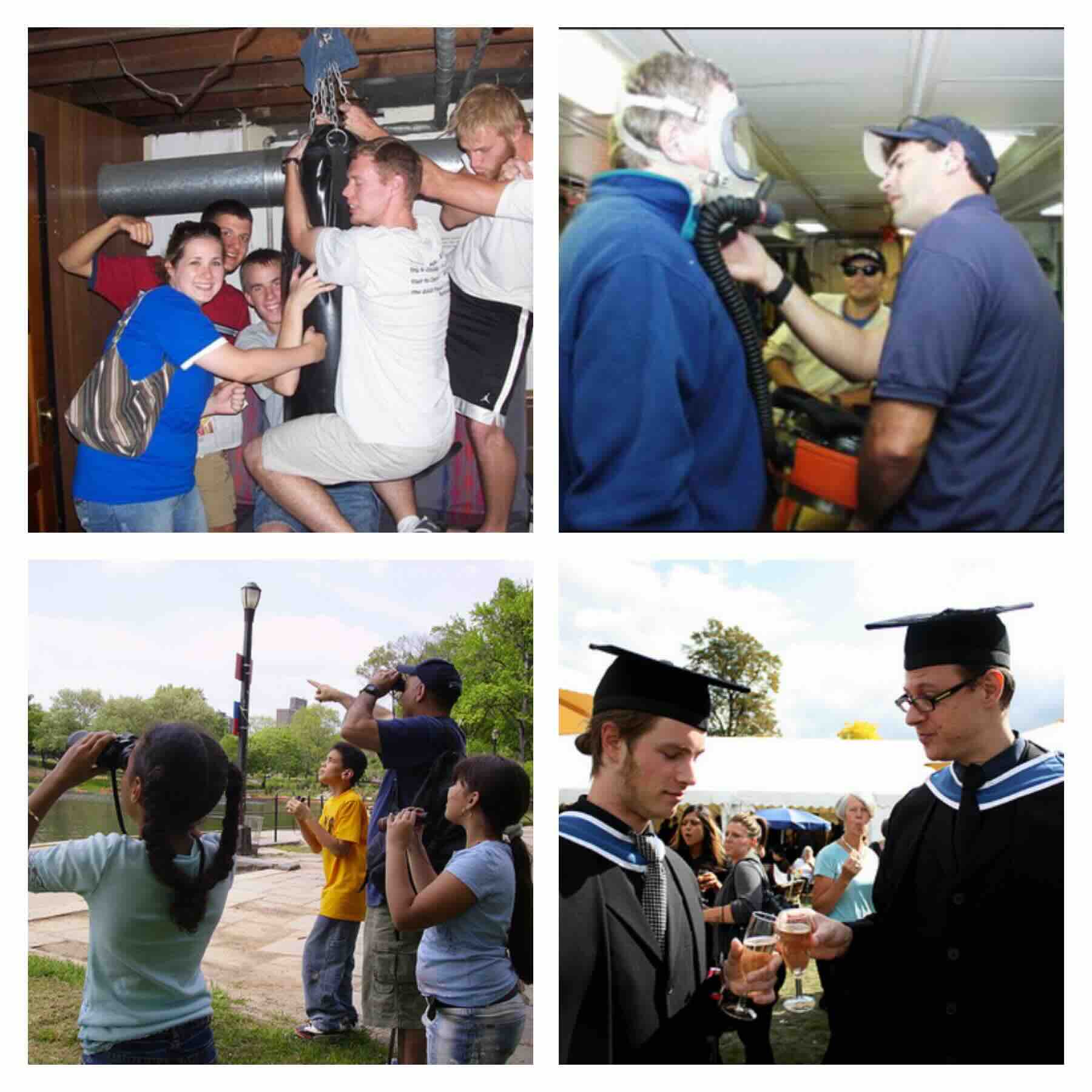} & Social Gatherings, Education and Learning, Sports and Recreation \\
      S2 LH 365 & \includegraphics[width=\linewidth]{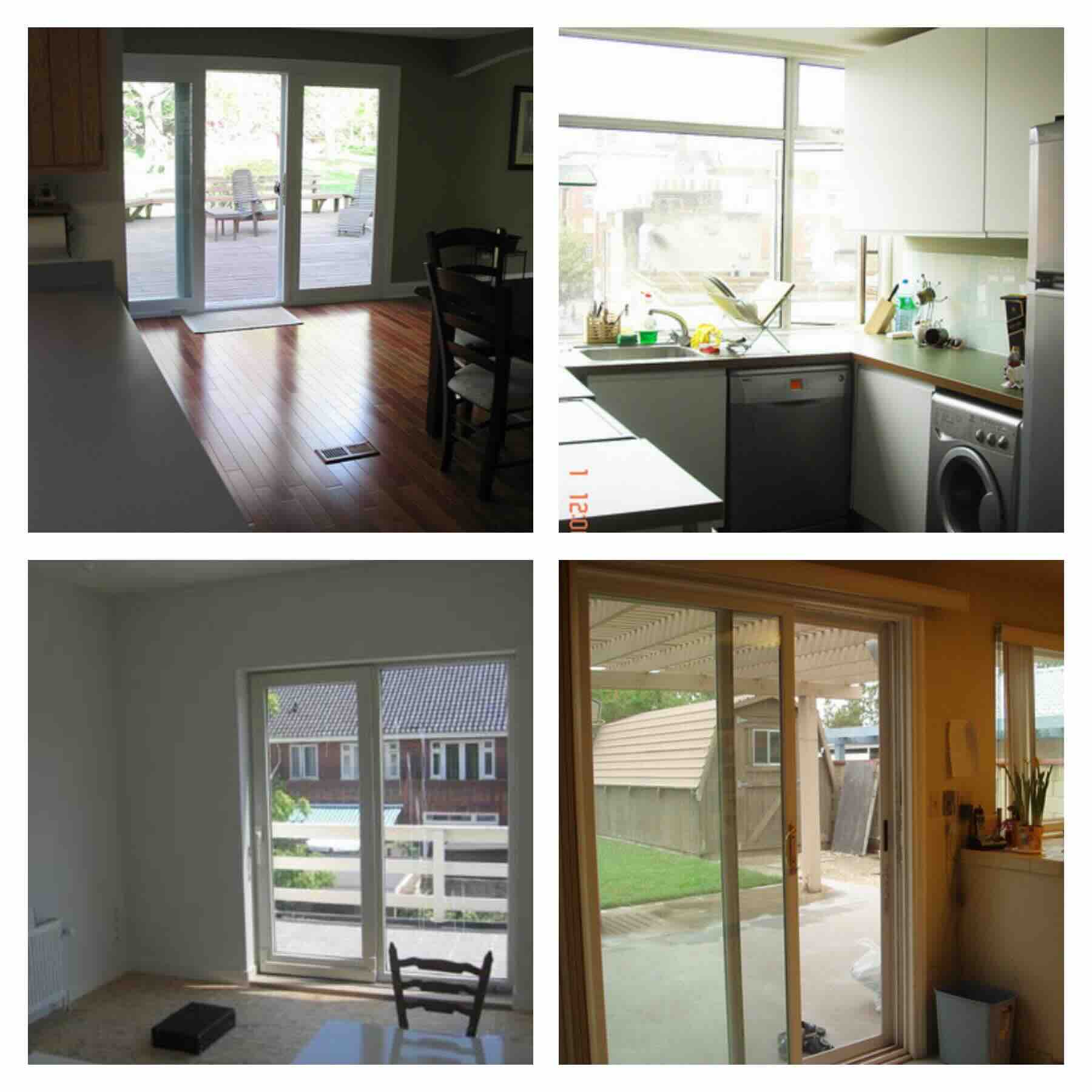} & Interior Design, Living Spaces, Furniture \\
      \bottomrule
    \end{tabular}
    \label{tab:sub2}
  \end{subtable}
  \hfill 
  \begin{subtable}[t]{0.45\textwidth}
    \centering
    \begin{tabular}{
    >{\raggedright\arraybackslash}p{0.22\textwidth}
    >{\centering\arraybackslash}m{0.25\textwidth}
    >{\raggedright\arraybackslash}m{0.38\textwidth}}
      \toprule
      Parcel & ImageNet top 4 & Caption \\
      \midrule
      S5 RH 476 & \includegraphics[width=\linewidth]{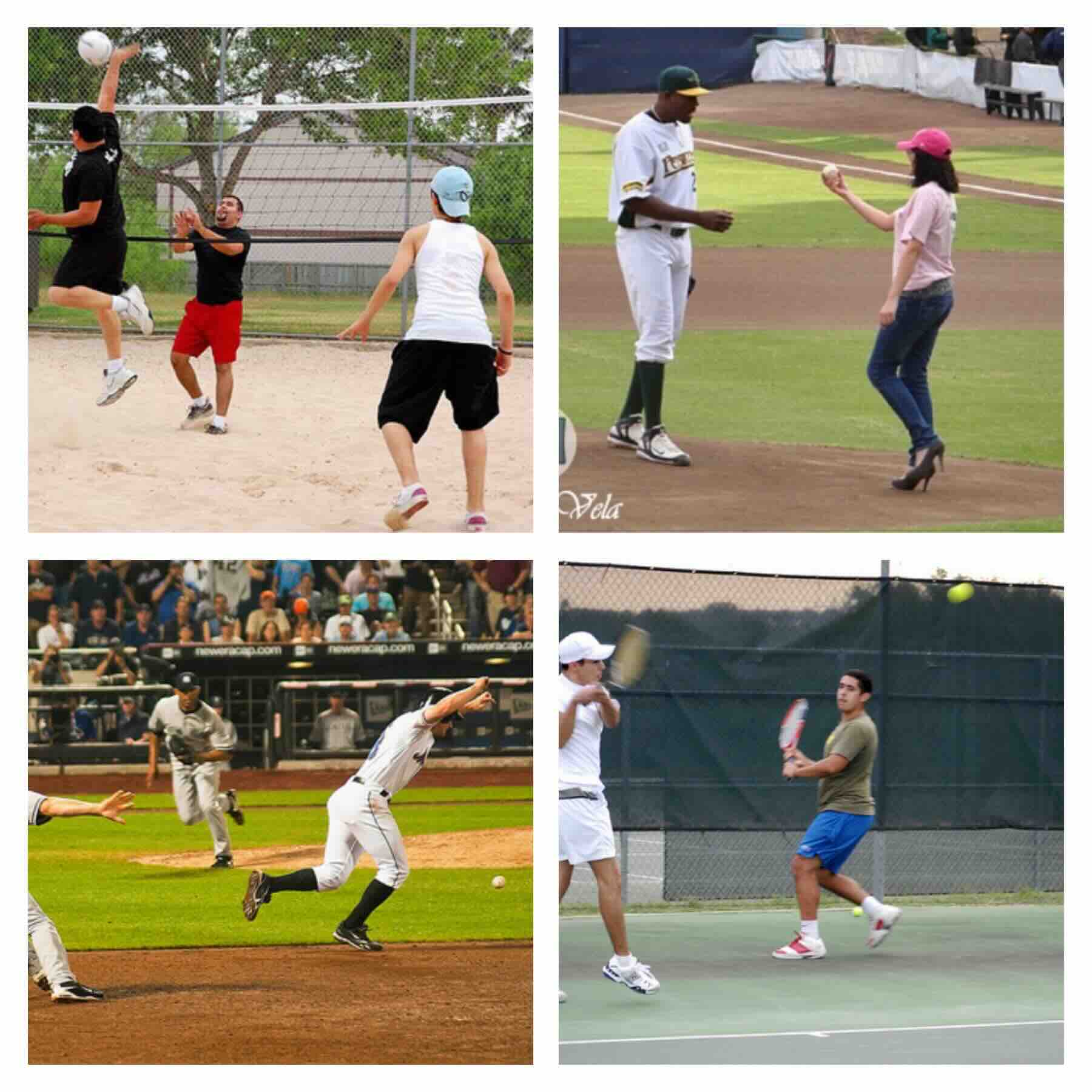} & Sports, Team sports, Outdoor activities \\
      S5 LH 239 & \includegraphics[width=\linewidth]{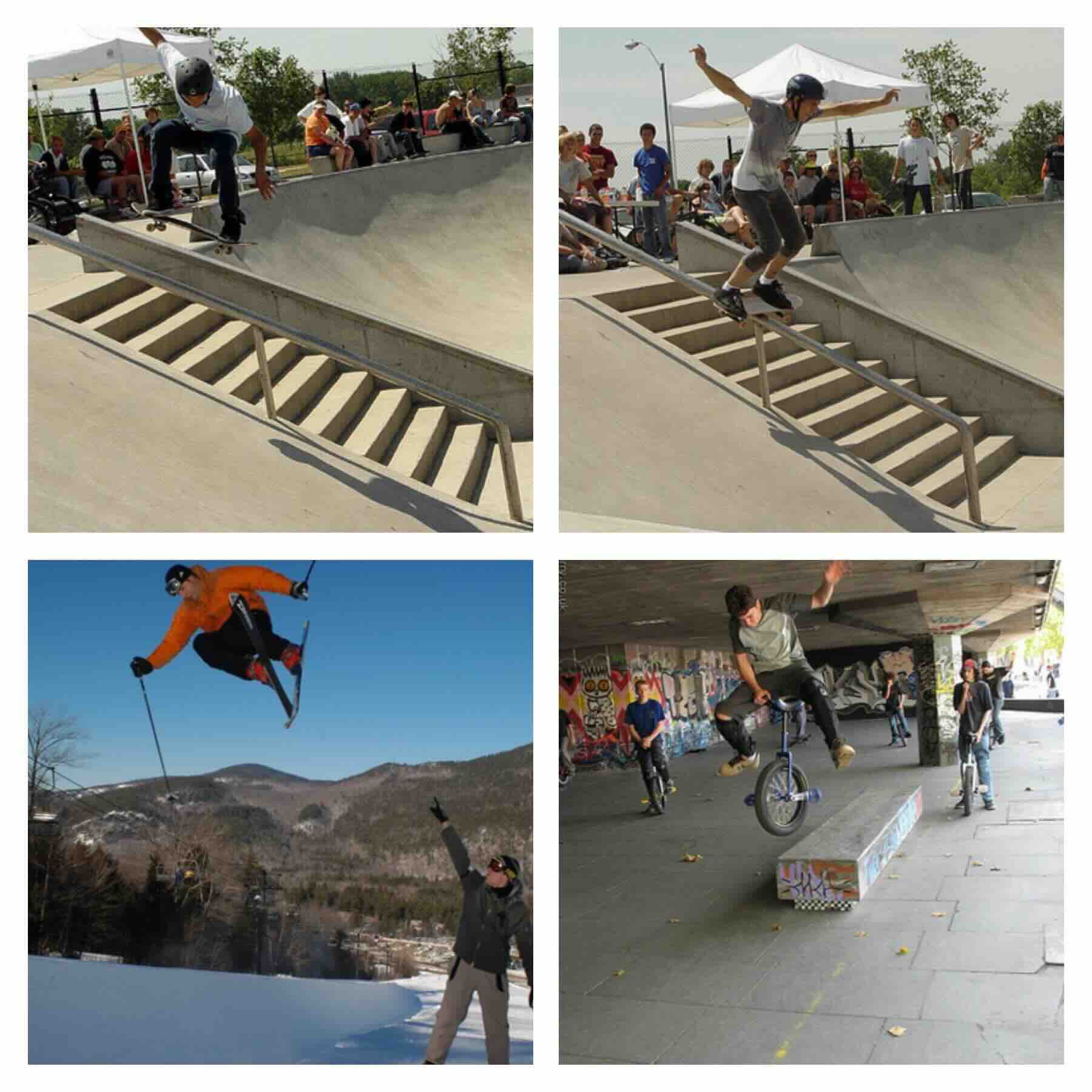} & Extreme sports, Skateboarding, BMX biking \\
      S5 LH 367 & \includegraphics[width=\linewidth]{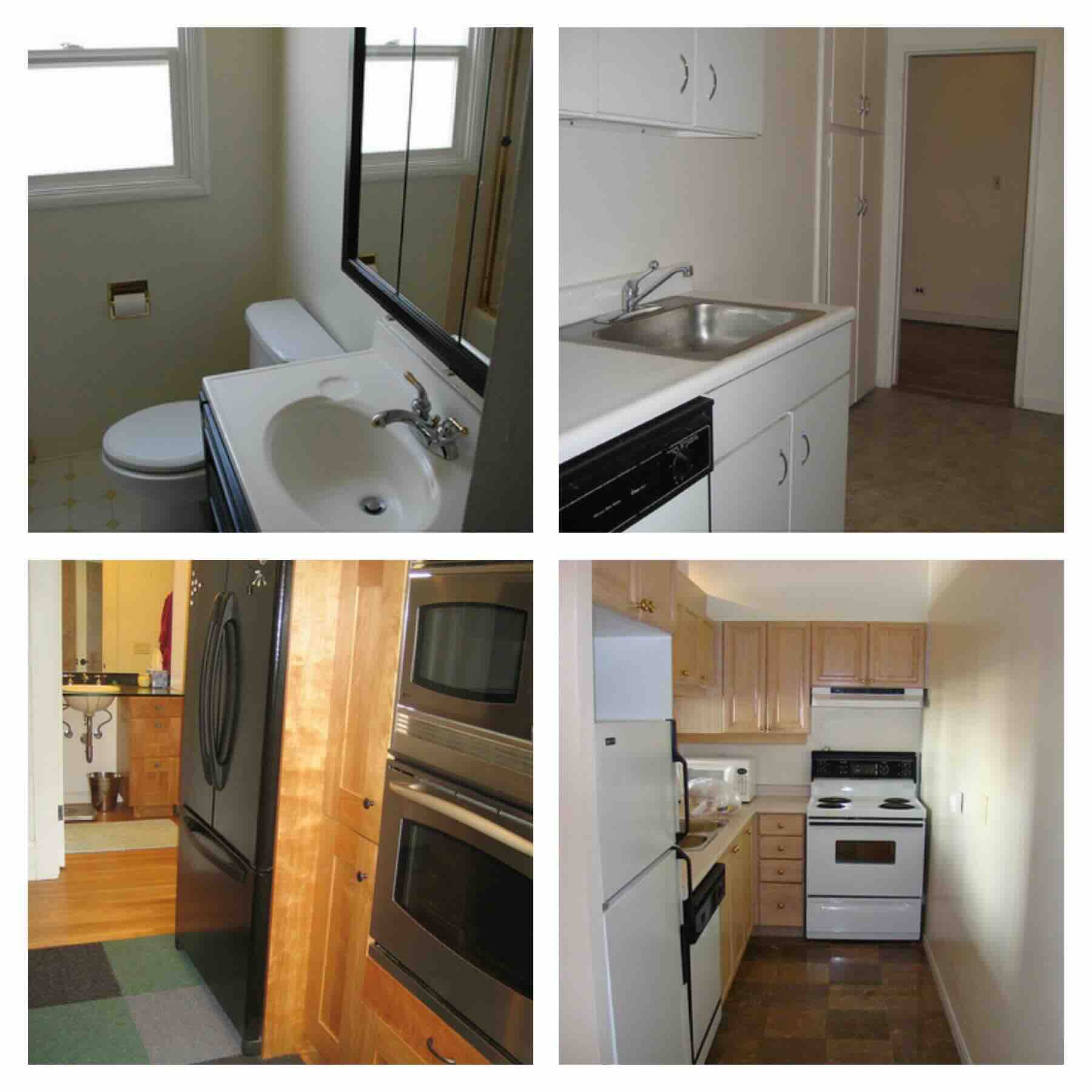} & Interior Design, Home Decor, Kitchen \\
      S5 RH 391 & \includegraphics[width=\linewidth]{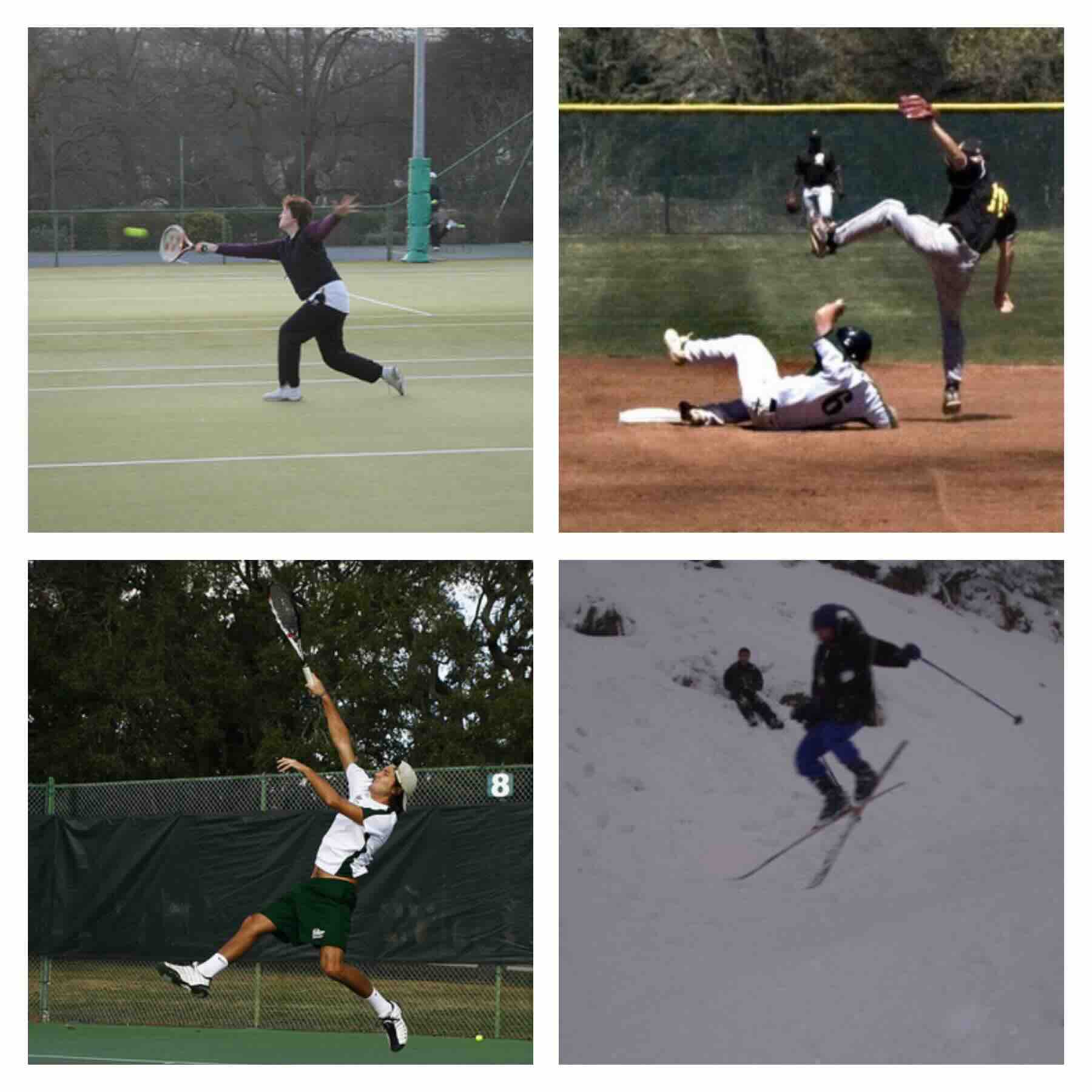} & Soccer, Ice hockey, Tennis \\
      S5 LH 167 & \includegraphics[width=\linewidth]{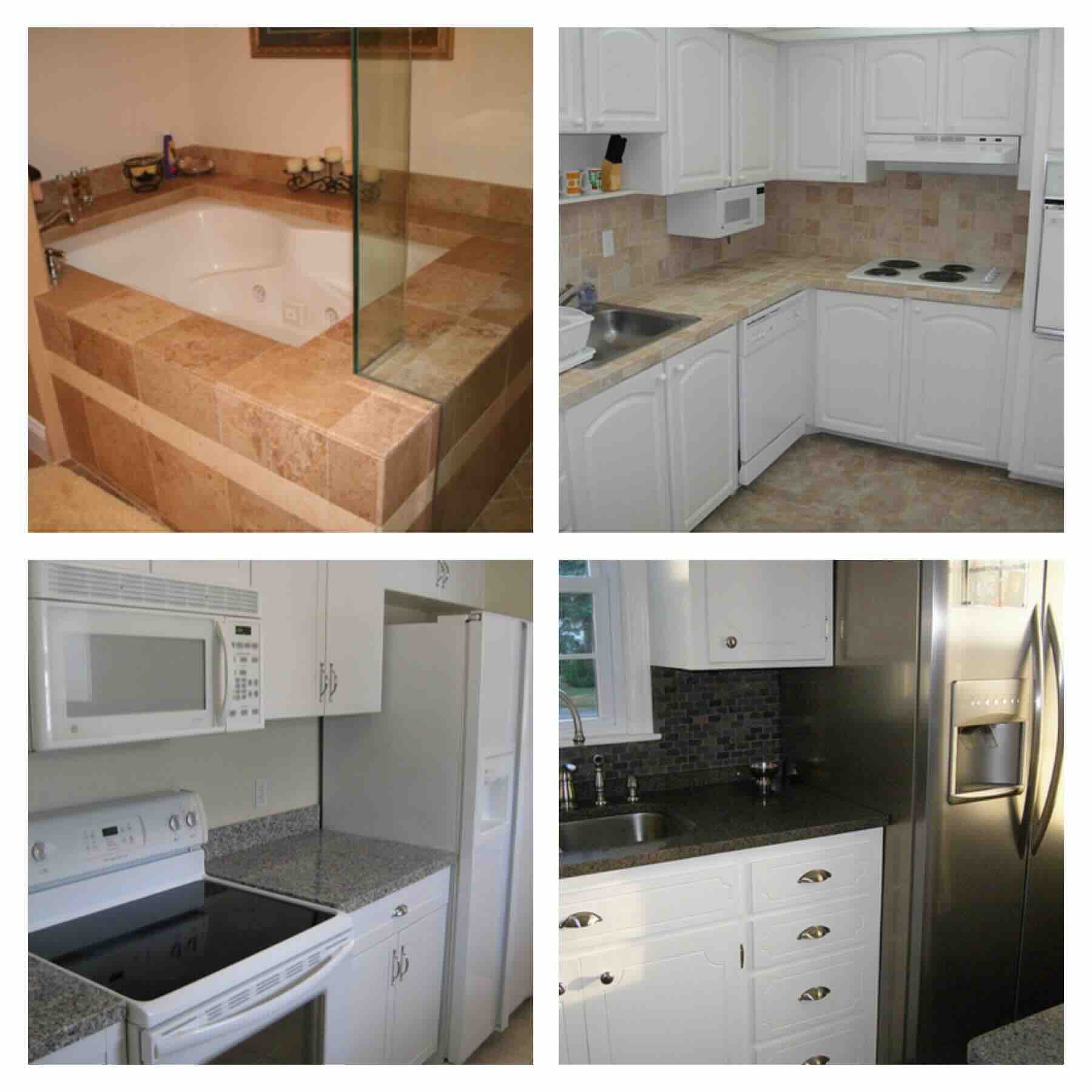} & Kitchen, Appliances, Cabinets \\
        \addlinespace[2pt]                
        \cmidrule[0.4pt]{1-3}    
        \addlinespace[2pt]
      S7 RH 474 & \includegraphics[width=\linewidth]{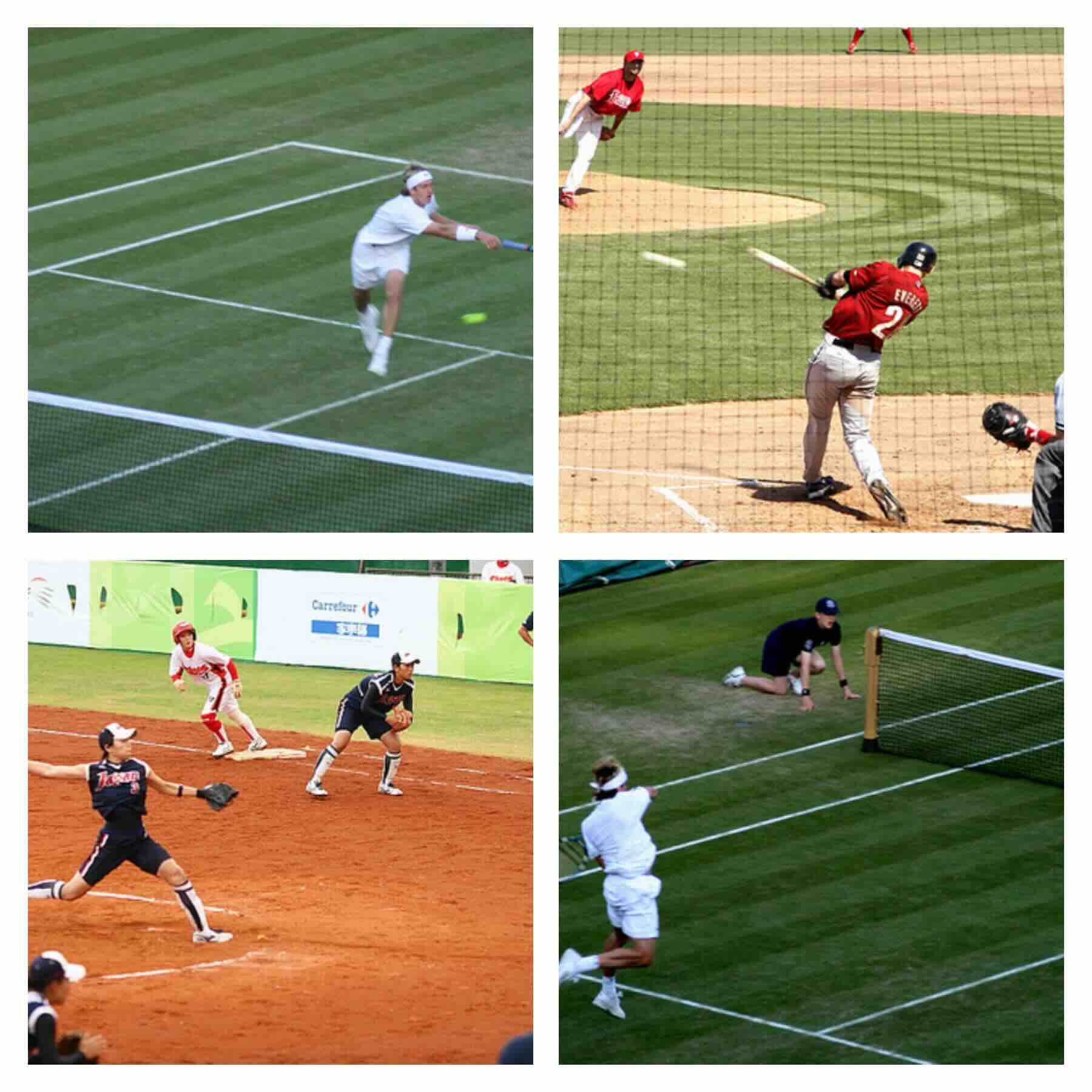}& Tennis, Baseball, Sports \\
      S7 RH 477 & \includegraphics[width=\linewidth]{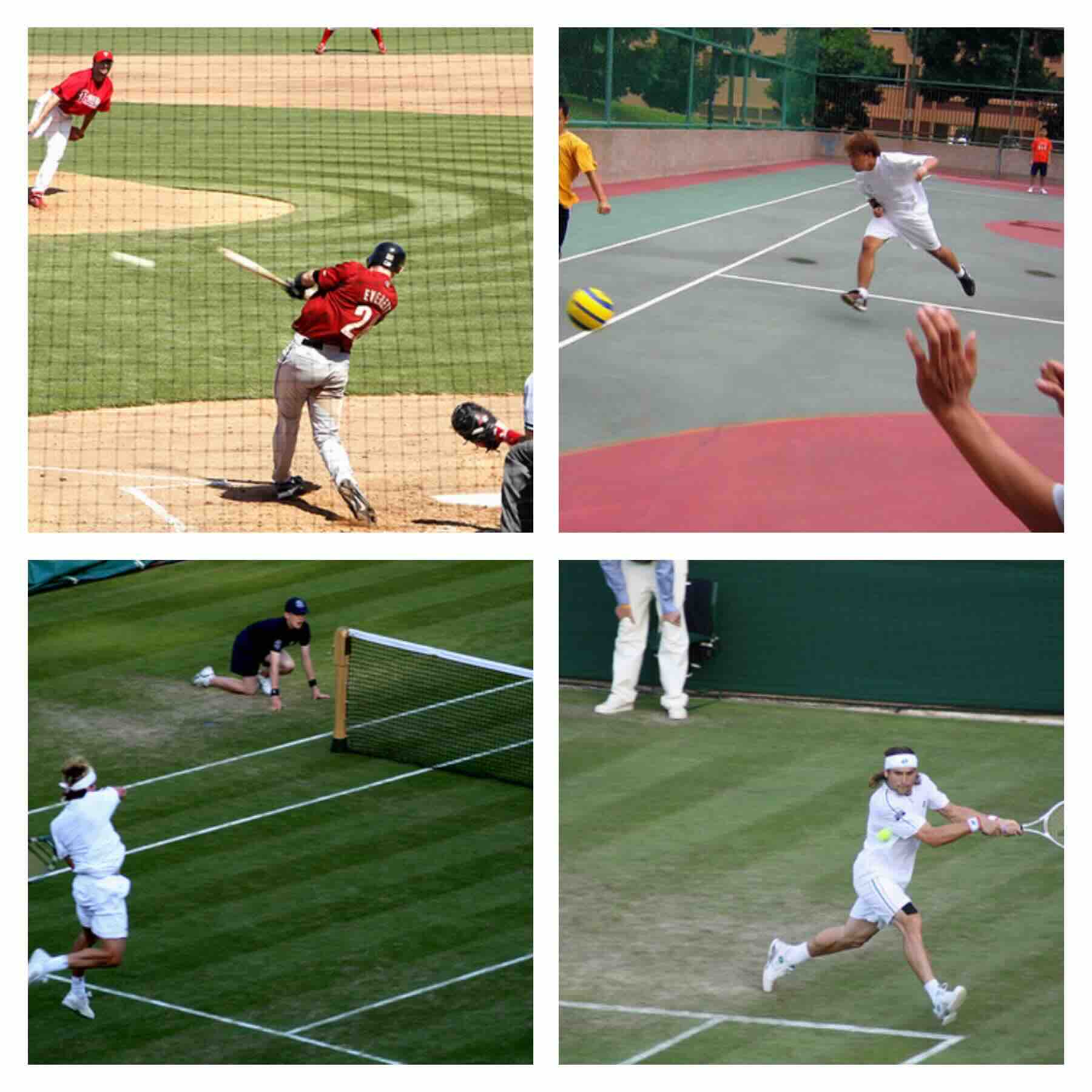} & Tennis, Baseball, Sports \\
      S7 LH 496 & \includegraphics[width=\linewidth]{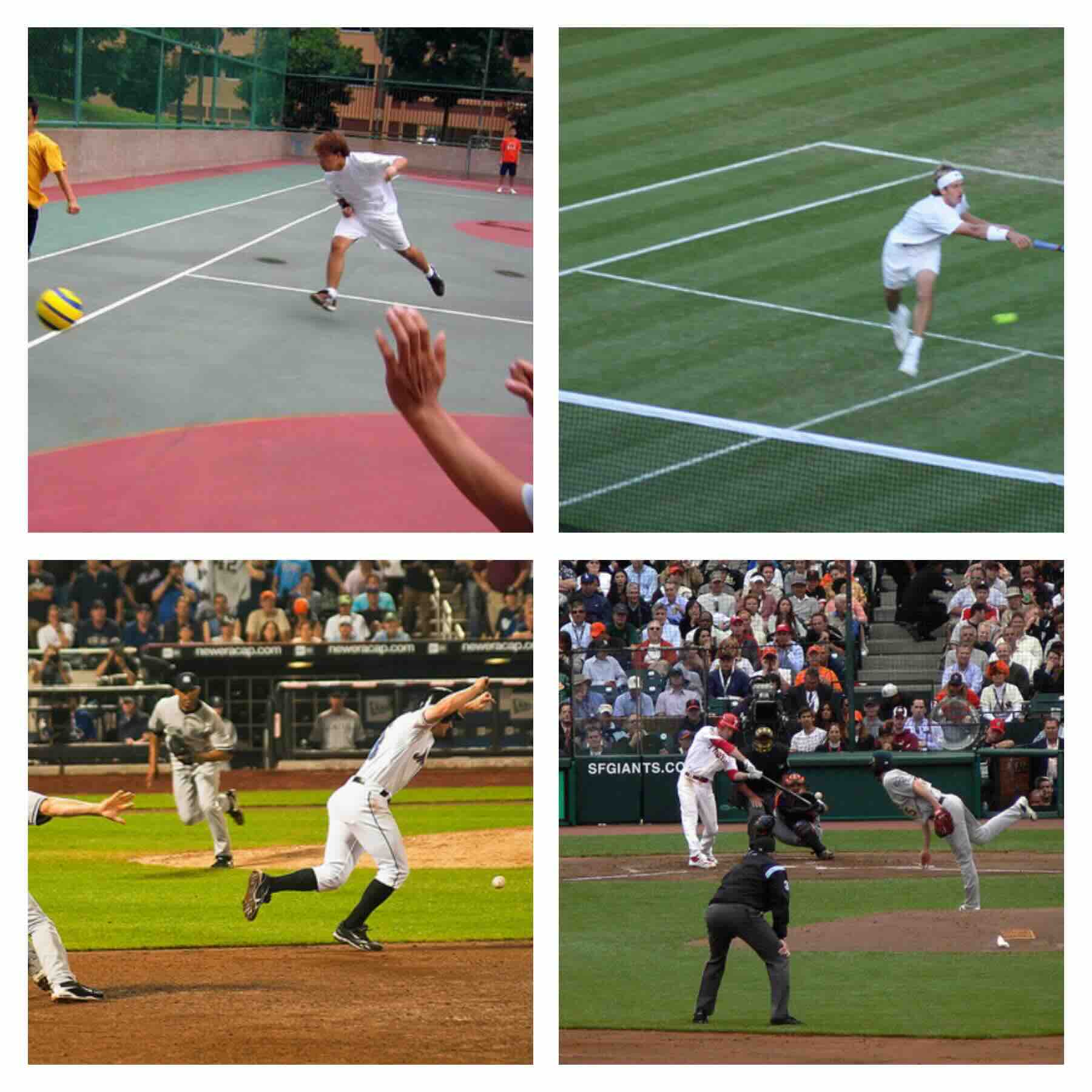} & Sports, Baseball, Tennis \\
      S7 LH 497 & \includegraphics[width=\linewidth]{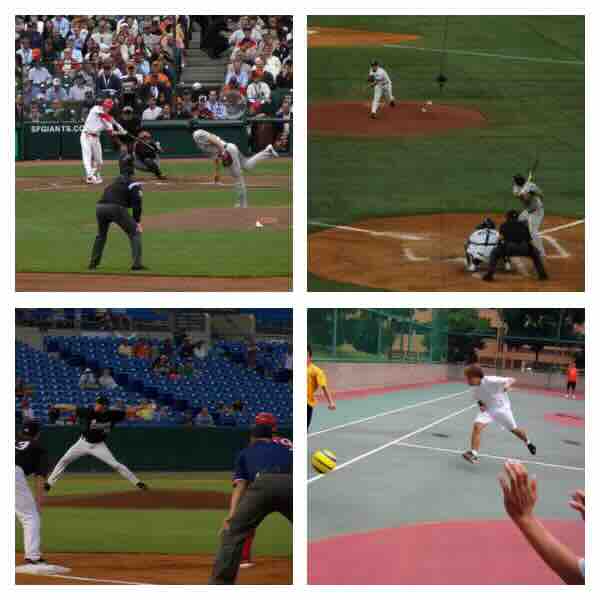} & Baseball, Pitching, Batting \\
      S7 RH 317 & \includegraphics[width=\linewidth]{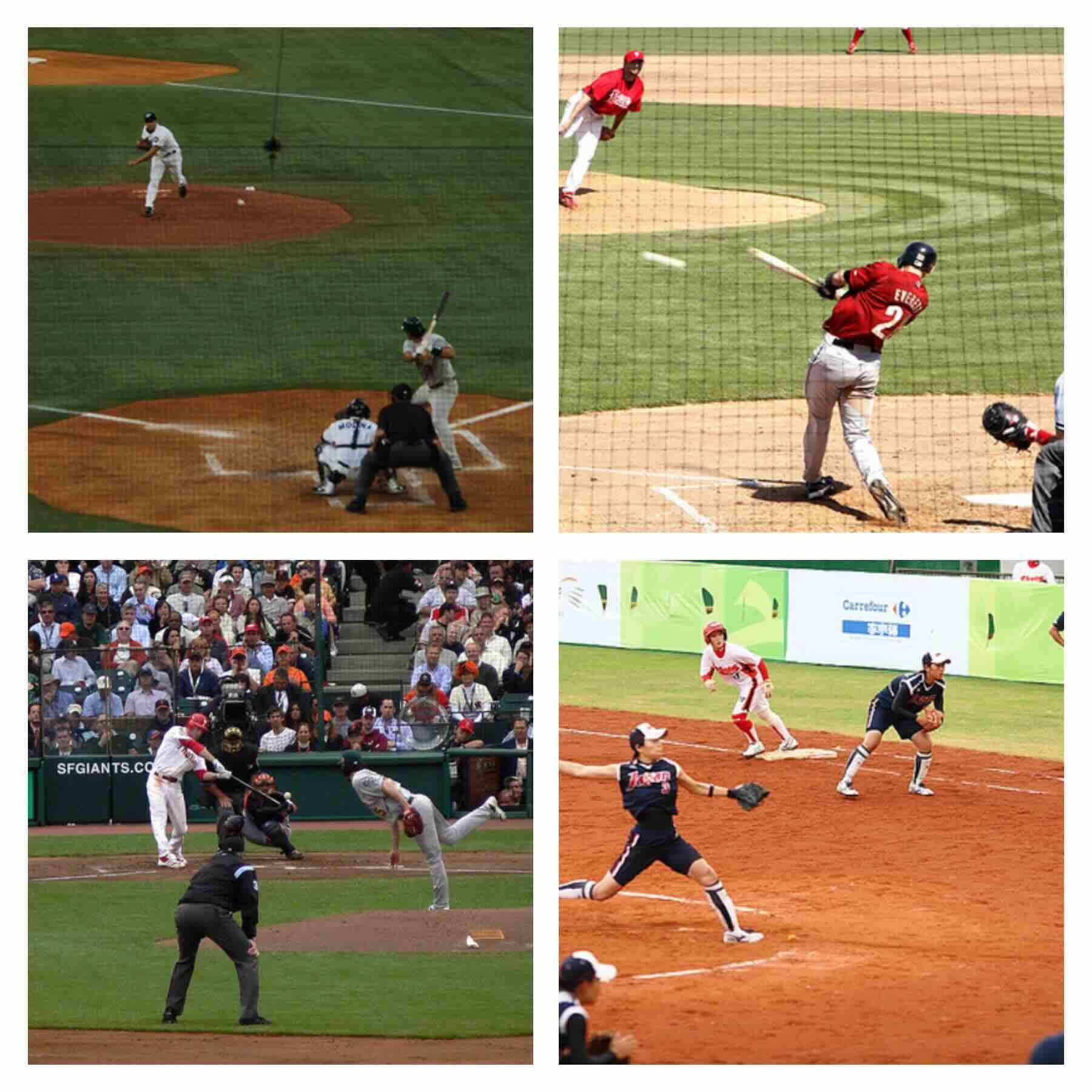} & Baseball, Pitching, Batting \\
      \bottomrule
    \end{tabular}
    \label{tab:sub7}
  \end{subtable}

  \caption{Maximally parcel-activating images (predicted) for unlabeled parcels.}
  \label{tab:all}
\end{table}

\clearpage
\subsection{Full collages for cross-subject tools parcel}
\label{app:tool-full-collage}

\begin{figure}[ht]
    \centering
    \begin{minipage}[b]{0.45\textwidth}
        \centering
        \includegraphics[width=\linewidth]{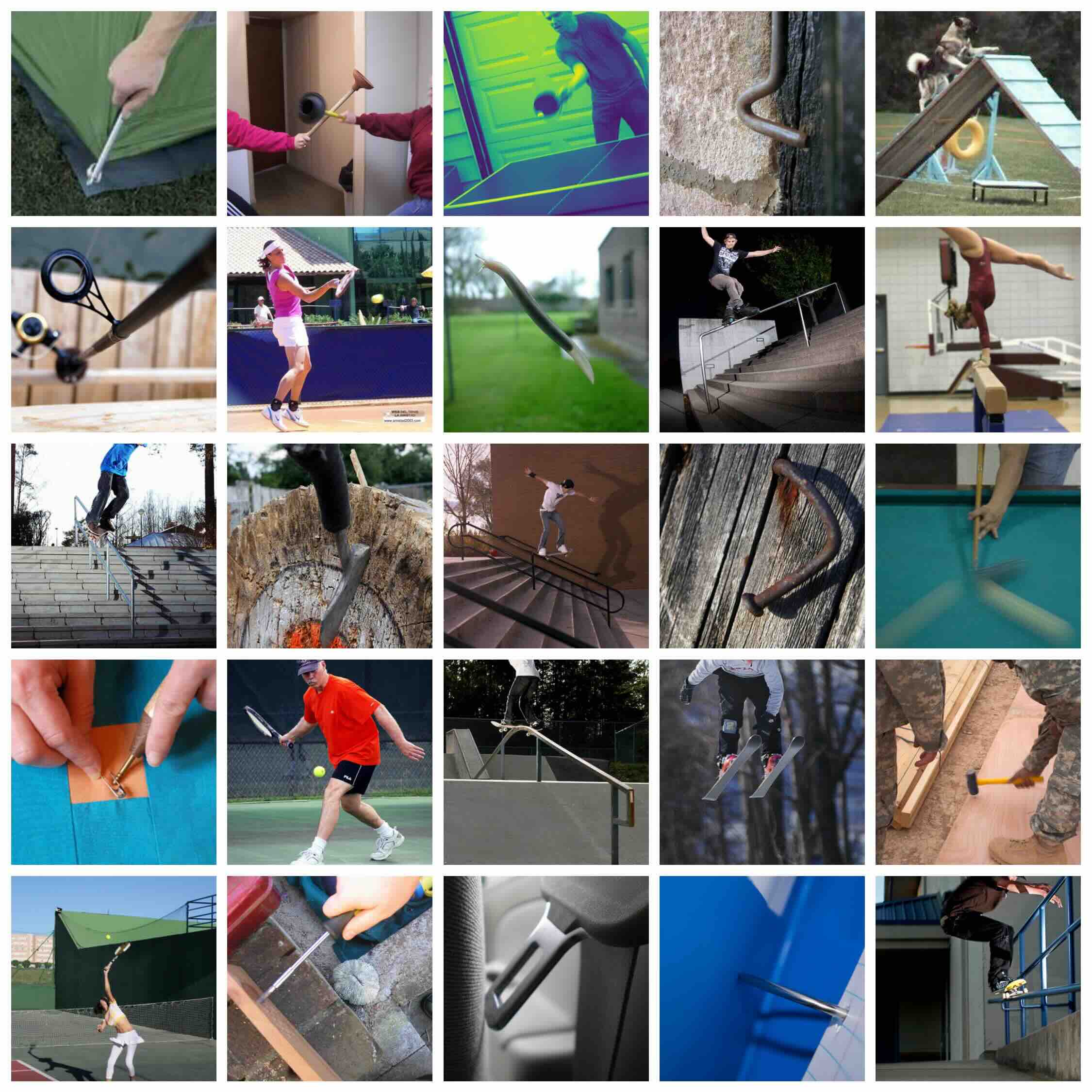}
        \caption*{(a) Subject 1}
    \end{minipage}
    \hfill
    \begin{minipage}[b]{0.45\textwidth}
        \centering
        \includegraphics[width=\linewidth]{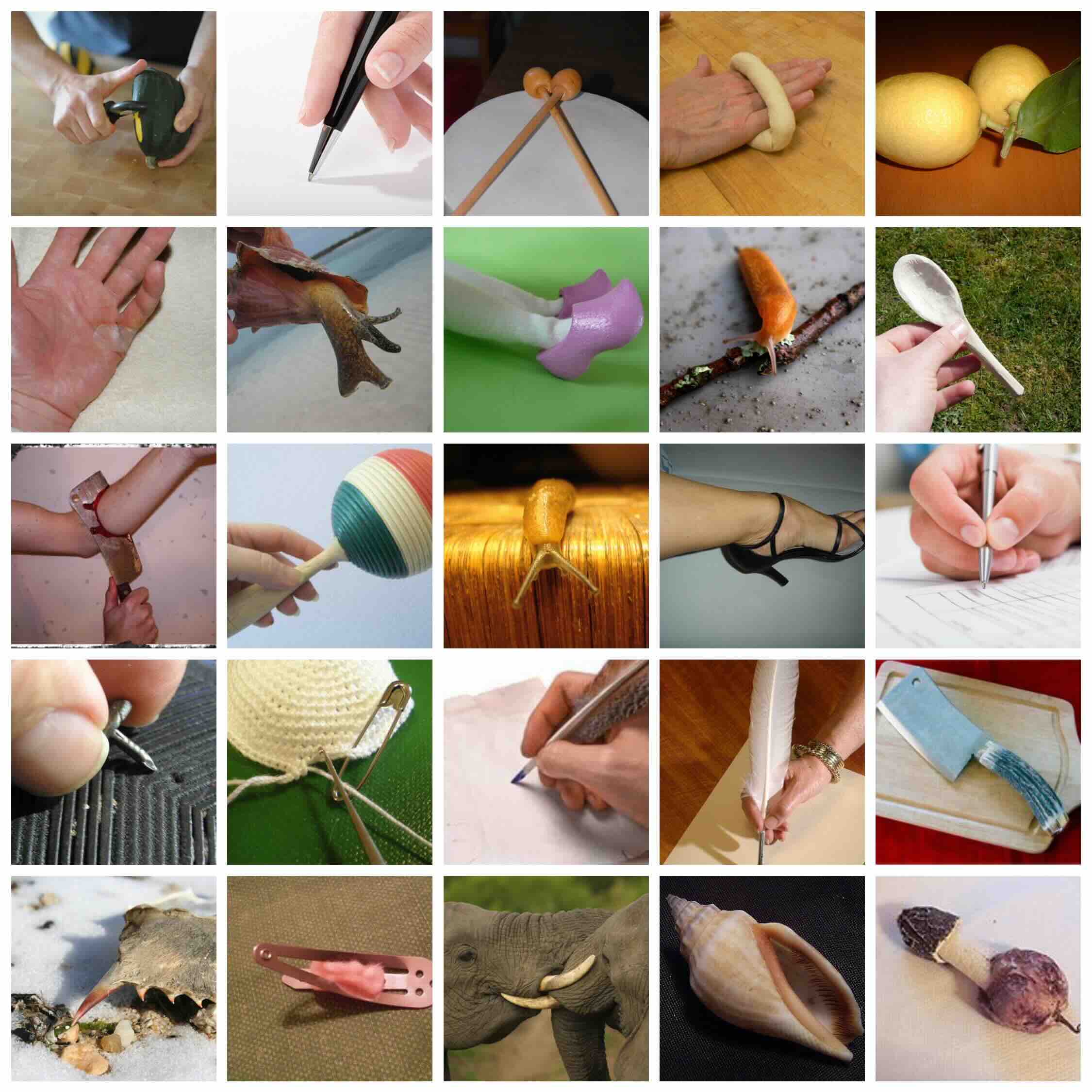}
        \caption*{(b) Subject 2}
    \end{minipage}
    
    \vspace{0.5cm}
    
    \begin{minipage}[b]{0.45\textwidth}
        \centering
        \includegraphics[width=\linewidth]{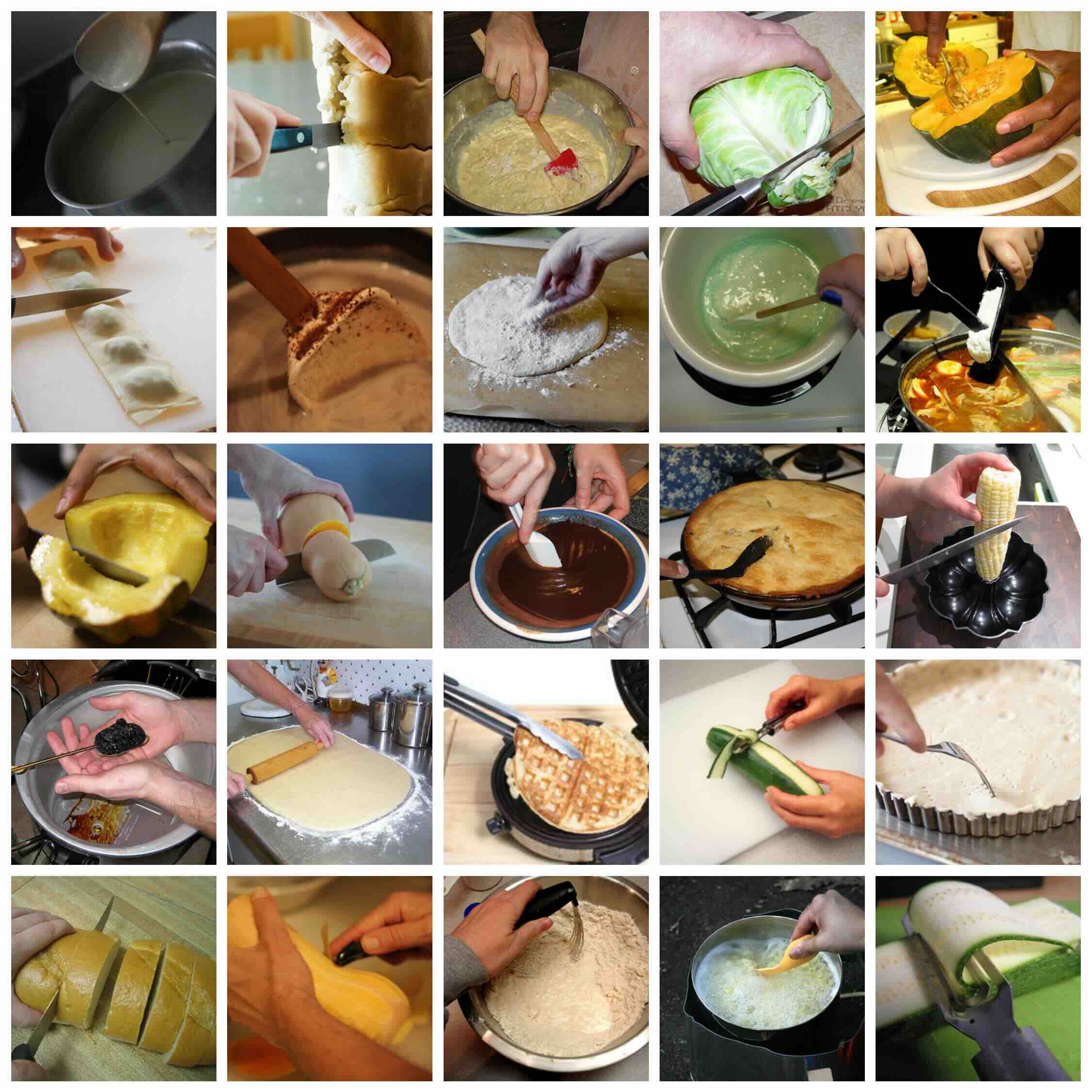}
        \caption*{(c) Subject 5}
    \end{minipage}
    \hfill
    \begin{minipage}[b]{0.45\textwidth}
        \centering
        \includegraphics[width=\linewidth]{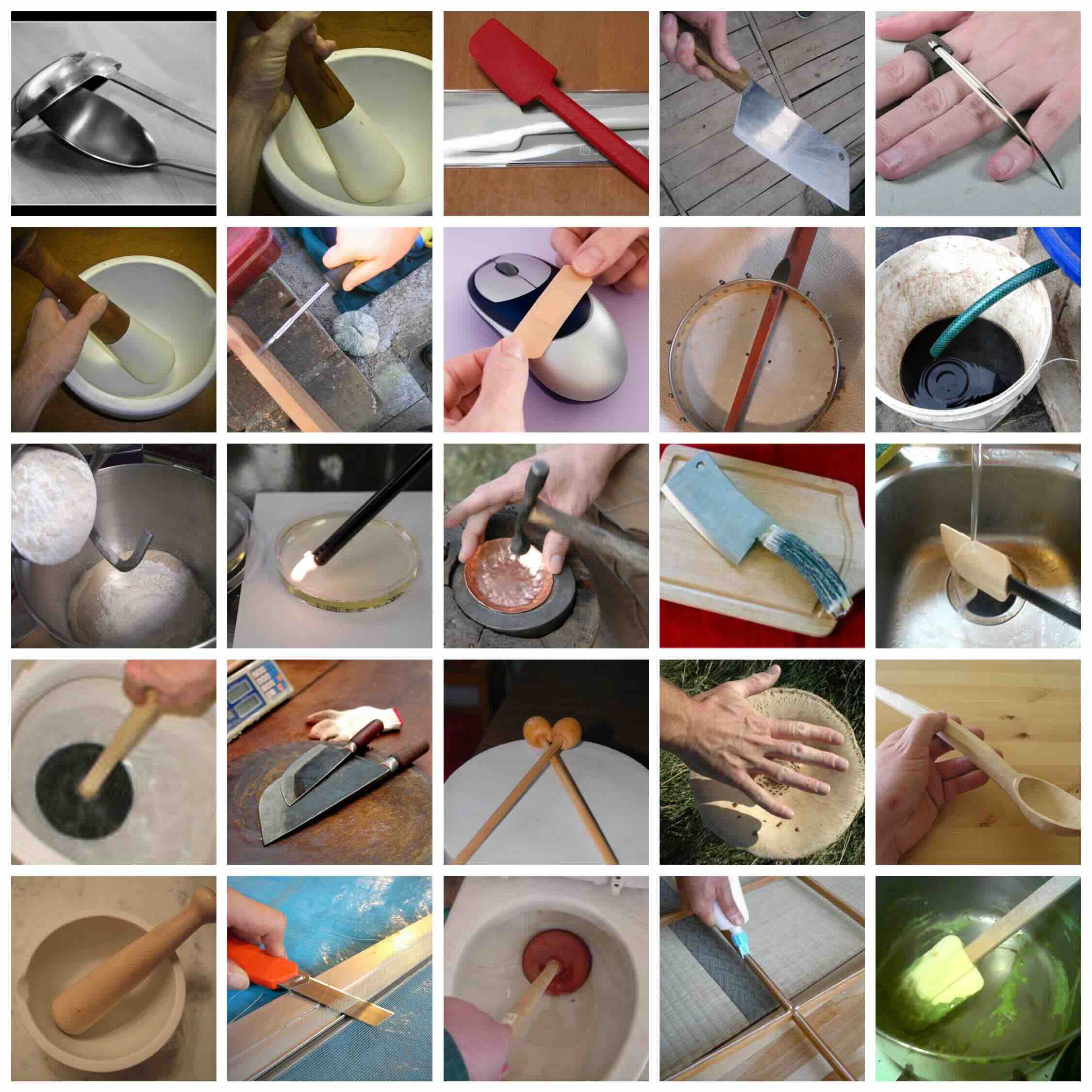}
        \caption*{(d) Subject 7}
    \end{minipage}
    
    \caption{Top 25 maximally parcel-activating ImageNet images (based on encoder predictions) for the cross-subject parcel in Figure~\ref{fig:cross_subj_217}.}
    \label{fig:p217_imgnet_grid}
\end{figure}

\subsection{Statistical tests for subject-specific parcel evaluation}
\label{app:stats-tests}

Here, we explain the details of the statistical tests performed in \ref{subsec:evaluating-our-labels}.

The underlying assumption in cognitive neuroscience is that for some parcels, there exists some category of visual stimuli that maximally activates the parcel.

Our goal is to assess whether the semantic content of the images our encoder identifies as highly activating aligns with the semantic content of NSD test images that actually elicit high ground-truth activations in that parcel. We ask whether a single “concept vector” derived from our encoder for a given parcel predicts parcel‐activation rank order in the NSD test set more accurately than the baseline model. 

We computed two concept vectors per parcel: one based on the top 32 maximally-activating ImageNet or BrainDIVE images predicted by our encoder, and one based on the top 32 maximally-activating NSD training images. We embedded each set of the images into CLIP space and then averaged the CLIP vectors to obtain a parcel-specific "concept vector" that reflects predicted parcel categorical selectivity. We then embedded each NSD test image into CLIP space, and ranked images by cosine similarity to the concept vector (encoder-derived or baseline). Since we expect that higher semantic similarity to the ``concept vector'' will correspond to higher parcel activation, we quantified alignment with the empirical activation ordering using Spearman's rank correlation between concept-vector similarity rank and ground-truth fMRI activity rank.

In Table~\ref{tab:rank_corr_vs_null}, for each parcel, we compared the rank correlation coefficient with a null distribution, bootstrapped with 10,000 random rankings of the NSD test images. We separate the results into most activating stimuli predicted by the model from BrainDIVE and ImageNet, as well as stimuli from the NSD training set (baseline).

\subsection{Statistical tests for cross-subject parcel evaluation}
\label{app:cross-subj-stats}

For a given parcel and dataset, we generated a selectivity hypothesis by averaging the CLIP vectors corresponding to the top 32 images according to the brain encoder, then used that CLIP vector to rank the NSD training images using cosine similarity. We calculate Spearman rank correlation between that ranking and the ranking determined by ground truth parcel activation, and compare this to a null distribution of correlation coefficients, which consists of 10,000 draws of random rankings.

Note that there are a fewer number of total parcels measured here since each subject has different parcels selected for experimentation (see Section~\ref{subsec:parcel-selection}), so cross-subject selectivity can only be tested for parcels selected for experimentation with every subject. 

\subsection{Cosine similarity analysis of labels for cross-subject parcels}

To demonstrate cross-subject consistency in semantic selectivity, we also compared directly the concept vectors generated by the encoder. Concept vectors generated for the same parcel, when analyzed across subjects, tend to be more similar (0.784 ± 0.069 for BrainDIVE) than the parcels coming from around the target parcel (0.710 ± 0.115 for BrainDIVE). Absolute cosine similarities are high because these parcels are selective for high-level visual concepts that share similar features, but our method can adjudicate between them using targeted stimulus generation.

\clearpage
\subsection{Head-to-head encoder model versus NSD train baseline}
\label{app:h2d-comp}

Table \ref{tab:h2h_nsd_train_baseline} expands upon the results shown in table \ref{tab:rank_corr_vs_null}, reporting the fraction of parcels for which the encoder-selected stimuli outperforms the NSD train baseline in Spearman's rank correlation coefficient, evaluated within the same subject.

\begin{table}[h]
  \centering
  \caption{Head to head comparison between our model and NSD train baseline: number of parcels where model rank correlation coefficient beats NSD train, evaluated within the same subject.}
  \label{tab:h2h_nsd_train_baseline}
  \begin{tabular}{lccccc}
    \toprule
    Comparison type & S1 & S2 & S5 & S7 \\
    \midrule
Our encoder w/ImageNet > NSD train: & $93/181$ & $94/192$ & $89/175$ & $62/196$ & \\
Our encoder w/BrainDIVE > NSD train: & $95/181$ & $123/192$ & $100/175$ & $76/196$ & \\
    \bottomrule
  \end{tabular}
\end{table}

Table~\ref{tab:h2h_nsd_train_baseline_cross_subj} expands upon results in Table~\ref{tab:rank_corr_vs_null_cross_subj}, reporting the number of parcels for which the rank correlation coefficient from the specified model is greater than that from the baseline model, when three subjects are used to predict a held-out subject.

\begin{table}[h]
  \centering
  \caption{Head to head comparison between our model and NSD train baseline: number of parcels where model rank correlation coefficient beats NSD train, evaluated with all subjects used to predict one held out.}
  \label{tab:h2h_nsd_train_baseline_cross_subj}
  \begin{tabular}{lccccc}
    \toprule
    Comparison type & S1 & S2 & S5 & S7 \\
    \midrule
Our encoder w/ImageNet > NSD train: & $30/49$ & $35/49$ & $36/49$ & $27/49$ & \\
Our encoder w/BrainDIVE > NSD train: & $29/49$ & $24/49$ & $30/49$ & $22/49$ & \\
    \bottomrule
  \end{tabular}
\end{table}

\subsection{Choosing parcels shared across subjects for future fMRI experimentation}
\label{app:choosing-shared-parcels}

\begin{figure}[h]
    \centering
    \includegraphics[width=\linewidth]{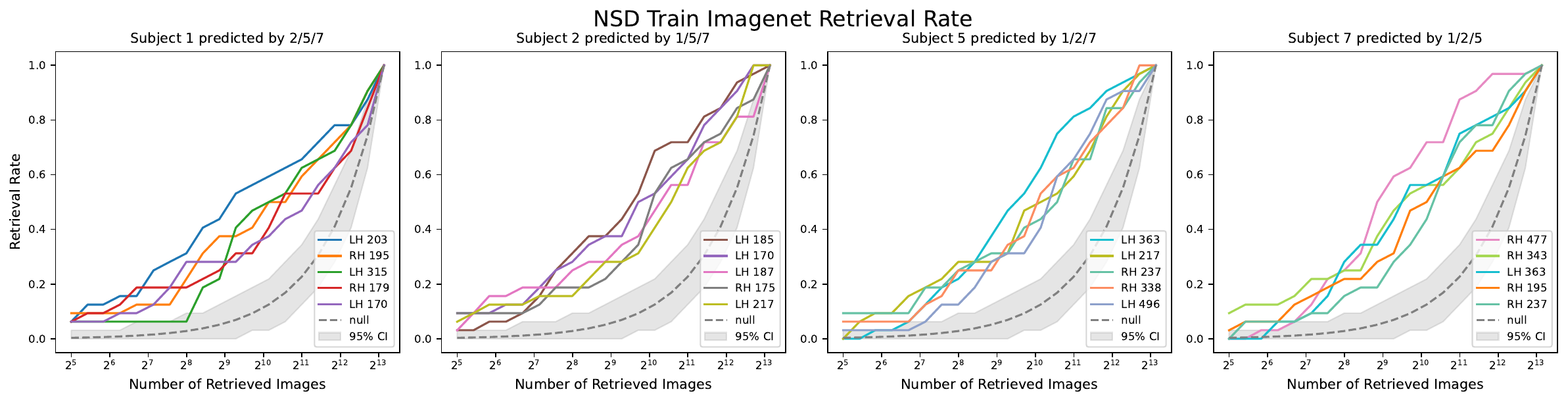}
    \caption{See caption from Figure~\ref{fig:roc_within_subj} for description. Each parcel shown is shared across subjects and the encoder train on three subjects is used to predict the selectivity of that same parcel on a held-out subject.} 
    \label{fig:roc_cross_subj}
\end{figure}

In Figure~\ref{fig:roc_cross_subj}, we report similar results as in Figure~\ref{fig:roc_within_subj}, except for top parcels shared across subjects.

\clearpage
\subsection{Interpreting statistical test results by comparing to known visual areas}
\label{app:intreting-statistical-test}

To quantify what rank correlation magnitude constitutes meaningful selectivity, we report the rank correlation the model achieves on known visual areas as a benchmark. Parcels with known high-level selectivity (EBA, FFA, FBA) tend to exhibit very high correlations, while parcels with known lower-level selectivity (V1–V4) tend to exhibit low correlations. The average Spearman's $\rho$ from our concept vectors are in the range of areas like PPA and RSC.

Using CLIP as our metric allows us to capture high-level features that can explain semantic selectivity. Given that fMRI studies tend to rely on experimenter-curated topics, the CLIP space is a move toward a data-driven approach, but ultimately any metric can be substituted.

\begin{figure}[h]
\centering
\caption{Spearman's $\rho$, actual vs. predicted activation ordering for subject 1. Top: high-level areas. Bottom: low-level areas.}
\label{tab:spearman_subject1}
\begin{subtable}[t]{0.8\textwidth}
\centering
\caption{High-level areas}
\begin{tabular}{lcc}
\toprule
Region & ImageNet & BrainDIVE \\
\midrule
Unlabeled parcel mean (ours) & 0.168 $\pm$ 0.106 & 0.163 $\pm$ 0.121 \\
EBA & 0.432 & 0.501 \\
FFA-1 & 0.217 & 0.243 \\
FFA-2 & 0.373 & 0.412 \\
FBA-2 & 0.373 & 0.401 \\
lateral & 0.323 & 0.339 \\
PPA & 0.166 & 0.128 \\
RSC & 0.170 & 0.162 \\
aTL-faces & 0.120 & 0.138 \\
mTL-words & 0.164 & 0.199 \\
\bottomrule
\end{tabular}
\end{subtable}

\vspace{1em}

\begin{subtable}[t]{0.8\textwidth}
\centering
\caption{Low-level areas}
\begin{tabular}{lcc}
\toprule
Region & ImageNet & BrainDIVE \\
\midrule
Unlabeled parcel mean (ours) & 0.168 $\pm$ 0.106 & 0.163 $\pm$ 0.121 \\
V1d & 0.010 & 0.027 \\
V1v & 0.024 & 0.036 \\
V2d & -0.032 & -0.035 \\
V2v & -0.030 & -0.002 \\
V3d & -0.004 & -0.018 \\
V3v & 0.007 & 0.009 \\
hV4 & 0.047 & 0.067 \\
early & -0.025 & -0.014 \\
\bottomrule
\end{tabular}
\end{subtable}
\end{figure}

Note that since we’re using the Schaefer-1000 parcellation, for each visual area, we average the rank correlation value across the top 3 Schaefer parcels with greatest overlap with that visual area.

\clearpage
\subsection{Varying top images used for statistical test}
\label{app:topk}

To confirm that the superior performance of ImageNet and BrainDIVE superstimuli stems from a broader diversity of concepts present—and not sampling near peak activation—we varied the top-K image threshold used to define the concept vector. We report $K=16, 8, 4, 2, 1$ versions for tables \ref{tab:rank_corr_vs_null}, \ref{tab:spearman_rank}, \ref{tab:rank_corr_vs_null_cross_subj}, and \ref{tab:spearman_rank_cross_subj}.

\begin{table}[h]
  \centering
  \caption{Fraction of parcels whose model-derived label predicts parcel activation rankings significantly better than chance ($p<0.05$, FDR corrected), varying number of top images for label generation.}
  \begin{tabular}{lccccc}
    \toprule
    Model type & S1 & S2 & S5 & S7 \\
    \midrule
        \multicolumn{5}{l}{\textbf{k=64}} \\
NSD train & $134/181$ & $156/192$ & $134/175$ & $154/196$ & \\
Our encoder w/ImageNet & $139/181$ & $166/192$ & $134/175$ & $149/196$ & \\
Our encoder w/BrainDIVE & $128/181$ & $168/192$ & $140/175$ & $152/196$ & \\
    \midrule\midrule
    \multicolumn{5}{l}{\textbf{k=32}} \\
NSD train & $150/181$ & $163/192$ & $136/175$ & $155/196$ & \\
Our encoder w/ImageNet & $139/181$ & $167/192$ & $130/175$ & $150/196$ & \\
Our encoder w/BrainDIVE & $135/181$ & $170/192$ & $139/175$ & $156/196$ & \\
    \midrule\midrule
    \multicolumn{5}{l}{\textbf{k=16}} \\
NSD train & $154/181$ & $169/192$ & $137/175$ & $157/196$ & \\
Our encoder w/ImageNet & $138/181$ & $166/192$ & $134/175$ & $151/196$ & \\
Our encoder w/BrainDIVE & $139/181$ & $171/192$ & $136/175$ & $152/196$ & \\
    \midrule\midrule
    \multicolumn{5}{l}{\textbf{k=8}} \\
NSD train & $144/181$ & $160/192$ & $131/175$ & $149/196$ & \\
Our encoder w/ImageNet & $138/181$ & $162/192$ & $134/175$ & $153/196$ & \\
Our encoder w/BrainDIVE & $135/181$ & $168/192$ & $136/175$ & $153/196$ & \\
    \midrule\midrule
    \multicolumn{5}{l}{\textbf{k=4}} \\
NSD train & $134/181$ & $146/192$ & $122/175$ & $140/196$ & \\
Our encoder w/ImageNet & $138/181$ & $164/192$ & $129/175$ & $148/196$ & \\
Our encoder w/BrainDIVE & $140/181$ & $166/192$ & $142/175$ & $151/196$ & \\
    \midrule\midrule
    \multicolumn{5}{l}{\textbf{k=2}} \\
NSD train & $131/181$ & $125/192$ & $109/175$ & $110/196$ & \\
Our encoder w/ImageNet & $134/181$ & $154/192$ & $122/175$ & $143/196$ & \\
Our encoder w/BrainDIVE & $141/181$ & $171/192$ & $136/175$ & $136/196$ & \\
    \midrule\midrule
    \multicolumn{5}{l}{\textbf{k=1}} \\
NSD train & $114/181$ & $120/192$ & $92/175$ & $86/196$ & \\
Our encoder w/ImageNet & $133/181$ & $145/192$ & $113/175$ & $130/196$ & \\
Our encoder w/BrainDIVE & $137/181$ & $162/192$ & $119/175$ & $131/196$ & \\
    \bottomrule
  \end{tabular}
\end{table}

\begin{table}[h]
  \centering
  \caption{Average parcel-level correlations (mean $\pm$ std) between model-derived labels and parcel activation rankings, varying number of top images for label generation.}
  \begin{tabular}{lccccc}
    \toprule
    Model type & S1 & S2 & S5 & S7 \\
    \midrule
\multicolumn{5}{l}{\textbf{k=64}} \\
NSD train & $0.145 \pm  0.098$ & $0.153 \pm  0.088$ & $0.141 \pm  0.096$ & $0.142 \pm  0.085$ & \\
Our encoder w/ImageNet & $0.167 \pm  0.106$ & $0.163 \pm  0.082$ & $0.143 \pm  0.092$ & $0.133 \pm  0.075$ & \\
Our encoder w/BrainDIVE & $0.154 \pm  0.119$ & $0.187 \pm  0.099$ & $0.153 \pm  0.095$ & $0.131 \pm  0.084$ & \\
\midrule\midrule
\multicolumn{5}{l}{\textbf{k=32}} \\
NSD train & $0.162 \pm  0.098$ & $0.164 \pm  0.091$ & $0.150 \pm  0.094$ & $0.148 \pm  0.086$ & \\
Our encoder w/ImageNet & $0.168 \pm  0.106$ & $0.163 \pm  0.082$ & $0.142 \pm  0.092$ & $0.133 \pm  0.075$ & \\
Our encoder w/BrainDIVE & $0.163 \pm  0.121$ & $0.190 \pm  0.099$ & $0.154 \pm  0.094$ & $0.133 \pm  0.083$ & \\
\midrule\midrule
\multicolumn{5}{l}{\textbf{k=16}} \\
NSD train & $0.167 \pm  0.096$ & $0.170 \pm  0.093$ & $0.149 \pm  0.096$ & $0.148 \pm  0.087$ & \\
Our encoder w/ImageNet & $0.169 \pm  0.106$ & $0.162 \pm  0.082$ & $0.142 \pm  0.091$ & $0.131 \pm  0.075$ & \\
Our encoder w/BrainDIVE & $0.168 \pm  0.120$ & $0.190 \pm  0.099$ & $0.155 \pm  0.092$ & $0.134 \pm  0.082$ & \\
\midrule\midrule
\multicolumn{5}{l}{\textbf{k=8}} \\
NSD train & $0.160 \pm  0.103$ & $0.164 \pm  0.094$ & $0.140 \pm  0.100$ & $0.139 \pm  0.091$ & \\
Our encoder w/ImageNet & $0.168 \pm  0.107$ & $0.159 \pm  0.081$ & $0.142 \pm  0.092$ & $0.128 \pm  0.073$ & \\
Our encoder w/BrainDIVE & $0.167 \pm  0.121$ & $0.188 \pm  0.099$ & $0.155 \pm  0.092$ & $0.133 \pm  0.082$ & \\
\midrule\midrule
\multicolumn{5}{l}{\textbf{k=4}} \\
NSD train & $0.151 \pm  0.112$ & $0.147 \pm  0.099$ & $0.128 \pm  0.097$ & $0.124 \pm  0.091$ & \\
Our encoder w/ImageNet & $0.165 \pm  0.108$ & $0.155 \pm  0.082$ & $0.140 \pm  0.091$ & $0.124 \pm  0.073$ & \\
Our encoder w/BrainDIVE & $0.165 \pm  0.120$ & $0.184 \pm  0.098$ & $0.151 \pm  0.092$ & $0.131 \pm  0.080$ & \\
\midrule\midrule
\multicolumn{5}{l}{\textbf{k=2}} \\
NSD train & $0.134 \pm  0.114$ & $0.127 \pm  0.106$ & $0.110 \pm  0.104$ & $0.106 \pm  0.092$ & \\
Our encoder w/ImageNet & $0.158 \pm  0.107$ & $0.147 \pm  0.082$ & $0.135 \pm  0.090$ & $0.118 \pm  0.071$ & \\
Our encoder w/BrainDIVE & $0.160 \pm  0.117$ & $0.176 \pm  0.094$ & $0.149 \pm  0.090$ & $0.127 \pm  0.080$ & \\
\midrule\midrule
\multicolumn{5}{l}{\textbf{k=1}} \\
NSD train & $0.121 \pm  0.120$ & $0.109 \pm  0.100$ & $0.088 \pm  0.105$ & $0.087 \pm  0.090$ & \\
Our encoder w/ImageNet & $0.148 \pm  0.101$ & $0.136 \pm  0.084$ & $0.125 \pm  0.091$ & $0.112 \pm  0.075$ & \\
Our encoder w/BrainDIVE & $0.148 \pm  0.115$ & $0.165 \pm  0.093$ & $0.140 \pm  0.092$ & $0.119 \pm  0.082$ & \\
    \bottomrule
  \end{tabular}
\end{table}

\begin{table}[h]
  \centering
  \caption{Fraction of parcels shared across subjects whose model-derived label predicts parcel activation rankings significantly better than chance ($p<0.05$, FDR corrected), varying number of top images for label generation.}
  \begin{tabular}{lccccc}
    \toprule
    Model type & S1 & S2 & S5 & S7 \\
\midrule
\multicolumn{5}{l}{\textbf{k=64}} \\
NSD train & $37/49$ & $44/49$ & $40/49$ & $40/49$ & \\
Our encoder w/ImageNet & $38/49$ & $45/49$ & $42/49$ & $39/49$ & \\
Our encoder w/BrainDIVE & $34/49$ & $40/49$ & $37/49$ & $38/49$ & \\
\midrule\midrule
\multicolumn{5}{l}{\textbf{k=32}} \\
NSD train & $37/49$ & $44/49$ & $40/49$ & \bm{$40/49$} & \\
Our encoder w/ImageNet & $38/49$ & $45/49$ & $42/49$ & $39/49$ & \\
Our encoder w/BrainDIVE & $37/49$ & $40/49$ & $38/49$ & $38/49$ & \\
\midrule\midrule
\multicolumn{5}{l}{\textbf{k=16}} \\
NSD train & $37/49$ & $42/49$ & $40/49$ & \bm{$40/49$} & \\
Our encoder w/ImageNet & $38/49$ & \bm{$46/49$} & $42/49$ & $39/49$ & \\
Our encoder w/BrainDIVE & $37/49$ & $40/49$ & $38/49$ & $38/49$ & \\
\midrule\midrule
\multicolumn{5}{l}{\textbf{k=8}} \\
NSD train & $36/49$ & $40/49$ & $41/49$ & $40/49$ & \\
Our encoder w/ImageNet & $37/49$ & $45/49$ & \bm{$43/49$} & $39/49$ & \\
Our encoder w/BrainDIVE & $37/49$ & $40/49$ & $38/49$ & $39/49$ & \\
\midrule\midrule
\multicolumn{5}{l}{\textbf{k=4}} \\
NSD train & \bm{$40/49$} & $41/49$ & $40/49$ & $38/49$ & \\
Our encoder w/ImageNet & $38/49$ & $44/49$ & $42/49$ & $40/49$ & \\
Our encoder w/BrainDIVE & $37/49$ & $40/49$ & $39/49$ & \bm{$41/49$} & \\
\midrule\midrule
\multicolumn{5}{l}{\textbf{k=2}} \\
NSD train & \bm{$40/49$} & $40/49$ & $40/49$ & $40/49$ & \\
Our encoder w/ImageNet & $38/49$ & $45/49$ & $42/49$ & $38/49$ & \\
Our encoder w/BrainDIVE & $37/49$ & $43/49$ & $39/49$ & $40/49$ & \\
\midrule\midrule
\multicolumn{5}{l}{\textbf{k=1}} \\
NSD train & $37/49$ & $35/49$ & $38/49$ & $40/49$ & \\
Our encoder w/ImageNet & $37/49$ & $44/49$ & $42/49$ & $37/49$ & \\
Our encoder w/BrainDIVE & $39/49$ & $40/49$ & $40/49$ & $40/49$ & \\
    \bottomrule
  \end{tabular}
\end{table}

\begin{table}[h]
  \centering
  \caption{Average parcel-level correlations (mean $\pm$ std) between model-derived labels and parcel activation rankings, varying number of top images for label generation.}
  \begin{tabular}{lccccc}
    \toprule
    Model type & S1 & S2 & S5 & S7 \\
    \midrule
\multicolumn{5}{l}{\textbf{k=64}} \\
NSD train & $0.066 \pm  0.068$ & $0.092 \pm  0.065$ & $0.091 \pm  0.073$ & $0.090 \pm  0.069$ & \\
Our encoder w/ImageNet & $0.072 \pm  0.072$ & $0.105 \pm  0.074$ & $0.105 \pm  0.078$ & $0.093 \pm  0.076$ & \\
Our encoder w/BrainDIVE & $0.075 \pm  0.076$ & $0.102 \pm  0.087$ & $0.102 \pm  0.090$ & $0.087 \pm  0.080$ & \\
\midrule\midrule
\multicolumn{5}{l}{\textbf{k=32}} \\
NSD train & $0.068 \pm  0.068$ & $0.094 \pm  0.065$ & $0.090 \pm  0.074$ & $0.091 \pm  0.070$ & \\
Our encoder w/ImageNet & $0.072 \pm  0.071$ & $0.105 \pm  0.073$ & $0.105 \pm  0.078$ & $0.093 \pm  0.075$ & \\
Our encoder w/BrainDIVE & $0.078 \pm  0.077$ & $0.103 \pm  0.087$ & $0.106 \pm  0.091$ & $0.089 \pm  0.079$ & \\
\midrule\midrule
\multicolumn{5}{l}{\textbf{k=16}} \\
NSD train & $0.068 \pm  0.067$ & $0.093 \pm  0.066$ & $0.089 \pm  0.072$ & $0.090 \pm  0.069$ & \\
Our encoder w/ImageNet & $0.073 \pm  0.070$ & $0.105 \pm  0.073$ & $0.107 \pm  0.079$ & $0.091 \pm  0.076$ & \\
Our encoder w/BrainDIVE & $0.081 \pm  0.078$ & $0.103 \pm  0.087$ & $0.106 \pm  0.093$ & $0.090 \pm  0.080$ & \\
\midrule\midrule
\multicolumn{5}{l}{\textbf{k=8}} \\
NSD train & $0.070 \pm  0.067$ & $0.095 \pm  0.064$ & $0.088 \pm  0.072$ & $0.088 \pm  0.067$ & \\
Our encoder w/ImageNet & $0.071 \pm  0.070$ & $0.104 \pm  0.074$ & $0.105 \pm  0.077$ & $0.091 \pm  0.075$ & \\
Our encoder w/BrainDIVE & $0.083 \pm  0.077$ & $0.102 \pm  0.086$ & $0.106 \pm  0.093$ & $0.093 \pm  0.079$ & \\
\midrule\midrule
\multicolumn{5}{l}{\textbf{k=4}} \\
NSD train & $0.068 \pm  0.065$ & $0.093 \pm  0.068$ & $0.090 \pm  0.072$ & $0.086 \pm  0.071$ & \\
Our encoder w/ImageNet & $0.071 \pm  0.070$ & $0.106 \pm  0.075$ & $0.104 \pm  0.075$ & $0.092 \pm  0.076$ & \\
Our encoder w/BrainDIVE & $0.083 \pm  0.078$ & $0.102 \pm  0.086$ & $0.110 \pm  0.093$ & $0.093 \pm  0.079$ & \\
\midrule\midrule
\multicolumn{5}{l}{\textbf{k=2}} \\
NSD train & $0.073 \pm  0.068$ & $0.093 \pm  0.068$ & $0.090 \pm  0.076$ & $0.082 \pm  0.071$ & \\
Our encoder w/ImageNet & $0.070 \pm  0.068$ & $0.108 \pm  0.073$ & $0.102 \pm  0.073$ & $0.088 \pm  0.073$ & \\
Our encoder w/BrainDIVE & $0.084 \pm  0.079$ & $0.103 \pm  0.086$ & $0.111 \pm  0.087$ & $0.094 \pm  0.081$ & \\
\midrule\midrule
\multicolumn{5}{l}{\textbf{k=1}} \\
NSD train & $0.062 \pm  0.069$ & $0.079 \pm  0.075$ & $0.081 \pm  0.076$ & $0.078 \pm  0.063$ & \\
Our encoder w/ImageNet & $0.068 \pm  0.070$ & $0.105 \pm  0.070$ & $0.104 \pm  0.076$ & $0.086 \pm  0.076$ & \\
Our encoder w/BrainDIVE & $0.084 \pm  0.080$ & $0.099 \pm  0.090$ & $0.109 \pm  0.087$ & $0.090 \pm  0.082$ & \\
    \bottomrule
  \end{tabular}
\end{table}

\clearpage
\subsection{Generating concept vectors using predicted activations}
\label{app:predicted-activation-cvs}
In the main text, we generated the baseline NSD train concept vector using the top maximally-activating images based on measured responses of the training set. Here, we generate concept vectors using top images based on predicted responses of the same training set—reproducing tables \ref{tab:rank_corr_vs_null}, \ref{tab:spearman_rank}, \ref{tab:rank_corr_vs_null_cross_subj}, and \ref{tab:spearman_rank_cross_subj}. Interestingly, the rankings generated by the model-predicted concept vector outperforms the concept vector generated by measured responses, and even our model in some cases. We suspect that using model-predicted responses may further reduce the noise compared to measured responses and be responsible for the increase in performance. We did not include the model-predicted concept vector performance in the main text since it's not quite a baseline (since it relies on the encoding model choice).

\begin{table}[h]
  \centering
  \caption{Fraction of parcels whose model-derived label predicts parcel activation rankings significantly better than chance ($p<0.05$, FDR corrected)}
  \label{tab:rank_corr_vs_null_predicted_acttivations}
  \begin{tabular}{lccccc}
    \toprule
    Model type & S1 & S2 & S5 & S7 \\
    \midrule
    NSD train (ground truth) & $150/181$ & $163/192$ & $136/175$ & $155/196$ \\
    NSD train (model-predicted) & \bm{$169/181$} & \bm{$176/192$} & \bm{$156/175$} & \bm{$168/196$} \\
    \bottomrule
  \end{tabular}
\end{table}

\begin{table}[h]
  \centering
  \small
  \setlength{\tabcolsep}{4pt}
  \caption{Spearman's $\rho$ (mean $\pm$ std) between the model-predicted and ground-truth activation rankings on the NSD test set, averaged across parcels.}
  \label{tab:spearman_rank_predicted_activation}
  \begin{tabular}{lcccc}
    \toprule
    Model type & S1 & S2 & S5 & S7 \\
    \midrule
    NSD train (ground truth) & $0.162 \pm  0.098$ & $0.164 \pm  0.091$ & $0.150 \pm  0.094$ & $0.148 \pm  0.086$ \\
    NSD train (model-predicted) & \bm{$0.191 \pm  0.092$} & \bm{$0.187 \pm  0.084$} & \bm{$0.174 \pm  0.087$} & \bm{$0.155 \pm  0.082$} \\
    \bottomrule
    \addlinespace[0.5em]
    \multicolumn{5}{@{}p{\linewidth}@{}}{\footnotesize
    \emph{Notes.} INet = ImageNet. BD = BrainDIVE.}
  \end{tabular}
\end{table}

\begin{table}[h]
  \centering
  \caption{Fraction of parcels shared across subjects whose model-derived label predicts parcel activation rankings significantly better than chance ($p<0.05$, FDR corrected)}
  \label{tab:rank_corr_vs_null_cross_subj_predicted_activation}
  \begin{tabular}{lccccc}
    \toprule
    Model type & S1 & S2 & S5 & S7 \\
    \midrule
    NSD train (ground truth) & $37/49$ & $44/49$ & $40/49$ & \bm{$40/49$} & \\
    NSD train (model-predicted) & \bm{$40/49$} & \bm{$46/49$} & \bm{$43/49$} & \bm{$40/49$}\\
    \bottomrule
  \end{tabular}
\end{table}

\begin{table}[!htbp]
  \centering
  \small
  \setlength{\tabcolsep}{4pt}
  \caption{Spearman's $\rho$ (mean $\pm$ std) between the model-predicted (from all subjects other than heldout) and ground-truth activation rankings on the NSD training set, averaged across parcels.}
  \label{tab:spearman_rank_cross_subj_predicted_activation}
  \begin{tabular}{lccccc}
    \toprule
    Model type & S1 & S2 & S5 & S7 \\
    \midrule
    NSD train (ground truth) & $0.068 \pm  0.068$ & $0.094 \pm  0.065$ & $0.090 \pm  0.074$ & $0.091 \pm  0.070$ \\
    NSD train (model-predicted) & \bm{$0.078 \pm  0.074$} & \bm{$0.110 \pm  0.068$} & \bm{$0.113 \pm  0.081$} & \bm{$0.101 \pm  0.074$}\\
    \bottomrule
    \addlinespace[0.5em]
    \multicolumn{5}{@{}p{\linewidth}@{}}{\footnotesize
    \emph{Notes.} INet = ImageNet. BD = BrainDIVE.}
  \end{tabular}
\end{table}

\clearpage
\subsection{LLM prompts used}
\label{app:prompts}

We used OpenAI's GPT-4o \citep{openai2024gpt4technicalreport} to caption the images selected by our model in order to obtain a linguistic representation of the selectivity of a parcel. The captions in Figure~\ref{fig:cross_subj_217} were generated with ChatGPT 4o (04/06/2025) with a collage of top-25 encoder-ranked ImageNet images shown, followed by the instruction ``Describe a theme present in most of the images presented.'' The captions in Table~\ref{tab:all} were generated by GPT-4o (05/12/2025, via the API) with the prompt ``Give keywords for the central concept or categories present in these images.''

\subsection{NSD dataset quirks}
\label{app:nsd-quirks}
\begin{figure}[h]
    \centering
\includegraphics[width=0.5\linewidth]{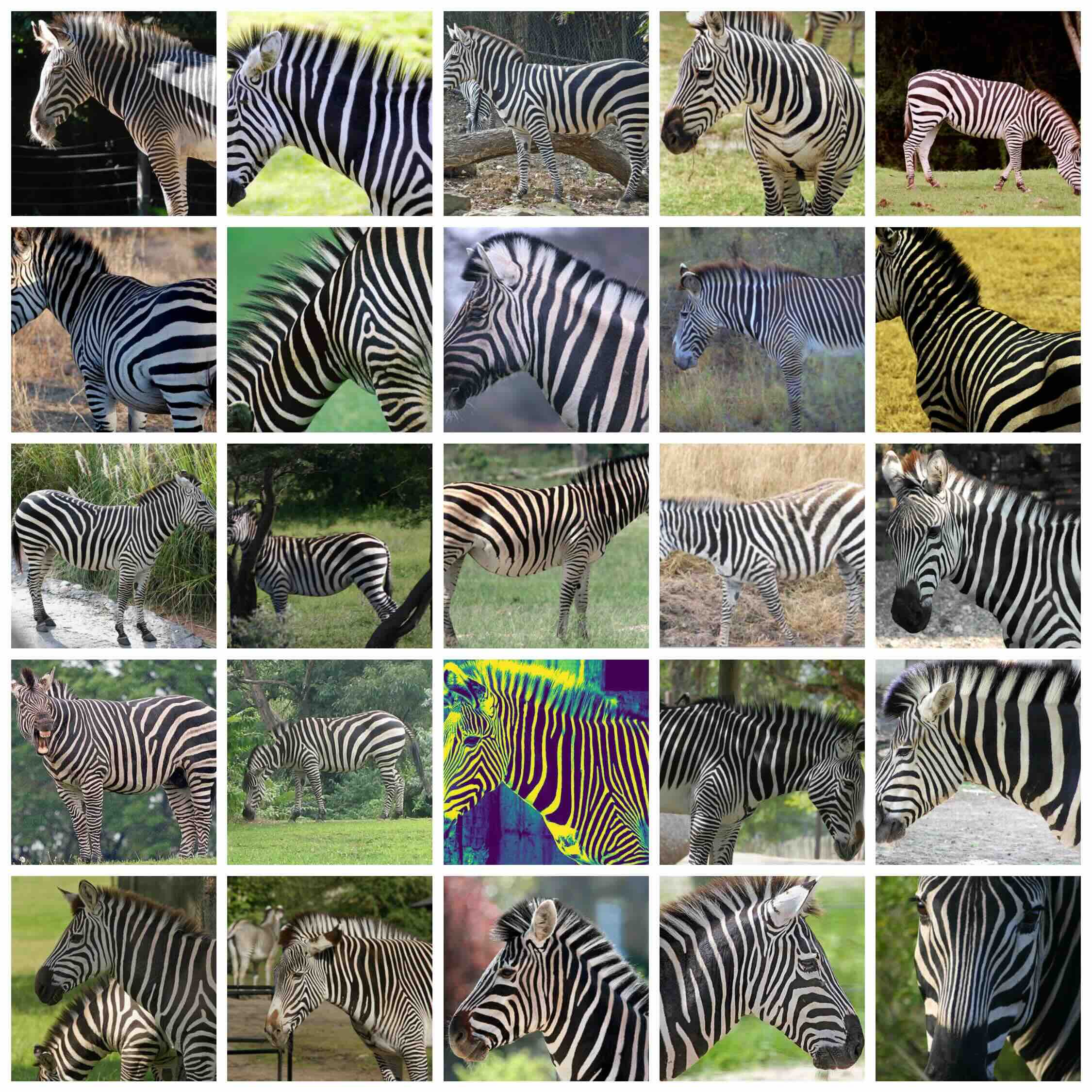}
    \caption{Top 25 maximally parcel-activating ImageNet images (based on encoder predictions) for a parcel in subject 2, left hemisphere.} 
    \label{fig:s2_top25_imgnet_p309}
\end{figure}

\begin{figure}[h]
    \centering
\includegraphics[width=0.5\linewidth]{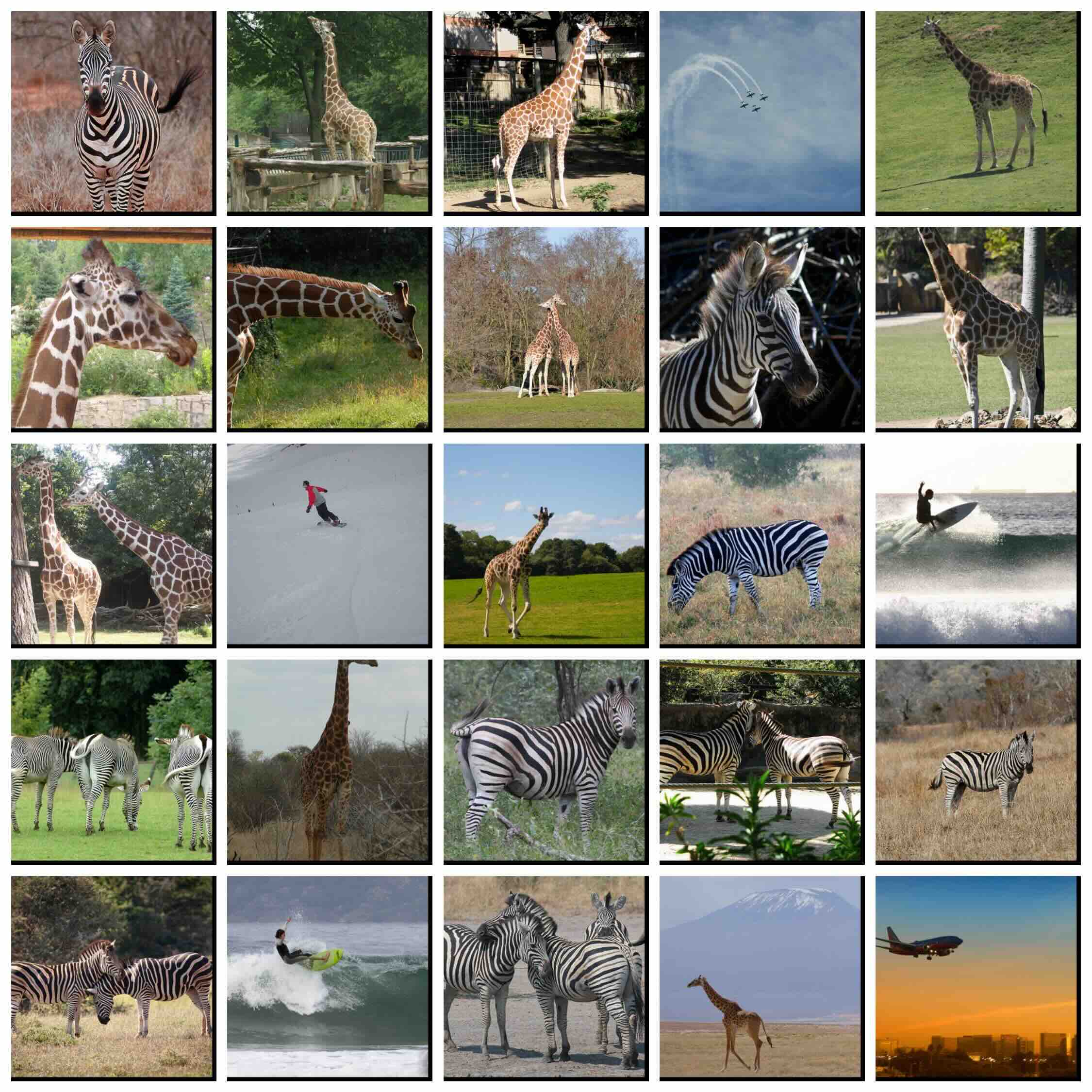}
    \caption{Top 25 maximally parcel-activating NSD images (based on ground truth) for a parcel in subject 2, left hemisphere.} 
    \label{fig:s2_top25_nsd_p309}
\end{figure}

\clearpage
\subsection{Compute used}
\label{app:compute}

We used CPU (AMD EPYC 7662), GPU (NVIDIA A40, L40), memory, and storage resources from an internal cluster. Storage for the entire project totals roughly 10TB. Training the model used roughly 3,000 GPU hours, 24,000 CPU core hours, and 32 GB per GPU hour. Running the remaining experiments used roughly 20,000 GPU hours, 160,000 CPU core hours, and 32 GB per GPU hour. The full project required more compute than these estimates due to failed experiments, experiments not included in the paper, and model iteration.

\end{document}